\documentclass{ws-procs9x6}

\usepackage{axodraw}

\def\beq{\begin{equation}}
\def\eeq#1{\label{#1}\end{equation}}
\def\eeqn{\end{equation}}
\def\beqa{\begin{eqnarray}}
\def\eeqa#1{\label{#1}\end{eqnarray}}
\def\eeqan{\end{eqnarray}}
\def\CR{\nonumber \\ }
\def\leqn#1{(\ref{#1})}

\newbox\charbox
\newbox\slabox
\def\sl#1{{      
        \setbox\charbox=\hbox{$#1$}
        \setbox\slabox=\hbox{$/$}
        \dimen\charbox=\ht\slabox
        \advance\dimen\charbox by -\dp\slabox
        \advance\dimen\charbox by -\ht\charbox
        \advance\dimen\charbox by \dp\charbox
        \divide\dimen\charbox by 2
        \raise-\dimen\charbox\hbox to \wd\charbox{\hss/\hss}
        \llap{$#1$}
}}

\def\mbar{\overline{\left|{\cal M}\right|}^2}
\def\heta{\hat{\eta}}
\def\stilde{\widetilde}

\def\stacksymbols #1#2#3#4{\def\theguybelow{#2}
    \def\vp{\lower#3pt}
    \def\sp{\baselineskip0pt\lineskip#4pt}
    \mathrel{\mathpalette\intermediary#1}}

\def\intermediary#1#2{\vp\vbox{\sp
     \everycr={}\tabskip0pt
     \halign{$\mathsurround0pt#1\hfil##\hfil$\crcr#2\crcr
              \theguybelow\crcr}}}

\def\gsim{\stacksymbols{>}{\sim}{2.5}{.2}}
\def\lsim{\stacksymbols{<}{\sim}{2.5}{.2}}

\begin{document}

\title{INTRODUCTION TO COLLIDER PHYSICS}

\author{MAXIM PERELSTEIN$^*$}

\address{Newman Laboratory of Elementary Particle Physics\\
	     Cornell University, Ithaca, NY 14853, USA \\
	     E-mail: mp325@cornell.edu}

\begin{abstract}
This is a set of four lectures presented at the Theoretical Advanced Study Institute (TASI-09) in June 2009. The goal of the lectures is to introduce students to some of the basic ideas and tools required for theoretical analysis of collider data. Several examples of Standard Model processes at electron-positron and hadron colliders are considered to illustrate these ideas. In addition, a general strategy for formulating searches for physics beyond the Standard Model is outlined. The lectures conclude with a brief survey of recent, ongoing and future searches for the Higgs boson and supersymmetric particles.

\end{abstract}


\bodymatter

\section{Introduction}

Our knowledge of the laws of physics in the sub-nuclear domain 
(at distance scales of about $10^{-13}$ cm and smaller) is for the most part derived from analyzing the outcomes of high-energy collisions of elementary particles. While the size and sophistication of each component of high-energy collision experiments have steadily grown, the basic experimental setup has remained unchanged since late 1960's. First, a particle accelerator uses a carefully designed combination of electric and magnetic fields to produce narrowly focused beams of energetic particles (typically electrons, protons, and their antiparticles). Then, two beams collide head-on, usually with equal and opposite momenta so that the center-of-mass frame of the colliding system coincides with the laboratory frame.\footnote{For about three decades before the advances in accelerator technology made it possible to steer two beams into a mid-air head-on collision, the principal technique was to accelerate a single beam and crash it into a stationary target. I will always consider a collider setup in these lectures, and all experimental data I will show are from colliders; however, in terms of theoretical interpretation, the main subject of these lectures, the differences between the collider and fixed-target setups are rather trivial.} The region where collisions occur (the ``interaction point") is surrounded by a set of particle detectors, which attempt to identify the particles coming out of the collision, and measure their energies and momenta. 

Since physics at subatomic distance scales is governed by laws of quantum mechanics, the outcome of each collision cannot, as a matter of principle, be known ahead of time; the best that any theory can do is to predict the {\it probabilities} of various possible outcomes. 
Modern collider experiments collect and analyze outcomes of huge number of collisions; the number of events with specified properties within the collected data set is proportional to the probability of such an event. Thus, the probability of a specific outcome of a collision (or a closely related quantity typically used in particle physics, the {\it cross section}) provide a natural bridge between theory and experiment. Narrowly speaking, one can say that a particle theorist's job is to infer laws of physics from experimentally measured cross sections. There is no known algorithm that can do this job. (This is lucky for us, since otherwise theorists could be replaced by computers!) There is, however, a well-developed formalism for predicting cross sections, given a Lagrangian quantum field theory. In practice, therefore, theoretical interpretation of collider data proceeds by picking a candidate theory, computing relevant cross sections within that theory, comparing with data, and (if the comparison does not work) moving on to the next candidate theory. For the last 30 years, the default ``leading candidate" theory has been the Standard Model (SM). As everyone knows, no statistically significant deviation from predictions of this theory  
has been observed so far, although at times mild inconsistencies with data have motivated theorists to try alternative candidates. It is, of course, equally well known to the TASI participants that strong theoretical reasons exist to expect that the SM hegemony will finally break down at the energy scales around a TeV, which will be explored experimentally for the first time by the Large Hadron Collider  
(LHC) in the next few years. Obtaining detailed quantitative predictions for the LHC experiments, both from the SM and from alternative candidate models, is a crucial task for theorists in the LHC era. The aim of these lectures is to introduce TASI students to some of the basic concepts and theoretical tools necessary to make such predictions.

\subsection{Definitions and Basics} 

Consider a collision of two elementary particles, $A$ and $B$, in the reference frame where the net momentum of the pair is zero. This frame is called the {\it center-of-mass frame}, or {\it c.o.m. frame} for short, since the center of mass of the system is at rest. In the case of $e^+e^-$ colliders, $A$ and $B$ are just the electron and the positron, and the c.o.m. frame coincides with the lab frame. For hadron colliders, $A$ and $B$ are partons (quarks or gluons), and the c.o.m. frame, which we will also call the {\it ``parton frame"} in this case, is generally moving along the collision axis with respect to the lab frame. In either case, we will neglect the masses of $A$ and $B$, since they are tiny compared to the energies we're interested in (of order 10 GeV and higher). By convention, we will choose the $z$ axis to lie along the direction of the $A$ momentum. The four-momenta of the colliding particles are 
\beq
p_A \,=\, (E,0,0,+E)\,,~~~p_B \,=\, (E,0,0,-E)\,.
\eeq{pin}
The total energy of the colliding system, the {\it center-of-mass energy}, is $E_{\rm cm} = 2E$. We will also frequently use the Mandelstam variable $s
=(p_A+p_B)^2 = E_{\rm cm}^2$. In the case of hadron colliders, the center-of-mass energy in a collision of two partons will be denoted by $\sqrt{\hat{s}}$, to distinguish it from the energy of the colliding hadron pair $\sqrt{s}$. 

In particle colliders, collisions actually take place between beams containing large number of particles. If two beams collide head-on, the number of collisions leading to a final state with particular characteristics (type of particles, their momenta, etc.) should be proportional to the number of particles in each beam, $N_A$ and $N_B$, and inversely proportional to the beams' cross-sectional area ${\cal A}$. The coefficient of proportionality is the {\it scattering cross section} for this particular final state:
\beq
\sigma \,=\, \frac{{\rm (Number~of~events)}\,\cdot {\cal A}}{N_AN_B}\,.
\eeq{xsec_def}
If beams collide at a frequency $f$ Hz, the rate $R$ (number of events of a particular kind recorded per second) can be written as
\beq
R \,=\, L \cdot \sigma\,,
\eeq{Rev}
where 
\beq
L \,=\, \frac{N_A N_B f}{A}\,
\eeq{Linst}
is the {\it instantaneous luminosity}. Simple as it looks, Eq.~\leqn{Rev} is a fundamental cornerstone of collider physics, and it is worth examining it more closely. The rate $R$ is measured directly by experimentalists.\footnote{Actually, what is measured is $R\cdot {\cal E}$, where ${\cal E}$ is the {\it detector efficiency}: the probability that an actual event with particular properties is identified as such by the detector. Efficiencies vary widely depending on the detector and the kind of process one is considering. In addition, the measured rate typically includes events that do {\it not} actually have the requested properties, but are mis-identified due to detector imperfections. In these lectures, we will mostly not be concerned with such detector effects, except for an occasional brief comment. An interested reader is referred to Eva Halkidakis' lectures at this school.}  The quantity $L$ (together with $E_{\rm cm}$) contains all the information about the accelerator needed to analyze the experiment. Experimental collaborations carefully monitor and record $L$, as a function of time. The experimentally measured value of the cross section is inferred from Eq.~\leqn{Rev}. This value can then be compared with the theoretically expected cross section.   

\renewcommand{\arraystretch}{1.4}
\begin{table}
\tbl{Recent and future energy-frontier particle colliders. (Parameters listed for the LHC and the ILC are design values.)}
{\begin{tabular}{|c|c|c|c|c|c|c|} \hline
  Name & Type & $\sqrt{s}$ (GeV) & $L_{\rm int}$ (pb$^{-1}$) & 
Years of & Detectors  & Location \\
 & & & & operation & & \\ \hline \hline
 LEP & $e^+e^-$ & 91.2 (LEP-1) & $\approx 200$ (LEP-1) & 1989-95 (LEP-1) & ALEPH, OPAL, & CERN \\
      &  & 130-209 (LEP-2) & $\approx 600$ (LEP-2) & 1996-2000 (LEP-2) & DELPHI, L3 & \\
 SLC & $e^+e^-$ & 91.2 & 20 & 1992-98 & SLD & SLAC \\
 HERA & $e^\pm p$ & 320 & 500 & 1992-2007 & ZEUS, H1 & DESY \\
Tevatron & $p\bar{p}$ & 1800 (Run-I) & 160 (Run-I) & 1987-96 (Run-I) & CDF, D{\O} & FNAL \\
& & 1960 (Run-II) & 6 K (Run-II, 06/09) & 2000-??? (Run-II) & & \\
LHC & $pp$ & 14000 & 10 K/yr ("low-L") & 2010? - 2013?&
ATLAS, CMS & CERN \\
& & & 100 K/yr ("high-L") &2013?? - 2016???& & \\ 
ILC & $e^+e^-$ & 500-1000 & 1 M??? & ??? & ??? & ??? \\
 \hline
\end{tabular} }
\label{tab:colliders}
\end{table}
\renewcommand{\arraystretch}{1.0}

Throughout the lectures, we will contrast theoretical predictions with data from recent and ongoing experiments at energy-frontier colliders. Table~\ref{tab:colliders} shows the basic parameters of these colliders, along with the upcoming LHC and the proposed next-generation electron-positron collider, the International Linear Collider (ILC).\footnote{For lack of time, I will not be able to discuss results from recent lower-energy, ``luminosity-frontier" collider experiments, such as CLEO, BaBar, and Belle.} It is important to keep in mind that, for hadron colliders, the listed center-of-mass energy corresponds to the colliding (anti)protons. Since high-energy processes are initiated by partons, which only carry a fraction of the proton momentum, the energy scales that can be probed at a hadron collider are substantially lower than this energy, typically by factors of $3-10$ depending on the process. Electron-positron colliders, on the other hand, are able to explore many reactions at energy scales extending all the way to their nominal $\sqrt{s}$. The luminosity values shown in the table are the {\it integrated} luminosities, $L_{\rm int}=\int L dt$ over the lifetime of the experiment. The table also lists the detectors at each collider. Detector names coincide with the names of collaborations of physicists operating them, and are frequently used to refer to the data published by these collaborations.    

Computing and interpreting cross sections will be our main focus. It is clear from its definition, Eq.~\leqn{xsec_def}, that the cross section has cgs units of cm$^2$. A unit typically used in experimental nuclear and particle physics is 1 {\it barn} $=10^{-24}$ cm$^2$. In ``theory units", $c=\hbar=1$, the natural unit for cross section is GeV$^{-2}$; the conversion factor is
\beqa
1~{\rm bn} &=& 2568~{\rm GeV}^{-2}\,,\CR
1~{\rm GeV}^{-2} &=& 3.894\cdot 10^{-4}\,~{\rm bn}.
\eeqa{bn_GeV_convert}
To get a very rough estimate of cross sections expected in particle physics experiments, we can use dimensional analysis: away from thresholds and resonances, the only energy scale in a collision of two massless particles is $E_{\rm cm}$, and we should expect the (total) scattering cross section to behave roughly as 
\beq
\sigma \sim \frac{1}{E_{\rm cm}^2}\,.
\eeq{xsec_scaling}
A similar result (larger by $\pi$) is obtained by replacing the colliding particles with classical ``billiard balls" of radius equal to their Compton wavelength $\lambda \sim 1/E$, and taking their geometric cross section as an estimate. The geometric cross section also coincides with the upper bound on the total inelastic cross section (assuming $s$-wave scattering) from unitarity considerations. The cross sections for specific processes are typically lower, by an order of magnitude or more, than this bound: For example, the $e^+e^-\to Z$ cross section on resonance ($\sqrt{s}=M_Z$) is about 40 nb, compared to $\sigma_{\rm geom} = \pi/M_Z^2 \approx 2500$ nb. 
The decrease of cross sections with energy has an important implication for accelerator design: Colliders operating at higher center-of-mass energies must also have higher luminosity, adding to the technical challenges of expanding the high-energy frontier. This trend is clear in Table~\ref{tab:colliders}. 

The ``master formula" for evaluating the cross section and kinematic distributions for a $2\to N$ scattering process is
\beq
d\sigma \,=\, \frac{1}{2s}\,\left( \prod_{i=1}^N\,\frac{d^3p_i}{(2\pi)^3}\,\frac{1}{2E_i}\right)\,\cdot\, (2\pi)^4 \delta^4(p_A+p_B-\sum p_i)\,\cdot\, \left| {\cal M} (p_A,p_B\to \{ p_i \} \right|^2\,,
\eeq{master}
where ${\cal M}$ is the invariant matrix element, a.k.a. {\it scattering amplitude}, and $p_i = (E_i, {\bf p}_i)$ are the 4-momenta of the final-state particles. Note that ${\cal M}$ contains all information specific for the process under consideration (such as coupling constant dependence, etc.), whereas all other ingredients are simply kinematic factors common for any $2\to N$ process. While Eq.~\leqn{master} is written in the center of mass frame of the colliding particles, it is in fact invariant under boosts parallel to the collision axis. This feature will be important when hadron collisions are considered. If the colliding beams are unpolarized, one needs to average the quantity $\left|{\cal M}\right|^2$ over all possible initial-state polarizations. If the beams are polarized (this was the case at the SLC, and may be implemented at the ILC), an appropriately weighted average should be computed instead. In addition, if the final-state particles have spin, $\left|{\cal M}\right|^2$ should typically be summed over all possible spin states, since no collider detector is capable of detecting spins of individual particles. (Exception occurs when the final-state particles decay promptly, in which case the angular distribution of their decay products may carry information about their polarization state.) The appropriately averaged and/or summed $|$scattering amplitude$|^2$
will be denoted by $\mbar$. 

The number of independent kinematic variables in a $2\to N$ process is $3N-4$. In practice, the initial state is always symmetric under rotation around the collision axis, and no physical observable can depend on the overall azimuthal coordinate, leaving $3N-5$ physical variables. The simplest case, most commonly encountered in practice, is $2\to 2$ scattering. The only observable not constrained by energy and momentum conservation is the scattering angle $\theta$, which by convention is defined as the angle between the 3-momenta of particles $A$ and $1$. The differential cross section is given by
\beq
\frac{d\sigma}{d\cos\theta}\,=\,\begin{cases} \frac{1}{16\pi}\,\frac{|{\bf p}_1|}{s^{3/2}}\, & \text{ if $\sqrt{s}> m_1+m_2$;} \\ 0 & \text{otherwise,} \end{cases}
\eeq{2to2}
where 
\beq
|{\bf p}_1|=\frac{1}{2}\sqrt{\frac{(s-m_1^2-m_2^2)^2-4m_1^2m_2^2}{s}}\,.
\eeq{momentum}
In the most common case of equal masses in the final state, $m_1=m_2=m$, this formula further simplifies (for $\sqrt{s}>2m$) to
\beq
\frac{d\sigma}{d\cos\theta}\,=\,\frac{1}{32\pi s}\,\sqrt{1-\frac{4m^2}{s}}\,\mbar\,.
\eeq{2to2m} 
Note that the square-root factor is simply the velocity of the final-state particles (in units of $c$). The quantity $\mbar$ is often expressed in terms of the {\it Mandelstam variables}, Lorentz-invariant (scalar) bilinears of the 4-momenta of incoming and outgoing particles. In the case of $2\to 2$ scattering, these are
\beqa
s &=& (p_A+p_B)^2\,,\CR
t &=& (p_A-p_1)^2\,,\CR
u &=& (p_A-p_2)^2\,.
\eeqa{Mandel}
They are not independent: it can be easily shown that $s+t+u=m_1^2+m_2^2$. The Mandelstam variables are related to the scattering angle: for example, if $m_1=m_2=0$, we simply have
\beq
t \,=\, - \frac{s}{2}\,(1-\cos\theta)\,,~~~u \,=\, - \frac{s}{2}\,(1+\cos\theta)\,. 
\eeq{Mtotheta}
The main advantage of using Mandelstam variables comes in applications of crossing symmetry to relate processes such as, for example,  electron-positron annihilation $e^+e^-\to\gamma\gamma$ and Compton scattering $e^-\gamma\to e^-\gamma$. They are also convenient for analyzing hadron collisions, being invariant under boosts connecting the parton and lab reference frames. 

\section{Electron-Positron Collisions}
\label{sec:epem}

In this Lecture, we will study a few examples of reactions initiated by electron-positron collisions, and use them to illustrate some fundamental concepts and issues central to the field. Since colliding particles are elementary, $e^+e^-$ collisions are somewhat easier to analyze than collisions between hadrons, which will be discussed in Lecture~\ref{sec:hadrons}.

\subsection{Muon Pair-Production}
\label{sec:eemumu}

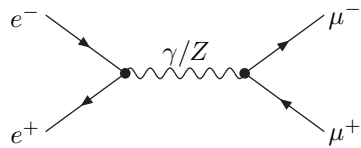
\begin{figure}
\begin{center}
\begin{picture}(130,65)(0,0)
\SetOffset(20,-65)
\ArrowLine(1,109)(31,87)
\ArrowLine(31,87)(1,65)
\Photon(31,87)(76,87){2}{6}
\ArrowLine(76,87)(106,109)
\ArrowLine(106,65)(76,87)
\Vertex(31,87){2}
\Vertex(76,87){2}
\Text(-1,109)[cr]{$e^-$}
\Text(-1,65)[cr]{$e^+$}
\Text(108,109)[cl]{$\mu^-$}
\Text(108,65)[cl]{$\mu^+$}
\Text(54,100)[tc]{$\gamma/Z$}
\end{picture}
\end{center}
\caption{Leading-order (tree-level) Feynman diagrams contributing to the process $e^+e^-\to\mu^+\mu^-$.}
\label{fig:eemm}
\end{figure}

We start with the process $e^+e^-\to\mu^+\mu^-$. At tree level, only two diagrams contribute, see Fig.~\ref{fig:eemm}, making it possibly the simplest $2\to 2$ reaction in the SM - the {\tt "hello world"} example of collider physics. 
Most (probably all) TASI students would have calculated the cross section of this reaction in their Quantum Field Theory (QFT) classes, probably using four-component (Dirac) notation and trace technology to perform spin sums. An alternative is to use two-component (Weyl) fermions, and to evaluate the scattering amplitudes for particles in definite helicity eigenstates. This method provides more insight into the physics of the process, and becomes especially valuable when weak interactions are considered. Let us outline the calculation.

\subsubsection{Muon Pair-Production in QED}

To begin with, let us only consider the diagram with the virtual photon exchange, ignoring the $Z$. (This would be a good approximation at low energies, $\sqrt{s}\ll M_Z$.) 
In two-component notation, electrons and  positrons are described by a pair of two-component (Weyl) spinor fields, $e_L$ and $e_R$. The subscript $L/R$ denotes the field's {\it chirality}, which determines its transformation properties under the Lorentz group. The QED Lagrangian in this notation is
\beq
{\cal L} \,=\, i e_R^\dagger \sigma^\mu D^R_\mu e_R \,+\,i e_L^\dagger \bar{\sigma}^\mu D^L_\mu e_L + m_e \left( e_R^\dagger e_L + e_L^\dagger e_R\right)\, .
\eeq{L_QED}
Here  $D^L_\mu=D^R_\mu=\partial_\mu - ieA_\mu$ are covariant derivatives, and 
\beq
\sigma\,=\,(1, \vec{\sigma}),~~~\bar{\sigma}\,=\,(1, -\vec{\sigma})\,,
\eeq{sigmas}
where $\vec{\sigma}$ is a three-vector consisting of the usual three Pauli matrices. 

A particle's {\it helicity} is defined as the projection of its spin on the direction of its motion. Since helicity of a free particle is conserved, we can choose to describe a scattering process in the basis of one-particle in- and out-states of definite helicity. For electrons and positrons, these states are $|e^\pm_h\rangle$, where the superscript denotes the particle's electric charge and the subscript $h=\pm 1/2$ its helicity:
\beq
\frac{{\bf p}\cdot {\bf S}}{\left|{\bf p}\right|} \,| e^\pm_h\rangle \,=\, h | e^\pm_h\rangle \,.
\eeq{hel_def}  
States of positive helicity $|e^\pm_+\rangle$ are often referred to as ``right-handed", while states of negative helicity $|e^\pm_-\rangle$ are called ``left-handed". We will use this nomenclature, and replace the subscripts $+\to r$, $-\to l$.

In the limit of zero electron mass, the Weyl fields $e_L$ and $e_R$ are completely decoupled in the Lagrangian~\leqn{L_QED}. After quantization, each field contains creation/annihilation operators for states of specific helicity only: for example, 
\beqa
e_L\,|0\rangle &\sim &|e^+_r\rangle\,,~~~ e_L^\dagger\,|0\rangle \,\sim\, |e^-_l\rangle\,,\CR
e_R\,|0\rangle &\sim &|e^+_l\rangle\,,~~~ e_R^\dagger\,|0\rangle \,\sim \,|e^-_r\rangle\,.
\eeqa{crann}
Note the relation between helicity of a state and chirality of the field creating it:
\beqa
{\rm Particle:} & & {\rm helicity}~=~{\rm chirality}\,;\CR
{\rm Antiparticle:} & & {\rm helicity}~= -~{\rm chirality}\,.
\eeqa{hel_chi}
It follows that helicity eigenstates are simply the solutions to Weyl equations of motion:
\beq
i\sigma^\mu \partial_\mu \psi_R = 0,~~~i\bar{\sigma}^\mu \partial_{\mu} \psi_L = 0\,.
\eeq{Weyleom}
Explicitly, the solutions have the form (up to normalization constants)
\beq
e_h^-  \sim e^{-ip\cdot x} \xi_h\,,~~e_h^+ \sim e^{+ip\cdot x} \xi_{-h}\,, 
\eeq{Weyl_soln}
where 
\beq
\xi_r=\exp\left[\frac{i}{2}\vec{\sigma}\cdot\vec{\omega} \right]\,\cdot\,\left( \begin{array}{c} 1\\0\end{array}\right)\,,~~~ \xi_l=\exp\left[\frac{i}{2}\vec{\sigma}\cdot\vec{\omega} \right]\,\cdot\,\left( \begin{array}{c} 0\\1\end{array}\right)\,,
\eeq{xis}
where $\vec{\omega}$ is the rotation from the $+z$ direction to the direction of the momentum ${\bf p}$. It is straightforward to obtain the Feynman rules in two-component language: For example, an incoming electron line of helicity $h$ gives $\sqrt{2E}\xi_h$,
an incoming positron of helicity $h$ gives $\sqrt{2E}\xi_{-h}^\dagger$,
the $e^-_r e^+_l\gamma$  vertex is $ie\sigma_\mu$, the 
$e^-_l e^+_r\gamma$  vertex is $ie \bar{\sigma}_\mu$, etc. Note that $e^-_r e^+_r$ and $e^-_l e^+_l$ vertices do not exist, since there is no coupling between $e_L$ and $e_R$ fields in the Lagrangian. 

The same construction describes the electromagnetic interactions of any other fermion, as long as its mass can be neglected: in particular,  it can be applied to the $e^+e^-\to\mu^+\mu^-$ scattering in the limit $\sqrt{s}\gg m_\mu$. Using the Feynman rules above, and the standard photon propagator, yields the {\it helicity amplitudes}
\beqa
{\cal M} (e^-_le^+_r\to\mu^-_l\mu^+_r) = {\cal M} (e^-_re^+_l\to\mu^-_r\mu^+_l) &=& -e^2\,(1-\cos\theta)\,,\CR
{\cal M} (e^-_le^+_r\to\mu^-_r\mu^+_l) = {\cal M} (e^-_re^+_l\to\mu^-_l\mu^+_r)&=& -e^2\,(1+\cos\theta)\,,
\eeqa{amps_mu}
with all other helicity configurations giving vanishing amplitudes. Using Eq.~\leqn{2to2m}, this yields the differential cross section
\beq
\frac{d\sigma}{d\cos\theta}\,=\,\frac{\pi\alpha}{2s}\,\left(1+\cos^2\theta \right)\,,
\eeq{eemm_xsec}
where we introduced the fine-structure constant $\alpha=e^2/(4\pi)$.

\subsubsection{Scalar Muon}

It is instructive to repeat the above calculation replacing the muon with a {\it scalar} (spin-0) particle $\tilde{\mu}$ of the same mass and electric charge. (Such ``scalar muons", or smuons, are actually predicted by supersymmetric theories.) The helicity amplitudes (in the limit $\sqrt{s}\gg m_{\tilde{\mu}}$) are 
\beq
{\cal M} (e^-_le^+_r\to\tilde{\mu}^-\tilde{\mu}^+) = {\cal M} (e^-_re^+_l\to\tilde{\mu}^-\tilde{\mu}^+) \,=\, -e^2\,\sin\theta\,.
\eeq{amps_smu} 

{\bf Homework Problem 1:} Derive the amplitudes~\leqn{amps_smu}.
~\\

\begin{figure}
\begin{center}

\begin{picture}(120,120)(0,0)
\SetOffset(20,20)
\LinAxis(0,0)(100,0)(4,5,3,0,1.0)
\Line(0,0)(0,100)
\Line(0,100)(100,100)
\Line(100,100)(100,0)
\Text(0,-2)[tc]{$-1$}
\Text(100,-2)[tc]{$+1$}
\Text(50,-4)[tc]{$\cos\theta$}
\DashCurve{(0,0)(10,27)(20,48)(30,63)(40,72)(50,75)(60,72)(70,63)(80,48)(90,27)(100,0)}{3}
\Curve{(0,75)(10,61.5)(20,51)(30,43.5)(40,39)(50,37.5)(60,39)(70,43.5)(80,51)(90,61.5)(100,75)}
\Text(12,64)[bl]{$\mu$}
\Text(10,25)[tl]{$\tilde{\mu}$}
\Text(-3,90)[cr]{$\frac{d\sigma}{d\cos\theta}$}
\end{picture}

\end{center}
\caption{Angular distributions for muons (solid line) and scalar muons (dashed line) produced in $e^+e^-$ collisions. (Normalization is arbitrary.)}
\label{fig:dists}
\end{figure}
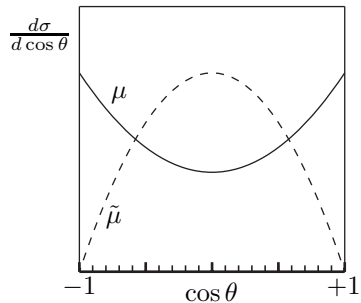

\noindent This yields the differential cross section
\beq
\frac{d\sigma}{d\cos\theta}\,=\,\frac{\pi\alpha}{4s}\,\sin^2\theta\,.
\eeq{eesmsm_xsec}
The angular distributions predicted for muons and smuons are plotted in Fig.~\ref{fig:dists}. Comparing these distributions with data allows one to distinguish between the two possible spin assignments, providing a determination of the muon spin. (Note that spins of elementary particles are {\it not} directly observed by the detector, so spins can in fact only be inferred by indirect means such as this.) No evidence for spin-0 muon has been seen in the data, indicating that supersymmetry, if it is indeed a symmetry of nature, must be broken to lift the muon-smuon mass degeneracy.

\begin{figure}
\begin{center}

\begin{picture}(300,65)(0,0)
\SetOffset(20,-50)
\Vertex(1,90){2}
\LongArrow(1,90)(31,90)
\Vertex(81,90){2}
\LongArrow(81,90)(51,90)
\DashLine(-10,90)(111,90){3}
\LongArrow(106,90)(111,90)
\Line(25,80)(10,80)
\Line(25,76)(10,76)
\Line(6,78)(14,82)
\Line(6,78)(14,74)
\Line(75,80)(60,80)
\Line(75,76)(60,76)
\Line(56,78)(64,82)
\Line(56,78)(64,74)
\Text(107,87)[tc]{$z$}
\Text(81,95)[br]{$e^+_r$}
\Text(3,95)[bl]{$e^-_l$}
\Text(41,67)[tc]{$J_z=-1$}

\Vertex(151,90){2}
\LongArrow(151,90)(181,90)
\Vertex(231,90){2}
\LongArrow(231,90)(201,90)
\DashLine(140,90)(271,90){3}
\LongArrow(266,90)(271,90)
\Line(175,80)(160,80)
\Line(175,76)(160,76)
\Line(179,78)(171,82)
\Line(179,78)(171,74)
\Line(225,80)(210,80)
\Line(225,76)(210,76)
\Line(229,78)(221,82)
\Line(229,78)(221,74)
\Text(267,87)[tc]{$z$}
\Text(231,95)[br]{$e^+_l$}
\Text(153,95)[bl]{$e^-_r$}
\Text(191,67)[tc]{$J_z=+1$}

\end{picture}

\end{center}
\caption{Helicity configurations of the $e^+e^-$ pair leading to non-vanishing tree-level scattering amplitudes for muon and smuon production. Thin arrows show the direction of particles' momenta, while thick arrows indicate their helicities.}
\label{fig:arrows}
\end{figure}
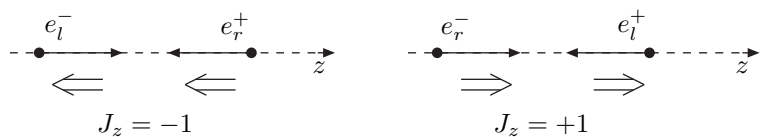

The most striking feature of the distributions in Fig.~\ref{fig:dists} is that the smuons cannot be produced in forward and backward directions ($\cos\theta=\pm 1$), while muons can. This can be easily traced to conservation of angular momentum. The $z$-component of the total angular momentum of the colliding system is $J_z=\pm 1$; all helicity amplitudes corresponding to $J_z=0$ vanish (see Fig.~\ref{fig:arrows}). Since $J_z$ is conserved, the same must be true in the final state; but for scalar particles, $J_z$ can only be contributed by the {\it orbital} angular momentum $L_z$. When particles move along the $z$ axis, $L_z=0$, and angular momentum cannot be conserved; thus, scattering along this direction is forbidden for smuons.

\subsubsection{Including the $Z$ Exchanges}
\label{sec:Zmuons}

Apart from the somewhat more complicated form of matrix elements, two conceptually important new features arise when the diagram involving the $Z$ boson in Fig.~\ref{fig:eemm} is included. 

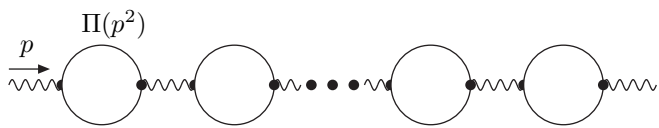
\begin{figure}
\begin{center}

\begin{picture}(260,55)(0,0)
\SetOffset(-40,-65)
\Photon(46,87)(66,87){2}{4}
\Vertex(66,87){2}
\BCirc(81,87){15}
\Vertex(96,87){2}
\Photon(96,87)(116,87){2}{4}
\Vertex(116,87){2}
\BCirc(131,87){15}
\Vertex(146,87){2}
\Photon(146,87)(156,87){2}{2}
\Vertex(160,87){2}
\Vertex(168,87){2}
\Vertex(176,87){2}
\Photon(180,87)(190,87){2}{2}
\Vertex(190,87){2}
\BCirc(205,87){15}
\Vertex(220,87){2}
\Photon(220,87)(240,87){2}{4}
\Vertex(240,87){2}
\BCirc(255,87){15}
\Vertex(270,87){2}
\Photon(270,87)(290,87){2}{4}
\LongArrow(46,93)(60,93)
\Text(53,98)[bc]{$p$}
\Text(86,105)[bc]{$\Pi(p^2)$}
\end{picture}

\end{center}
\caption{Vacuum polarization corrections to the $Z$ propagator. Each circle denotes the sum of one-loop diagrams with quarks, leptons, $W$ and Higgs bosons circulating in the loop.}
\label{fig:Zprop}
\end{figure}

\begin{figure}
\begin{center}
\psfig{file=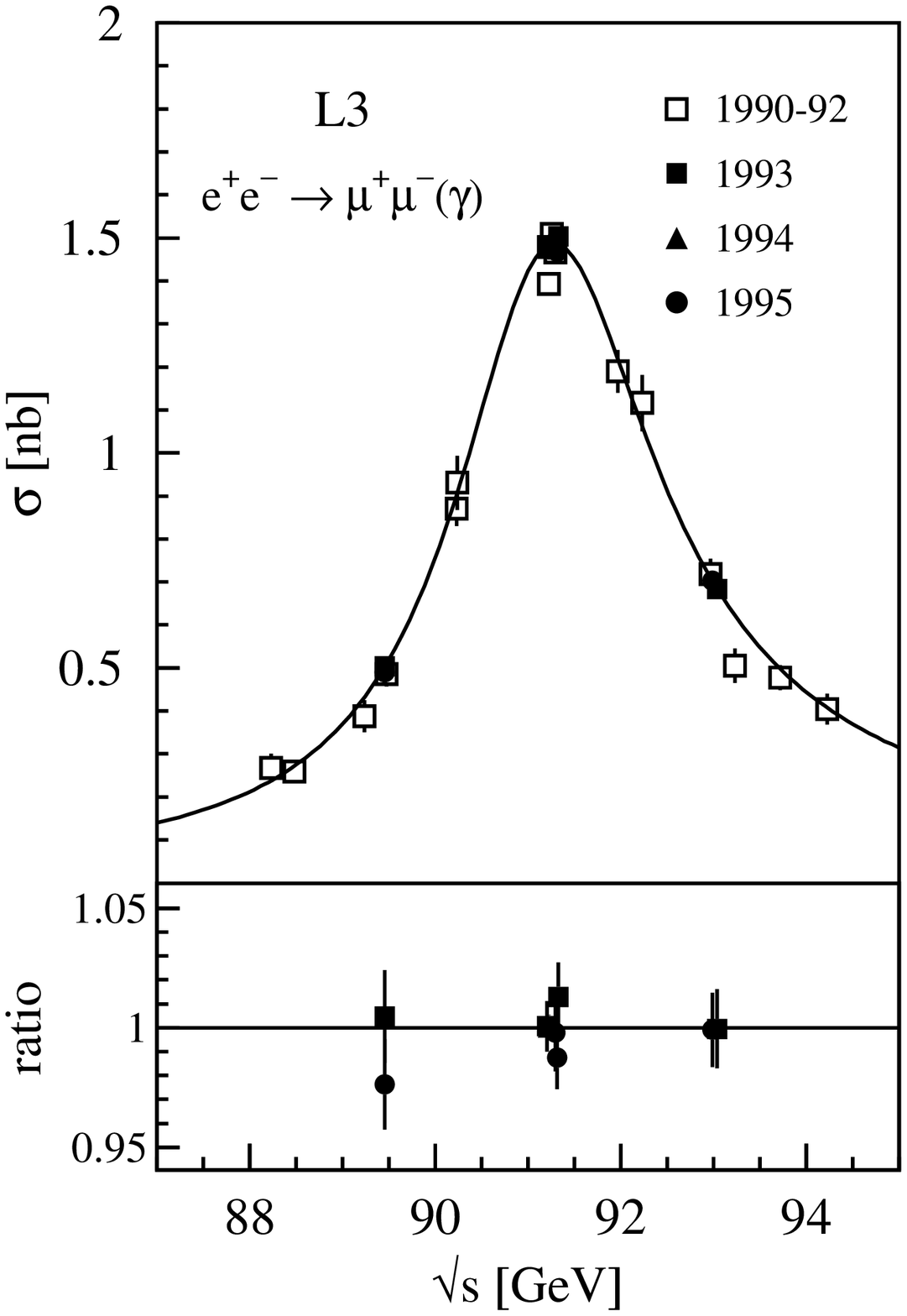,width=7.5cm}
\hskip1cm
\psfig{file=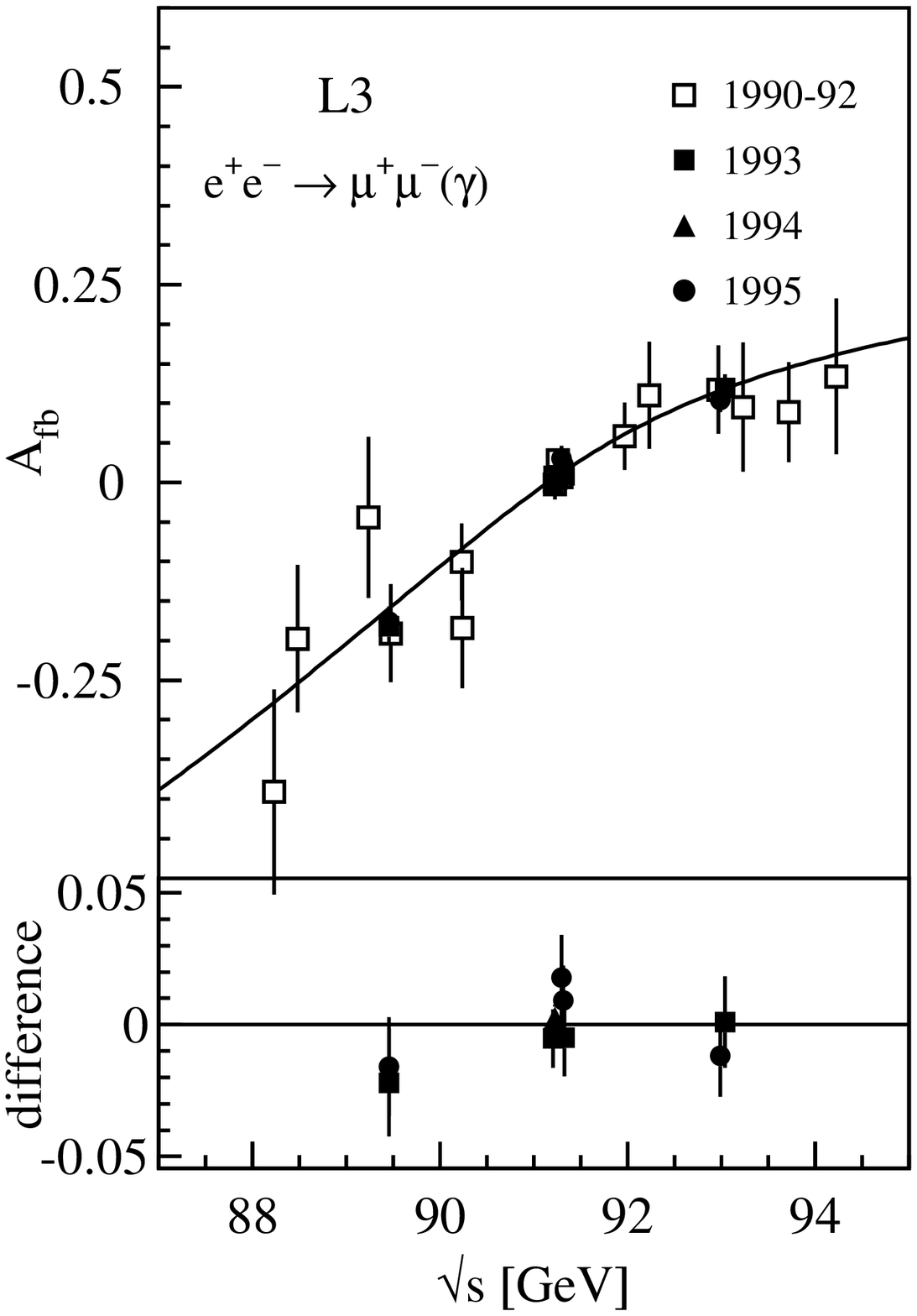,width=7.5cm}
\end{center}
\vskip-0.5cm
\caption{Cross section (left panel) and forward-backward asymmetry (right panel) of muon-pair production at LEP, measured by the L3 collaboration, compared with the SM prediction (solid lines). From Ref.~\cite{L3_mumu}.}
\label{fig:ALEPH1}
\end{figure}

First, the $Z$ propagator is proportional to $(s-M_Z^2)^{-1}$, which is infinite when $\sqrt{s}=M_Z$ (as was the case at LEP-1 and SLC). This infinity is in fact a feature of leading-order (``tree-level") perturbation theory, and is automatically removed when radiative corrections (``quantum loops") are included. More specifically, consider the vacuum polarization diagrams shown in Fig.~\ref{fig:Zprop}. They can be included by replacing
\beq
\frac{1}{p^2-M_Z^2} \,\longrightarrow\,\frac{1}{p^2-M_{Z,0}^2-\Pi(p^2)}
\eeq{Zprop1} 
in the $Z$ propagator, where $\Pi(p^2)$ is the one-loop vacuum polarization. We introduced the notation $M_{Z,0}$ for the bare (Lagrangian) $Z$ mass, to distinguish it from the physical (pole) mass $M_Z$. Note that to derive Eq.~\leqn{Zprop1}, an infinite series of diagrams in Fig.~\ref{fig:Zprop} had to be included and summed; this is not surprising, since the infinity in the tree-level calculation should be interpreted as a sign of a breakdown in perturbation theory. The physical $Z$ mass is the solution to $M_Z^2 - M_{Z,0}^2 -
~{\rm Re}~\Pi(M_Z^2)=0$. For $p^2\approx M_Z^2$, we can expand
\beqa
p^2-M_{Z,0}^2-\Pi(p^2) &\approx& p^2-M_{Z,0}^2-\left(~{\rm Re}~\Pi(M_Z^2) + \frac{d}{dp^2}~{\rm Re}~\Pi(M_Z^2)\,\,(p^2-M_Z^2)\right) +i~{\rm Im}~\Pi(M_Z^2)\,\CR &=& \left( 1 + \frac{d}{dp^2}~{\rm Re}~\Pi(M_Z^2)\,\right)\,(p^2-M_Z^2) + i~{\rm Im}~\Pi(M_Z^2) \CR &=& {\cal Z}^{-1}(p^2-M_Z^2) + i~{\rm Im}~\Pi(M_Z^2)\,,
\eeqa{prop_exp}
where in the last line we introduced the field strength renormalization factor ${\cal Z}$. Note that the propagator no longer blows up at $s=M_Z^2$, as long as the imaginary part of the vacuum polarization is non-zero. This is in fact guaranteed by the optical theorem, which relates Im~$\Pi$ to the {\it total decay width} $\Gamma_Z$ of the $Z$ boson:
\beq
{\rm Im}~\Pi(M_Z^2)\,=\,-\frac{{\cal Z}^{-1}}{2}\,\sum_{f_1,f_2} \int \frac{d^3p_1}{(2\pi)^3}\frac{1}{2E_1}\frac{d^3p_2}{(2\pi)^3}\frac{1}{2E_2}\,\overline{|{\cal M}(Z\to f_1 f_2)|}^2\,=\,-{\cal Z}^{-1} M_Z \Gamma_Z\,,
\eeq{optical} 
where the sum runs over all possible two-body decay channels of the $Z$. To summarize, the $Z$ propagator near the pole (in the 't Hooft-Feynman gauge) can be approximated by 
\beq
\frac{-ig_{\mu\nu}{\cal Z}}{p^2-M_Z^2+iM_Z\Gamma_Z}\,.
\eeq{Zprop2}
If the diagram with the photon is ignored (which is a reasonable approximation near the $Z$ pole), the cross section is proportional to
\beq
\left| \frac{1}{s-M_Z^2+iM_Z\Gamma_Z} \right|^2 = \frac{1}{(s-M_Z^2)^2 + M_Z^2 \Gamma_Z^2}\,.
\eeq{BW}
This is the famous {\it Breit-Wigner function}. Since the above discussion did not depend on any specific features of the $Z$ boson, the same function can be used to approximate the dependence of any cross section on $s$ in the neighborhood of a resonance of a given mass and width. This behavior is in excellent agreement with experiment, as shown in Fig.~\ref{fig:ALEPH1}. (The data is in fact precise enough that the contribution of the photon exchange diagram must be included even at the $Z$ peak; this has been done in the figure.)

The second interesting feature brought in by the $Z$ is violation of two discrete symmetries, charge conjugation (C) and parity (P). In the two-component language, $Z$ coupling to electrons has the form
\beq
{\cal L} \,=\, \left( g_L e_L^\dagger \bar{\sigma}^\mu e_L + g_R e_R^\dagger \sigma^\mu e_R \right) \,Z_\mu\,,
\eeq{Zcoupl}
with
\beq
g_L = \frac{e}{c_ws_w}\,\left( -\frac{1}{2}+s_w^2\right),~~~g_R = \frac{es_w}{c_w}\,,
\eeq{Zcoupl1}
where $s_w, c_w$ are the sine and cosine of the Weinberg angle ($s^2_w=0.231$). The fact that $g_L\not=g_R$ signals C and P non-conservation. The $Z$ coupling to the muon has an identical structure, and the $Z$-exchange contribution to the helicity amplitudes 
for $e^+e^-\to\mu^+\mu^-$ is given by
\beqa
{\cal M} (e^-_le^+_r\to\mu^-_l\mu^+_r) &=& -g_L^2 (1+\cos\theta)\,f_{BW}(s)\,,\CR
{\cal M} (e^-_re^+_l\to\mu^-_r\mu^+_l) &=& -g_R^2 (1+\cos\theta)\,f_{BW}(s)\,,\CR
{\cal M} (e^-_le^+_r\to\mu^-_r\mu^+_l) &=& {\cal M} (e^-_re^+_l\to\mu^-_l\mu^+_r) \,=\, -g_L g_R
(1-\cos\theta)\,f_{BW}(s)\,,
\eeqa{Zamps}
where
\beq
f_{BW}(s) \,=\, \frac{s}{s-M_Z^2 + iM_Z \Gamma_Z}\,.
\eeq{fBW}
Summing and averaging over spins (and ignoring, for simplicity, the photon exchange contribution) yields the cross section
\beq
\frac{d\sigma}{d\cos\theta}\,=\,\frac{|f_{BW}|^2}{32\pi s}\,\left[
\left(g_L^4 + g_R^4 \right) (1+\cos\theta)^2 \,+\, 2g_L^2g_R^2 (1-\cos\theta)^2\right]\,.
\eeq{Zxsec}
The cross section contains a term proportional to $\cos\theta$, so that the angular distribution is no longer symmetric under the parity transformation $\theta\to\pi-\theta$. A useful way to quantify this is to introduce the {\it forward-backward asymmetry} via
\beq
A_{FB} \,=\, \frac{\sigma(\cos\theta>0)-\sigma(\cos\theta<0)}{\sigma(\cos\theta>0)+\sigma(\cos\theta<0)}\,.
\eeq{Afb_def}
It is easy to show using Eq.~\leqn{Zxsec} that $A_{FB}\propto 
(g_L^2-g_R^2)^2$. Once the photon exchange diagrams are included, 
$A_{FB}$ becomes a function of $s$, since the relative contributions of the P-conserving photon exchange and P-violating $Z$ exchange depend on $s$. Experimental measurement of $A_{FB}$ at LEP, along with the SM expectation, are shown in Fig.~\ref{fig:ALEPH1}.

\subsection{Initial-State Radiation}
\label{sec:ISR}


Not all collisions occurring at an electron-positron collider with the nominal c.o.m. energy $\sqrt{s}$ in fact have that energy; in some cases, the energy is lowered by an emission of a photon (or multiple photons) by the electron, the positron, or both, just ahead of the collision. This process is called {\it initial state radiation}, and it is worth a closer look: In addition to being important in its own right, it serves as a portal into the topic of treatment and interpretation of infrared (soft and collinear) divergences in applications of quantum field theory to collider physics. These divergences are behind many of the most challenging issues in the field, from both conceptual and technical standpoints.  

\begin{figure}
\begin{center}

\begin{picture}(290,75)(0,0)
\SetOffset(20,-45)
\Line(1,109)(16,98)
\Line(16,98)(31,87)
\Photon(16,98)(38,120){2}{6}
\Line(31,87)(1,65)
\Photon(31,87)(76,87){2}{6}
\Vertex(16,98){2}
\Line(76,87)(106,109)
\Line(106,65)(76,87)
\Vertex(31,87){2}
\Vertex(76,87){2}
\Text(-1,109)[cr]{$e^-$}
\Text(-1,65)[cr]{$e^+$}
\Text(40,120)[cl]{$\gamma$}
\Text(108,109)[cl]{$\mu^-$}
\Text(108,65)[cl]{$\mu^+$}
\Text(54,100)[tc]{$\gamma/Z$}
\Text(54,54)[cc]{(a)}

\Line(151,109)(181,87)
\Photon(166,76)(188,54){2}{6}
\Line(166,76)(181,87)
\Line(166,76)(151,65)
\Photon(181,87)(226,87){2}{6}
\Vertex(166,76){2}
\Line(226,87)(256,109)
\Line(256,65)(226,87)
\Vertex(181,87){2}
\Vertex(226,87){2}
\Text(204,54)[cc]{(b)}

\end{picture}
\begin{picture}(290,75)(0,0)
\SetOffset(20,-45)

\Line(1,109)(31,87)
\Line(31,87)(1,65)
\Photon(31,87)(76,87){2}{6}
\Vertex(91,98){2}
\Line(76,87)(91,98)
\Line(91,98)(106,109)
\Photon(91,98)(121,98){2}{6}
\Line(106,65)(76,87)
\Vertex(31,87){2}
\Vertex(76,87){2}
\Text(54,54)[cc]{(c)}

\Line(151,109)(181,87)
\Photon(241,76)(271,76){2}{6}
\Line(151,65)(181,87)
\Photon(181,87)(226,87){2}{6}
\Vertex(241,76){2}
\Line(226,87)(256,109)
\Line(256,65)(226,87)
\Vertex(181,87){2}
\Vertex(226,87){2}
\Text(204,54)[cc]{(d)}

\end{picture}

\end{center}
\caption{Leading-order (tree-level) Feynman diagrams contributing to the process $e^+e^-\to\mu^+\mu^-\gamma$. Diagrams (a) and (b) describe initial-state radiation (ISR), while (c) and (d) describe final-state radiation (FSR).}
\label{fig:ISRmu}
\end{figure}
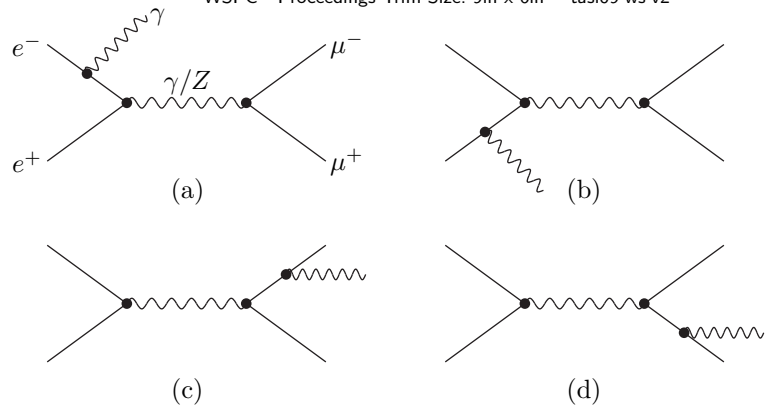

Consider a simple extension of the muon pair-production process: the reaction in Fig.~\ref{fig:ISRmu}, $e^+e^-\to\mu^+\mu^-\gamma$. We will first focus on initial-state radiation (ISR), described by diagrams (a) and (b); we will consider diagrams (c) and (d), describing the final-state radiation (FSR), in Section~2.3. Naively, this process should be much more rare than the muon pair-production: 
\beq
\sigma_{2\to3} \sim \frac{\alpha}{\pi}\sigma_{2\to 2} \sim \frac{1}{300}\sigma_{2\to 2}\,.
\eeq{mmgamma_naive}
However, this estimate is missing an important point. Consider, for example, the diagram~\ref{fig:ISRmu} (a). The matrix element is proportional to the electron propagator:
\beq
{\cal M}_{2\to3} \propto \frac{1}{(p_A-p_\gamma)^2-m_e^2} = \frac{1}{-2p_A \cdot p_\gamma}\,.
\eeq{M3} 
Neglecting the electron mass, the electron and photon four-momenta can be written as
\beq
p_A = (E,0,0,E);~~~p_\gamma = (zE,{\bf p}_\perp, \sqrt{z^2E^2 - {\bf p}_\perp^2})\,,
\eeq{4momenta}
where we introduced the photon energy fraction $z=E_\gamma/E$ and the {\it transverse momentum} ${\bf p}_\perp$. Then,
\beq
p_A \cdot p_\gamma = zE^2 \left( 1-\sqrt{1-\frac{{\bf p}_\perp^2}{z^2E^2}}\right)\,.
\eeq{dotprod}
There are two limits in which the dot product vanishes, so that the matrix element in Eq.~\leqn{M3} blows up:
\beqa
p_A \cdot p_\gamma \to 0~~{\rm when}~& & z \to 0~~~{\rm ``soft ~singularity"}\,, \CR
{\rm or}~& & \frac{{\bf p}_\perp}{z} \to 0~~~{\rm ``collinear~singularity"}\,.
\eeqa{singularities}
There is also a double-singular region where both conditions are satisfied. 

What is the meaning of these singularities? First, we should note that both are artifacts of approximations we have made: the collinear singularity is in fact removed by including the mass of the electron,  
while the soft divergence in the tree-level diagram cancels an infinity encountered in the one-loop correction to the QED vertex function. So, the full theory gives finite answers. Still, the apparent divergence does have physical meaning: it signals a large enhancement in the probability of emission of soft and collinear photons, above the naive perturbative expectation of Eq.~\leqn{mmgamma_naive}. This enhancement is even more pronounced in the case of gluon emission in QCD, where the coupling constant is larger. Understanding this effect quantitatively is crucial for making successful predictions.

In an experiment, very soft and very collinear ISR photons are unobservable. Photons emerging at a small angle to the beam line do not get detected, since even the most hermetic detectors must have small holes around the beam line to let the beams in. Soft photons leave very small energy deposits in the calorimeter, which are drowned by noise, e.g. from thermal effects in readout electronics. Thus, experimental conditions set minimal values of energy and angle for which the photon is registered. To simplify the discussion, let us assume that a photon must have a minimal transverse momentum $Q$ to get detected. In other words, if $p_\perp>Q$, we register the event as a $2\to 3$ reaction, whereas if $p_\perp<Q$, the photon is not observed and the event gets recorded as an ordinary $2\to 2$ muon pair-production. The ``observable" $2\to3$ cross section defined in this way is automatically finite; the soft and collinear singularities are hidden in the correction to the $2\to2$ cross section. Let us estimate this correction, focusing for concreteness on the collinear singularity.

The matrix element corresponding to the diagram~\ref{fig:ISRmu} (a) has the form
\beq
{\cal M}_{2\to3} \,=\, \bar{v}_B \gamma^\mu \frac{i(\sl{p}_A-\sl{p}_\gamma )}{-2p_A\cdot p_\gamma}\,\cdot(-ie\gamma^\alpha)\,u_A\,\epsilon^a_\alpha(p_\gamma) \times \left[ \ldots \right]\,,
\eeq{M3_full}
where the dots in square brackets denote ``the rest of the diagram" (in this case, the photon propagator and the muon Dirac string; however, as we're about to show, the ISR calculation is independent of the particular structure of those terms). In the small-$p_\perp$ limit, $p_\gamma\approx z p_A$ up to terms linear in $p_\perp$. The momentum flowing through the electron propagator is then $p_A-p_\gamma \approx (1-z) p_A$, so that $(p_A-p_\gamma)^2\approx 0$ -- the electron propagator is nearly on-shell! This suggests replacing the numerator of the propagator with a spin sum: 
\beq
\sl{p}_A-\sl{p}_\gamma \approx \sum_s u^s\left((1-z)p_A\right) \bar{u}^s\left((1-z)p_A\right).
\eeq{spinsum}
The matrix element becomes
\beqa
{\cal M}_{2\to3} &\approx& \bar{v}_B \gamma^\mu \frac{\sum_s u^s\left((1-z)p_A\right) \bar{u}^s\left((1-z)p_A\right)}{-2(p_\perp^2/2z)}\,\cdot(e\gamma^\alpha)\,u_A\,\epsilon^a_\alpha(p_\gamma) \times \left[ \ldots \right]\,\CR &=& -\frac{z}{p_\perp^2}\,\cdot\sum_s
\Bigl[ e\,\bar{u}^s\left((1-z)p_A\right) \gamma^\alpha u_A  \epsilon^a_\alpha(p_\gamma) \Bigr] \cdot \Bigl[ \bar{v}_B \gamma^\mu u^s\left((1-z)p_A\right) \Bigr] \times \left[\ldots \right] \CR
&=& -\frac{z}{p_\perp^2}\,\cdot\sum_s {\cal S}^{s,a}_{e^-\to e^-\gamma}(z)\cdot {\cal M}^s_{2\to 2}\left((1-z)p_A, p_B\to p_1,p_2\right)\,,
\eeqa{M3_expand}
where we defined the {\it splitting amplitude} 
\beq
{\cal S}^{s,a}_{e^-\to e^-\gamma}(z)\,=\,e \bar{u}^s(p_A(1-z))\gamma^\alpha u_A  \epsilon^a_\alpha(p_\gamma)\,.
\eeq{split}
The $2\to 3$ matrix element is a product of the splitting amplitude, which describes collinear photon emission, and the $2\to 2$ matrix element, describing muon pair-production in the electron-positron collision, with the electron energy reduced by photon emission. This behavior of the matrix element is called {\it factorization}. It is clear from the derivation that the splitting amplitude is {\it universal}: it does not depend on what particular reaction the electron enters after the photon has been emitted. The derivation can also be easily generalized to include soft photons. Factorization of matrix elements in soft and collinear limits is a general property of QED. It holds even in processes where a photon can be emitted off virtual particles in addition to external legs, since the diagrams with photon radiation off internal propagators are non-singular in the collinear and soft regimes, and their contributions are subleading.

Physically, factorization occurs because the soft/collinear photon emission and the $2\to2$ scattering process involve different length scales. Collinear photon emission typically occurs at a distance of order $m_e^{-1}$ from the interaction point, while the $2\to 2$ scattering process involves fluctuations of fields at a much shorter scale of order $1/\sqrt{s}$. The separation of scales suppresses quantum interference between the two, so they can in fact be considered as independent, sequential events, and total probability (proportional to the cross section) is simply a product of two probabilities. We will see other examples of such factorization later in these lectures.

Returning to our derivation, it can be easily shown that collinear photon emission does not change the electron's helicity. The splitting amplitudes are only non-vanishing when $s=h$, where $h$ is the helicity of the incoming electron:
\beq
{\cal M}_{2\to3} \approx -\frac{z}{p_\perp^2}\,\cdot {\cal S}^{h,a}_{e^-\to e^-\gamma}(z)\cdot {\cal M}^h_{2\to 2}\left((1-z)p_A, p_B\to p_1,p_2\right)\,.
\eeq{M3_final} 
Plugging this form into the master formula for the differential cross section, Eq.~\leqn{master}, yields
\beqa
d\sigma_{2\to3} &\approx& \frac{1}{2s}\,d\Pi_\gamma\,\left(\frac{z}{p_\perp^2}\right)^2\,\frac{1}{2}
\sum_{h,a} \left|{\cal S}^{h,a}(z)\right|^2\,\CR &\times& d\Pi_1 d\Pi_2 
\left| {\cal M}^h_{2\to 2}\left((1-z)p_A, p_B\to p_1,p_2\right)\right|^2 \,\cdot \delta^{(4)} (p_A+p_B-p_\gamma-p_1-p_2)\,,
\eeqa{23xsec}
where we defined 
\beq
d\Pi_i = \frac{d^3p_i}{(2\pi)^3}\frac{1}{2E_i}.
\eeq{Pi_def}
It is straightforward to show that
\beq
\frac{1}{2}
\sum_{a} \left|{\cal S}^{h,a}(z)\right|^2 = \frac{2e^2 p_\perp^2}{z(1-z)}\,\Bigl[ \frac{1+(1-z)^2}{z} \Bigr]\,,
\eeq{Ssum}
independent of $h$.

\vskip.3cm
\noindent {\bf Homework Problem 2:} Derive the formula~\leqn{Ssum}.
\vskip.3cm

\begin{figure}
\begin{center}
\psfig{file=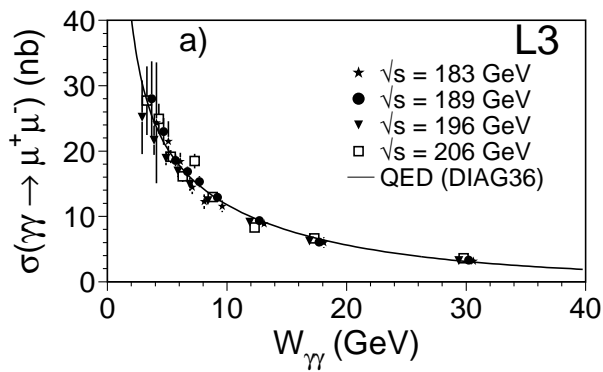,width=8cm}
\end{center}
\caption{Comparison of data from the L3 experiment at LEP-2 (points) and theoretical prediction (solid line) for the cross section of $\gamma\gamma\to\mu^+\mu^-$ as a function of $\gamma\gamma$ center-of-mass energy. From Ref.~\cite{L3_ggmumu}.}
\label{fig:ggmumu}
\end{figure}

\noindent Using this result, Eq.~\leqn{23xsec} can be rewritten as follows:
\beqa
d\sigma_{2\to3} &\approx& d\Pi_\gamma\,\left(\frac{2e^2}{p_\perp^2}\right)\,\left[ 1 + (1-z)^2 \right]
\CR \times 
\frac{1}{2s(1-z)} & & \hskip-0.3cm \sum_{h}  d\Pi_1 d\Pi_2\left| {\cal M}^h_{2\to 2}\left((1-z)p_A, p_B\to p_1,p_2\right)\right|^2 \delta^{(4)} ((1-z) p_A+p_B-p_1-p_2)\,.
\eeqa{23xsec1}
The second line of this equation is precisely the differential cross section for $2\rightarrow 2$ scattering (in our example, $e^+e^-\to\mu^+\mu^-$) at a reduced center-of-mass energy $\sqrt{(1-z)s}$. Thus, we can simply write
\beq
d\sigma_{2\to3} \approx d\Pi_\gamma\,\left(\frac{2e^2}{p_\perp^2}\right)\,\left[ 1 + (1-z)^2 \right]
\,\cdot d\sigma^{\rm LO}_{2\to2}\left( (1-z)s\right)\,,
\eeq{23xsec2}
where the superscript ``LO" serves as a reminder that this cross section is computed at tree level, or leading order in perturbation theory.
Factorization, which was previously shown at the level of matrix elements, persists at the cross section level: The cross section is a product of the cross section
for the reaction with no ISR photon, but at an energy reduced by the photon emission, and a universal factor describing the photon emission. Let us now work out this factor. In the collinear limit, the photon phase space element is
\beq
d\Pi_\gamma \,=\, \frac{d^3p_\gamma}{(2\pi)^3}\,\frac{1}{2E_\gamma}\,=\,
\frac{d^2p_\perp\cdot d(zE_A)}{(2\pi)^3}\,\frac{1}{2zE_A}\,=\,\frac{p_\perp dp_\perp}{8\pi^2}\,\frac{dz}{z}\,,
\eeq{phase_space}  
so that
\beq
d\sigma_{2\to3} \,\approx \,\frac{\alpha}{2\pi}\,\frac{dp_\perp}{p_\perp}\,dz\,\left[ \frac{1 + (1-z)^2}{z} \right]
\,\cdot d\sigma^{\rm LO}_{2\to2}\left( (1-z)s\right)\,.
\eeq{23xsec3}
To compute the correction to the observed $2\to2$ reaction rate, we need to integrate over the part of the phase space where the photon is unobservable, namely, $p_\perp\in [0, Q]$. This integral is logarithmically divergent at the low end. As already mentioned, this divergence in QED is not physical, but is due to the fact that we've set the electron mass to zero 
throughout the calculation. Restoring this mass is equivalent\footnote{The equivalence works in the limit $Q\gg m_e$; that is, the formula~\leqn{WW} captures the leading (logarithmic) behavior in this limit, but not the subleading corrections.} to imposing a lower cutoff of $m_e$ on the $p_\perp$ integral in~\leqn{23xsec3}, which regulates the divergence:
\beq
\frac{d\sigma^{\rm ISR}_{2\to3}}{dz} \,\approx \,\frac{\alpha}{2\pi}\,\left[ \frac{1 + (1-z)^2}{z} \right] \,\log\frac{Q}{m_e}
\,\cdot d\sigma^{\rm LO}_{2\to2}\left( (1-z)s\right)\,.
\eeq{WW}
This formula has a simple physical interpretation. A supposedly monochromatic beam of electrons of energy $E=\sqrt{s}/2$ in fact  contains electrons that lost some of their energy due to ISR, but were not deflected since the emitted photon was collinear, and so may collide with the positrons and initiate reactions such as muon pair-production. 
The probability of finding an electron carrying energy between $xE$ and $(x+dx)E$, where $x\in[0,1)$, can be read off from Eq.~\leqn{WW}: since $x=1-z$, we obtain
\beq
f_e(x)dx = \frac{\alpha}{2\pi}\,\left[ \frac{1 + x^2}{1-x} \right] \,\log\frac{Q}{m_e}\,dx\,.
\eeq{fe}
Of course, the total probability of emitting an ISR photon is less than one, so $f_e(x)$ must also contain a term proportional to $\delta(1-x)$. 
In addition, since the collinear ISR photons travel in essentially the same direction as the original electron, they may also collide with the positrons  (initiating, for example, a Compton scattering process $\gamma+e^+ \to
\gamma + e^+$), so it is quite reasonable to consider them a part of the beam as well.
The probability of finding a photon with energy between $zE$ and $(z+dz)E$ is given by 
\beq
f_\gamma(z) = f_e (1-z).
\eeq{fgamma} 
Since both electron and positron beam contain photons, there is even a possibility of a photon-photon collision: Fig.~\ref{fig:ggmumu} shows the cross section for $\gamma\gamma\to\mu^+\mu^-$, measured by the L3 collaboration at LEP. 

The functions $f_e$ and $f_\gamma$ describe the composition of the electron, in a way similar to how the parton distribution functions (pdf's) of the parton model describe the composition of the nucleon (more on pdf's in Lecture~\ref{sec:hadrons}). When ISR is included, an ``electron beam" in fact consists of electrons and photons. The probabilities of finding electrons and photons in the beam depend on $Q$: This should not be surprising, since $Q$ sets the boundary in $p_T$ below which the ISR photons would not be observed and so are included in the beam. In particular, if all photons were observed, corresponding to $Q= m_e$, we would get $f_e=\delta(1-x)$, $f_\gamma=0$, consistent with the naive picture of the beam. 

The distribution $f_e(x,Q)$ is singular when $x\to 1$, while $f_\gamma(z,Q)$ blows up when $z\to 0$. This is the region where the collinear ISR photons are also soft, {\it i.e.} overlapping soft and collinear divergence. As I already mentioned, this divergence is canceled when next-to-leading order (NLO) contribution to the $2\to 2$ cross section, containing the interference term between the tree-level and one-loop matrix elements, is taken into account. A convenient way to see this cancellation is to introduce a fictitious photon mass $\mu$, which regulates the $z\to 0$ divergence in~\leqn{23xsec3} in the same way as the electron mass regulates the $p_\perp\to 0$ singularity. The NLO contribution to the $2\to 2$ cross section also has logarithmic dependence on $\mu$, coming from an infrared singularity in the one-loop correction to the matrix element which is regulated by the photon mass. The dependence on $\mu$ cancels in the sum of ISR and NLO cross sections, giving a finite and well-defined answer for $f_e$ and $f_\gamma$. However, the probability of emitting a single soft photon is typically not small, indicating that processes with emission of {\it multiple} soft photons are important.\footnote{This is another example of a breakdown in perturbation theory in a particular region in phase space, similar to its breakdown in the neighborhood of an $s$-channel resonance discussed above.} The reason for this is that while formally each extra photon in the final state costs a factor of $\alpha/\pi \sim 1/300$ in the cross section, in reality collinear photon emission is only suppressed by 
$(\alpha/\pi)\log(Q/m_e)$, with an additional logarithmic enhancement in the soft-collinear double-singular region. The logarithmic factors can be large, slowing down the convergence of the perturbation theory. 
To obtain accurate predictions, terms of higher order in $\alpha$ must be included, at least those containing the highest power of the large logarithms possible at each order. Luckily, this problem turns out to be tractable. To all orders in $\alpha$, leading-logarithm accuracy, the distribution functions obey a system of integro-differential equations known as {\it Gribov-Lipatov (GL) equations}. For example,
\beq
\frac{\partial f_\gamma(x,Q)}{\partial\log Q}\,=\,\frac{\alpha}{\pi}\,
\int_x^1 \,\frac{dz}{z} \,\left[ P_{e\to\gamma}(z) \left( f_{e^-}\left(\frac{x}{z},Q\right) + f_{e^+}\left(\frac{x}{z},Q\right)\right)
+ P_{\gamma\to\gamma}(z) f_\gamma\left(\frac{x}{z},Q\right)\right]\,,
\eeq{GL} 
where $f_{e^+}$ is the positron distribution function. (Positrons must be present in the ``electron beam" due to processes such as $e^-\to e^-e^-e^+$, which appear beyond the leading order in perturbation theory.) The {\it splitting functions} are given by
\beqa
P_{e\to\gamma} &=& \frac{1+(1-z)^2}{z}\,,\CR
P_{\gamma\to\gamma} &=& -\frac{2}{3}\,\delta(1-z)\,.
\eeqa{Ps}
Note that $P_{e\to\gamma}$ is essentially the leading-order ISR photon emission probability computed above, while $P_{\gamma\to\gamma}$ is proportional to a delta function, reflecting the impossibility of a photon splitting into a photon pair at leading order in QED. (The photon can, however, split into an $e^+e^-$ pair, hence the coefficient of the delta function is not unity.) The physical meaning of this equation is as follows: if we shift $Q\to Q+\delta Q$, the photons with $p_\perp \in [Q, Q+\delta Q]$ that were previously counted as ``detectable" should now be counted as ``part  of the beam" and included in $f_\gamma$. The beam at $Q$ consists of electrons, positrons, and photons; each of these particles, if its energy is above $xE$, can emit an ISR photon with energy $xE$ and 
$p_\perp \in [Q, Q+\delta Q]$. The right-hand side of the equation simply sums up the probabilities of such emissions. Solving Eq.~\leqn{GL}, together with the corresponding equations for $f_{e^\pm}$, with the boundary conditions 
\beq
f_{e^-}(x,Q=m_e)=\delta(1-x), f_\gamma(Q=m_e)=f_{e^+}(Q=m_e)=0,
\eeq{BCs}
gives the desired distribution functions. For more details, as well as for a derivation of the GL equations, the interested reader is encouraged to read Section 17.5 of Peskin and Schroeder's textbook~\cite{PS}.

\subsection{Final State Radiation}
\label{sec:FSR}

Final-state radiation diagrams, Fig.~\ref{fig:ISRmu} (c) and (d), can be treated in exactly the same way as the ISR. Again, the cross section is dominated by photons that are either soft or collinear with the muon or antimuon.
One slight difference experimentally is that collinear (but not soft) FSR photons can be observed, so instead of using $p_\perp$ to separate $2\to3$ and $2\to2$ events, we simply assume that the photon gets detected once its energy is above $E_{\rm min}$. Following the same steps as in the derivation of Section~\ref{sec:ISR}, we obtain the total {\it observable} $2\to 3$ cross section:
\beq
\sigma(\mu^+\mu^-\gamma)\,=\,\frac{\alpha}{\pi}\,\log \frac{s}{E^2_{\rm min}} \,\log\frac{s}{m_\mu^2}\,\cdot\,\sigma^{\rm LO}(\mu^+\mu^-)\,,
\eeq{mmgamma}
where we assumed $\sqrt{s}\gg m_\mu$ and $\sqrt{s}\gg E_{\rm min}$, and ignored terms not enhanced by large logarithms. The factors of $\log \frac{s}{E^2_{\rm min}}$ and $\log\frac{s}{m_\mu^2}$ come from the soft and collinear singularities, respectively; their product is referred to as the {\it Sudakov double logarithm}. To obtain the correction to the observed $2\to 2$ rate, we must combine the contribution from FSR with $E_\gamma<E_{\rm min}$ with the NLO correction to the $2\to 2$ cross section. The result (again in the leading-log approximation) is
\beq
\sigma^{\rm obs}(\mu^+\mu^-)\,=\,\left(1- \frac{\alpha}{\pi}\,\log \frac{s}{E^2_{\rm min}} \,\log\frac{s}{m_\mu^2}\right)\,\cdot\,\sigma^{\rm LO}(\mu^+\mu^-)\,. 
\eeq{mmnogamma}
Notice that, up to terms with no large logarithms,  $\sigma^{\rm obs}(\mu^+\mu^-)+\sigma(\mu^+\mu^-\gamma)=
\sigma^{\rm LO}(\mu^+\mu^-)$. This result is independent of $E_{\rm min}$. This observation will play an important role when FSR in QCD (gluon radiation off quarks) is considered below.

\subsection{Hadronic Final States}
\label{sec:ee_hadrons}

\begin{figure}
\begin{center}
\begin{picture}(100,70)(0,0)
\SetOffset(20,-50)
\ArrowLine(1,109)(31,87)
\ArrowLine(31,87)(1,65)
\Photon(31,87)(76,87){2}{6}
\ArrowLine(76,87)(106,109)
\ArrowLine(106,65)(76,87)
\Vertex(31,87){2}
\Vertex(76,87){2}
\Text(-1,109)[cr]{$e^-$}
\Text(-1,65)[cr]{$e^+$}
\Text(108,109)[cl]{$q$}
\Text(108,65)[cl]{$\bar{q}$}
\Text(54,100)[tc]{$\gamma/Z$}
\end{picture}
\end{center}
\caption{Leading-order (tree-level) Feynman diagrams contributing to the process $e^+e^-\to q\bar{q}$.}
\label{fig:qqbar}
\end{figure}
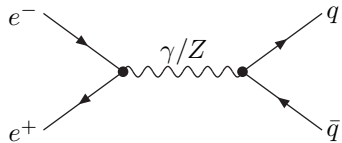

Since quarks are electrically charged and couple to the $Z$, quark-antiquark pairs can be produced in $e^+e^-$ collisions in the same way as muon pairs, see Fig.~\ref{fig:qqbar}. A qualitatively new feature in this case is that the quarks themselves are not observed: they turn into color-neutral hadrons, which are then detected by the tracker (if electrically charged) and the hadronic calorimeter. A complete theoretical description of the scattering process must include a model of {\it quark hadronization}. Since hadronization is inherently non-perturbative, involving momentum transfers comparable to the QCD confinement scale $\Lambda_{\rm QCD}\sim 200$ MeV, it is very challenging to analyze this part of the reaction from first principles, and in fact all existing approaches rely on phenomenological models. It is hard to quantify their accuracy. In general, the most useful theoretical predictions involving hadrons are those that are least dependent on the details of hadronization. This typically involves sums or averages over a large class of final states. Observables involving such sums are often referred to as {\it inclusive}, while those where a particular final state is specified are called {\it exclusive}. An extreme example of a fully inclusive observable is the total $e^+e^-\to$~hadrons rate considered below: At collision energies large compared to $\Lambda_{\rm QCD}$ (which we will assume throughout this section), it can be predicted, up to corrections suppressed by powers of $\Lambda_{\rm QCD}/\sqrt{s}$, with no recourse to a hadronization model. On the other extreme are fully exclusive quantities, e.g. $\sigma(e^+e^-\to 2\pi+2K+3\rho)$, which can only be calculated within a hardonization model. Trustworthy quantitative predictions for such observables are currently out of reach. Luckily, there exists a large class of useful observables that are sufficiently inclusive to allow for precise predictions, and at the same time carry much more information than the total hadronic rate. We will consider an example in Section~\ref{sec:jets}.

\subsubsection{$e^+e^-\to$ Hadrons}
\label{sec:R}

\begin{figure}
\begin{center}
\psfig{file=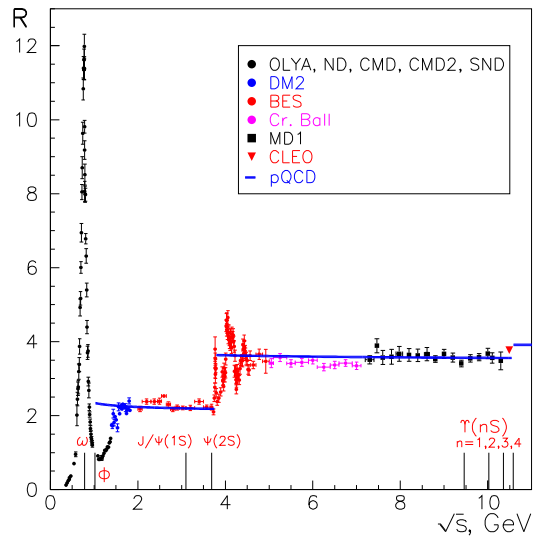,width=8cm}
\end{center}
\caption{Total hadronic $e^+e^-$ scattering cross section, normalized to $\sigma(e^+e^-\to\mu^+\mu^-)$, as a function of center-of-mass energy, at energies well below the $Z$ peak. The solid blue line is the SM prediction. From Ref.~\cite{Rhadron}.}
\label{fig:R}
\end{figure}

\begin{figure}
\begin{center}
\psfig{file=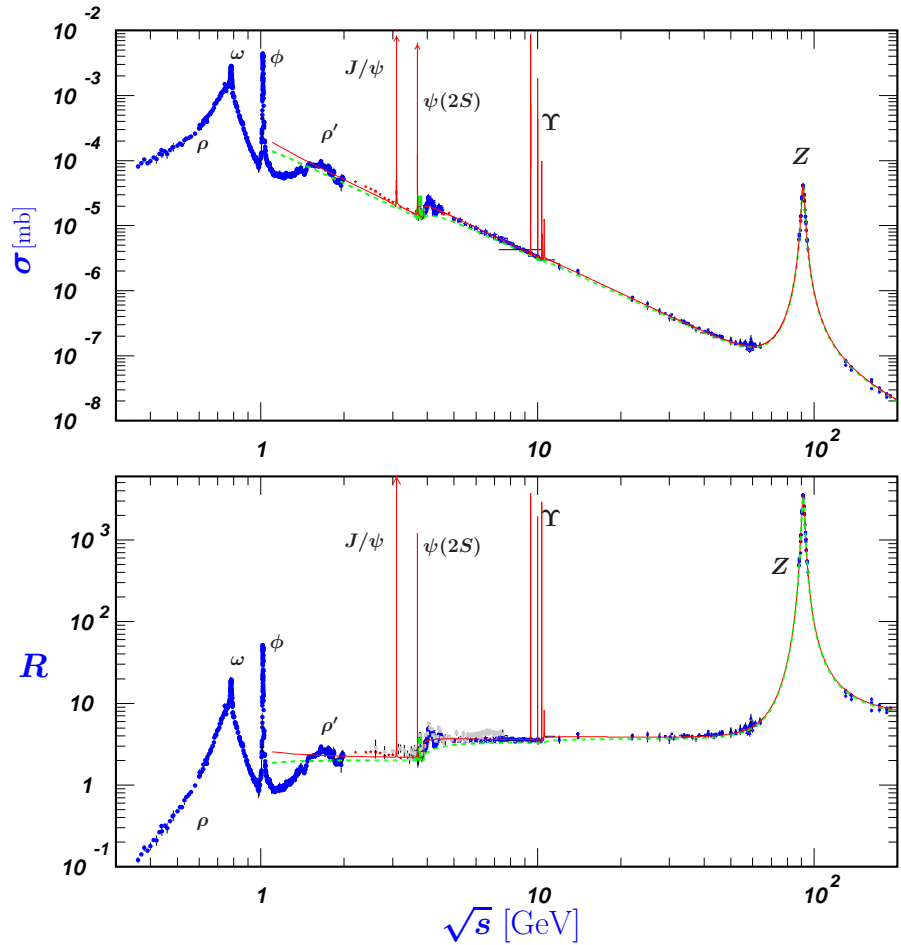,width=12cm}
\end{center}
\caption{Total hadronic $e^+e^-$ scattering cross section as a function of center-of-mass energy, at all currently available energies. Adopted from the Particle Data Group~\cite{PDG}.}
\label{fig:R_PDG}
\end{figure}

The simplest inclusive observable is the {\it total hadronic} cross section, $\sigma(e^+e^-\to$~hadrons$)$. If collision energy is well above $\Lambda_{\rm QCD}$, it is reasonable to assume that there is no quantum interference between the short-distance process of quark pair creation (characteristic length scale $1/\sqrt{s}$), and the long-distance 
hadronization process (typical scale $\sim\Lambda^{-1}_{\rm QCD}$)\footnote{This is another example of factorization due to separation of scales, see comments under Eq.~\leqn{split}.}. In other words, 
\beq
\sigma(e^+e^-\to{\rm~hadrons}) \,=\,\sigma(e^+e^-\to q\bar{q})\times~{\rm Prob}~(q\bar{q}\to {\rm~hadrons}) \,,
\eeq{voila}
up to corrections of order $\Lambda_{\rm QCD}/\sqrt{s}$ and $\alpha_s$. 
Since each quark must hadronize in one way or another, the probability factor is equal to unity. A simple calculation then gives (for collision energies well below $m_Z$)
\beq
\sigma(e^+e^-\to{\rm~hadrons}) \,=\, 3\,\cdot \sum_q Q_q^2 \cdot \sigma(e^+e^-\to \mu^+\mu^-)\,, 
\eeq{hadrons}
where $Q_q$ are the quarks' electric charges, and only quarks with $m_q\ll \sqrt{s}$ should be included in the sum. (In threshold regions, $\sqrt{s}\approx 2m_q$, the cross section behavior is somewhat more complicated due to quark mass effects and the presence of bound states.) 
It is customary to present experimental data on this process in terms of a dimensionless ratio
\beq
R = \frac{\sigma(e^+e^-\to{\rm~hadrons})}{\sigma(e^+e^-\to \mu^+\mu^-)}\,.
\eeq{R}
At energies well below the $Z$ peak, and away from thresholds, the formula~\leqn{hadrons} is in excellent agreement with experiment, see Fig.~\ref{fig:R}. This gives firm support for the parton model and the factorization assumption. The calculation can be easily extended to include $Z$ exchanges. This result is in good agreement with data up to the highest energies reached at LEP-2, see Fig.~\ref{fig:R_PDG}. Radiative corrections must be included to match the high experimental precision available for $\sqrt{s}\approx M_Z$ (see section~\ref{sec:jets}).

\subsubsection{Hadronization}
\label{sec:string}

To predict cross sections for specific hadronic final states (e.g. $\pi^+\pi^-$ or $K^+K^-\pi^0\pi^0$), one must model the hadronization process. In practice, this is done by using a {\it Monte Carlo (MC) generator}, a computer program implementing one of the available phenomenological hadronization models. Popular hadronization models include the {\it string fragmentation} model, implemented in {\tt PYTHIA}~\cite{pythia}, and the {\it cluster hadronization} model, implemented in {\tt HERWIG}~\cite{herwig}. The models are quite complicated, and contain a large number of free parameters that must be fitted to data. A detailed description of hadronization is beyond the scope of these lectures. Still, just to give a flavor of the physics involved, let me briefly consider the basic picture behind the string fragmentation model.

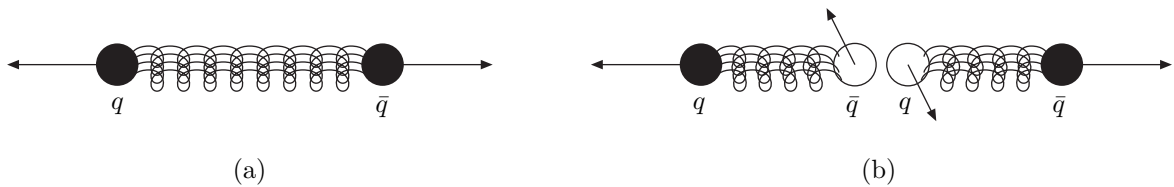
\begin{figure}
\begin{center}
\begin{picture}(450,80)(0,0)

\SetOffset(70,-40)
\Vertex(-20,90){8}
\Vertex(80,90){8}
\Gluon(-15,90)(75,90){4}{8}
\Gluon(-15,93)(75,93){4}{8}
\Gluon(-15,87)(75,87){4}{8}
\Gluon(-15,84)(75,84){4}{8}
\LongArrow(-20,90)(-60,90)
\LongArrow(80,90)(120,90)
\Text(-20,78)[tc]{$q$}
\Text(80,78)[tc]{$\bar{q}$}
\Text(30,55)[tc]{(a)}

\Vertex(200,90){8}
\BCirc(258,90){8}
\Gluon(205,90)(250,90){4}{4}
\Gluon(205,93)(252,93){4}{4}
\Gluon(205,87)(252,87){4}{4}
\Gluon(205,84)(253,84){4}{4}
\LongArrow(200,90)(160,90)
\LongArrow(258,90)(248,110)
\Text(200,78)[tc]{$q$}
\Text(258,78)[tc]{$\bar{q}$}
\Vertex(336,90){8}
\BCirc(278,90){8}
\Gluon(286,90)(331,90){4}{4}
\Gluon(284,93)(331,93){4}{4}
\Gluon(284,87)(331,87){4}{4}
\Gluon(283,84)(331,84){4}{4}
\LongArrow(336,90)(376,90)
\LongArrow(278,90)(288,70)
\Text(336,78)[tc]{$\bar{q}$}
\Text(278,78)[tc]{$q$}
\Text(268,55)[tc]{(b)}

\end{picture}
\end{center}
\caption{String fragmentation model of hadronization. (a): Quark-antiquark pair, connected by a color flux tube, immediately after the collision. (b): After the first string breakdown. Filled circles denote primary partons, while open circles denote secondary partons. Primary parton momenta are $\sqrt{s}$, while the typical secondary parton momenta are of order $\Lambda_{\rm QCD}\ll \sqrt{s}$.}
\label{fig:string}
\end{figure}

Consider a quark-antiquark pair created in an $e^+e^-$ collision. Let us refer to these $q$ and $\bar{q}$ as ``primary partons". 
To hadronize, the primary quark must form a bound state with an antiquark; however, the primary partons are rapidly moving away from each other. Their invariant mass $\sqrt{s}$ is large compared to the binding energy $\sim\Lambda_{\rm QCD}$, so they typically do not bind. Instead, quark hadronization proceeds by a spontaneous creation of new ``secondary" quark-antiquark pair(s) from the vacuum, and formation of bound states between the primary quark and a secondary anti-quark. The spontaneous pair-creation process is possible because the primary partons create color (gluon) field in the surrounding space; it is this field that supplies the needed energy. Color field lines originate at the quark and end on the antiquark. In the string model, the color field immediately after the collision is modeled as a {\it color flux tube}, or string, connecting the primary partons: see Fig.~\ref{fig:string}. The string tension is of order $\Lambda_{\rm QCD}^2 \sim 1$~GeV/fm, and its width is of order $\Lambda^{-1}_{\rm QCD}$, corresponding to typical transverse momenta $\sim\Lambda_{\rm QCD}$. As the primary partons move apart after the collision, the string between them is stretched, and at some point it breaks down. A secondary $q\bar{q}$ pair is spontaneously created at the breakdown point, so that the original string splits into two string fragments. The secondary pair has typical momentum (both along and transverse to the string) of order $\Lambda_{\rm QCD}$, so that the invariant mass of each string fragment after the breakdown is of order $\sqrt{E_q \Lambda_{\rm QCD}}\sim s^{1/4}\Lambda^{1/2}_{\rm QCD}\ll \sqrt{s}$. Each string fragment then breaks down again, further reducing the invariant mass. The process is iterated until the invariant mass of all fragments is about 1 GeV. At this point, each string fragment is associated with a meson. The probabilities for a string to turn into a particular meson are phenomenological parameters, to be fitted to data. A similar, though a bit more complicated, picture is used to describe baryon formation. 

It is clear that the above model is heuristic, and a large number of input parameters is needed: string tension, breakdown probability, secondary pair momentum distributions, relative probabilities of nucleating $u\bar{u}$, $d\bar{d}$ and $s\bar{s}$ pairs, {\it etc.} A reasonable agreement with data can be reached by adjusting these parameters; however, it is difficult to quantify how uncertain the model predictions really are, especially for highly-exclusive observables. 

\subsubsection{$e^+e^-\to$ Jets}
\label{sec:jets}

Since secondary quarks and antiquarks have typical momenta of order $\Lambda_{\rm QCD}\ll \sqrt{s}$, most of the momentum carried by the hadrons is due to the primary partons. This can be seen explicitly in the string model outlined above, but is in fact independent of the details of the hadronization model. The hadrons come out in two collimated {\it jets}, one approximately collinear with the quark and the other with the antiquark. (Experimentally, a ``jet-finding algorithm" is used to find clusters of approximately collinear hadrons in the event and identify them as jets, and the precise definition of a jet depends somewhat on the details of this algorithm.) The sum of the four-momenta of all hadrons in each jet is equal to the momentum of the corresponding primary parton. Corrections to this expectation, and the opening angle of each jet, are of order\footnote{This estimate applies only to effects of hadronization. As we will see below, there are additional corrections from higher-order perturbative QCD; however, these are still small, and in addition are (at least in principle) fully calculable.} $\Lambda_{\rm QCD}/\sqrt{s}$. While these corrections depend on the hadronization model, they are small. Ignoring them, jet production cross sections and kinematic distributions are identical to those of the primary partons, which can be calculated within perturbation theory. These cross sections and distributions provide a wide class of observables that can be used to test the SM. 

As an example, consider the angular distributions of jets produced in $e^+e^-\to q\bar{q}$. The differential cross section $d\sigma/d\cos\theta$ can be easily computed: at tree level, it has the same form as the muon cross section in Eq.~\leqn{Zxsec}, but with
\beqa
g_L^u &=& \frac{e}{c_ws_w}\left(\frac{1}{2}-\frac{2}{3}s_w^2 \right),~~~g_R^u \,=\, -\frac{2}{3}\frac{es_w}{c_w}\,;\CR
g_L^d &=& \frac{e}{c_ws_w}\left(-\frac{1}{2}+\frac{1}{3}s_w^2\right),~~~g_R^d \,=\, +\frac{1}{3}\frac{es_w}{c_w}\,,
\eeqa{quark_gs}
where the superscript refers to up-type and down-type quarks. Comparing this prediction with data 
at the $Z$ resonance could be used to test the SM prediction for quark couplings to the $Z$ boson, in the same way as the measurement of the muon forward-backward asymmetry discussed in Section~\ref{sec:eemumu} tests the muon couplings. However, there is an additional complication: Jets originated by $u, d, s$ quarks {\it and} their antiparticles are indistinguishable experimentally, since it is impossible to determine which of the partons making up the hadrons in a given jet was the primary. This ambiguity completely washes out the forward-backward asymmetry, which has opposite sign for particles and antiparticles. The situation is better if the primary is a heavy quark, $c$ or $b$: since the probability to nucleate heavy quark pairs from the vacuum is very small, the jet typically contains just a single heavy quark. This heavy quark can be detected 
using the fact that it travels a small but macroscopic distance before decaying ($c\tau\approx 500~\mu$m for $b$, $100-300~\mu$m for $c$), producing a secondary vertex in the inner tracker detector. If the decay is semileptonic, the charged lepton inside the jet can serve as an additional signature. A combination of these techniques is known as {\it flavor-tagging}: For example, a jet with an identified $b$ quark is said to be ``b-tagged". In addition, the charge of the lepton from a semileptonic decay can be used to infer the charge of the heavy quark (e.g. $b\to \ell^-$ vs. $\bar{b}\to \ell^+$; make sure you understand why!), giving a ``charge tag". This allowed for a direct measurement of 
the forward-backward asymmetry in $e^+e^-\to b\bar{b}$ at LEP-2 and SLC. The result is $A_{\rm FB}^b = 0.0992\pm 0.0016$, about 2.5 standard deviation below the SM prediction. This is the largest deviation from the SM among precision electroweak observables, although not large enough to be taken seriously as an indication of new physics. 

\begin{figure}
\begin{center}
\begin{picture}(300,80)(0,0)
\SetOffset(20,-40)
\Line(1,109)(31,87)
\Line(31,87)(1,65)
\Photon(31,87)(76,87){2}{6}
\Vertex(91,98){2}
\Line(76,87)(91,98)
\Line(91,98)(106,109)
\Gluon(91,98)(121,98){2}{6}
\Line(106,65)(76,87)
\Vertex(31,87){2}
\Vertex(76,87){2}
\Text(123,98)[cl]{$g$}
\Text(-1,109)[cr]{$e^-$}
\Text(-1,65)[cr]{$e^+$}
\Text(108,109)[cl]{$q$}
\Text(108,65)[cl]{$\bar{q}$}
\Text(54,54)[cc]{(a)}

\Line(151,109)(181,87)
\Gluon(241,76)(271,76){2}{6}
\Line(151,65)(181,87)
\Photon(181,87)(226,87){2}{6}
\Vertex(241,76){2}
\Line(226,87)(256,109)
\Line(256,65)(226,87)
\Vertex(181,87){2}
\Vertex(226,87){2}
\Text(204,54)[cc]{(b)}

\end{picture}
\end{center}
\caption{Feynman diagrams contributing to the process $e^+e^-\to q\bar{q}g$ at the tree level.}
\label{fig:3jets}
\end{figure}
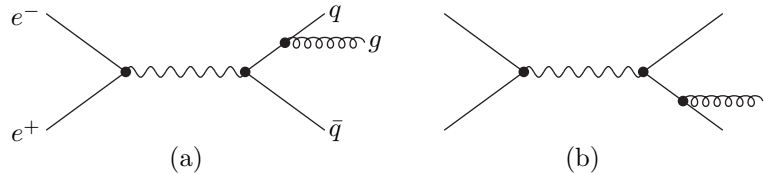

At leading order in the QCD coupling constant $\alpha_s$, all hadronic events in $e^+e^-$ collisions result in two jets. At the next-to-leading order, an additional gluon can be emitted, see Fig.~\ref{fig:3jets}.
While formally suppressed by $\alpha_s/\pi\sim 1/30$, this process is enhanced by singularities for soft and collinear gluon emission, analogous to those encountered in the photon FSR analysis of Sec.~\ref{sec:FSR}. 
The emitted gluon hadronizes, creating an additional jet. The experimental signature of this jet depends on its direction. If the gluon is approximately collinear with the (anti-)quark, the jet-finding algorithm will merge the hadrons originating from the (anti-)quark and the gluon into a single jet, identifying the final state as 2-jet. If the gluon is emitted at a large angle, the extra jet will be reconstructed separately, resulting in a 3-jet event. Thus the 2-jet and 3-jet rates depend strongly on the details of the jet-finding algorithm, and the value of parameters used to define a jet.  
However, just as in the case of photon FSR, the {\it sum} of the two rates is completely independent of these details. A complete NLO-QCD calculation gives\footnote{A reader interested in the derivation of this formula is encouraged to work through the Final Project at the end of Part I of Peskin and Schroeder's textbook~\cite{PS}.}
\beq
\sigma^{\rm NLO}(e^+e^-\to\geq2~{\rm jets})\,=\,\sigma^{\rm LO}(e^+e^-\to q\bar{q})\,\cdot\,\left[ 1\,+\,\frac{3\alpha_s}{4\pi}\right]\,.
\eeq{2plus3jet}
Comparing this result with the cross section at the $Z$ resonance precisely measured by LEP experiments yields one of the most precise determinations of the QCD coupling constant: $\alpha_s(M_Z)=0.123\pm 0.004$. 

Numerically large value of $\alpha_s$ and collinear enhancement mean that most jets contain multiple collinear gluons in addition to the primary parton. These gluons can in turn split into $q\bar{q}$ or $gg$ pairs, 
which can emit additional gluons {\it etc.} This phenomenon is known as
{\it parton showering}. The internal structure of the jets, {\it i.e.} observables such as the jet opening angle, is dominantly determined by the parton showering, which plays a more important role than hadronization. In addition, depending on the parameters used to define jets, parton showers may contribute significantly to the observed $n$-jet rates for $n>2$. Parton showering can in principle be completely described in terms of perturbative QCD, although resummation of large logarithms to all orders in perturbation theory is necessary. In practice, most phenomenological and experimental analyses rely on the numerical implementation of parton showers in Monte Carlo programs such as {\tt PYTHIA} and {\tt HERWIG}.  

\section{Hadron Collisions}
\label{sec:hadrons}

\begin{figure}
\begin{center}
\begin{picture}(200,60)(0,0)
\SetOffset(10,-60)
\LongArrow(1,90)(31,90)
\Text(15,94)[bc]{$P_1$}
\Oval(41,90)(20,10)(0)
\ArrowLine(51,90)(91,90)
\Vertex(91,90){4}
\ArrowLine(131,90)(91,90)
\LongArrow(91,90)(81,110)
\LongArrow(91,90)(101,110)
\LongArrow(91,90)(81,70)
\LongArrow(91,90)(101,70)
\Oval(141,90)(20,10)(0)
\LongArrow(181,90)(151,90)
\Text(166,94)[bc]{$P_2$}
\Text(116,94)[bc]{$x_2P_2$}
\Text(66,94)[bc]{$x_1P_1$}
\Text(91,112)[bc]{$Y$}

\end{picture}

\end{center}
\caption{Parton model picture of a hadron-initiated process $p(P_1)+p(P_2)\to Y+X$.}
\label{fig:partons}
\end{figure}
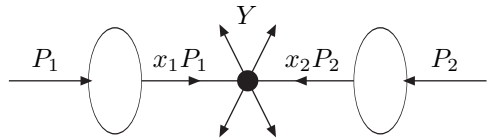

Theoretical analysis of hadron collisions ($pp$ or $p\bar{p}$) proceeds along the same lines as for $e^+e^-$. The main new difficulty is that the initial state consists of composite particles. According to the {\it parton model}, production of final states with total invariant mass large compared to $\Lambda_{\rm QCD}$ is initiated by a pair of partons, with the rest of colliding hadrons serving as spectators, as represented graphically in Fig.~\ref{fig:partons}. Quantitatively, for a proton-proton collision, this picture implies that
\beq
d\sigma(p(P_1)+p(P_2)\to Y+X) \,=\, \int_0^1 dx_1 \, \int_0^1 dx_2 
\, \sum_{i_1, i_2} f_{i_1}(x_1) \,f_{i_2}(x_2) d\sigma(i_1(x_1P_1) + i_2(x_2P_2) \to Y)\,.
\eeq{parton_model}
Here, $Y$ denotes the high-invariant-mass final state of interest; $X$ denotes anything else (including the re-hadronized remnants of the colliding protons); $P_1$ and $P_2$ are the momenta of the incoming protons; $x_1$ and $x_2$ are the fractions of those momenta carried by the partons that initiate the reaction; and the sum runs over the types of partons (quark and antiquark flavors and gluons). The functions $f_i(x)$ are the {\it parton distribution functions (pdf's)}; we will discuss them in more detail below. The analogous formula for cross sections in $p\bar{p}$ collision is obtained by replacing $f_{i_2}(x_2) \to \bar{f}_{i_2}(x_2) = f_{\bar{i}_2}(x_2)$, where $\bar{g}=g$ by definition. The parton-level cross section on the right-hand side, $\sigma(i_1(x_1P_1) + i_2(x_2P_2) \to Y)$, is calculated within perturbation theory. 

While the center of mass of the colliding hadron pair is at rest in the lab frame, the {\it parton frame}, {\it i.e.} the frame in which the center of mass of the colliding partons is at rest, is moving with respect to the lab frame with velocity
\beq
\beta\,=\,\frac{x_1-x_2}{x_1+x_2}
\eeq{beta} 
along the collision axis. (Note that the parton-level cross section in Eq.~\leqn{parton_model} can always be evaluated in the parton frame, since the differential cross section is invariant with respect to boosts along the collision axis.) Conservation of energy-momentum implies that the invariant mass of the state $Y$ has to be equal to the parton center-of-mass energy, $\sqrt{\hat{s}}$, where
\beq
\hat{s}=x_1x_2 s\,.
\eeq{shat}
Since $x_1$ and $x_2$ are not the same  in each event, but instead are picked from a distribution dictated by the pdf's and cross sections, the values of $\beta$ and $\hat{s}$ in each event are a priori unknown. If the final state $Y$ is fully reconstructed (all particles are detected and their energies and momenta measured), these quantities can be experimentally determined on an event-by-event basis. If, however, $Y$ includes invisible particles (neutrinos in the SM, neutralinos in supersymmetric models, {\rm etc.}), their values in each event remain unknown. This lack of kinematic information complicates the analysis of such final states at a hadron collider.  

\subsection{Parton Distribution Functions}

Building qualitative intuition about hadron collisions requires some familiarity with the parton distribution functions. Conceptually, the pdf's are 
close cousins of the electron and photon distribution functions in an ``electron beam", encountered in Section~\ref{sec:ISR}. Part of their job  is exactly the same: They account for emission of collinear ISR gluons, whose transverse momentum $p_T$ is too small for them to be detected individually, and for further splitting of these gluons into collinear $q\bar{q}$ and $gg$ pairs. Just as in the photon ISR case, these effects depend on the minimal transverse momentum $Q$ for which a gluon (or a quark from gluon splitting) {\it can} be resolved as an additional jet; thus, the pdf's are really functions of two arguments, $x$ and $Q$. The dependence of the pdf's on $Q$ is encoded in the {\it Altarelli-Parisi (AP) equations}, which extend the Gribov-Lipatov equations of Section~\ref{sec:ISR} to QCD. For example, the counterpart of Eq.~\leqn{GL} is
\beq
\frac{\partial f_g(x,Q)}{\partial\log Q}\,=\,\frac{\alpha_s}{\pi}\,
\int_x^1 \,\frac{dz}{z} \,\left[ P_{q\to g}(z) \left( f_{q}\left(\frac{x}{z},Q\right) + f_{\bar{q}}\left(\frac{x}{z},Q\right)\right)
+ P_{g\to g}(z) f_\gamma\left(\frac{x}{z},Q\right)\right]\,,
\eeq{AP}
where the QCD coupling constant $\alpha_s$ should be evaluated at the scale $Q$. The splitting functions are given by
\beqa
P_{q\to g} &=& \frac{4}{3}\,\Bigl[ \frac{1+(1-z)^2}{z}\Bigr] \,,\CR
P_{\gamma\to\gamma} &=& 6 \,\Bigl[ \frac{z}{(1-z)_+} + \frac{1-z}{z} + z(1-z) + \left(\frac{11}{12}-\frac{n_f}{18}\right)\,\delta(1-z) \Bigr]\,,
\eeqa{APsplit}
where $n_f$ is the number of quark flavors with masses $m_q \ll Q$. 
(For derivation, as well as definition of the ``+ prescription'' used to regulate the denominator of the first term in $P_{g\to g}$ at $z\to 1$, see Section 17.5 of Peskin and Schroeder~\cite{PS}.) 
The physical meaning is the same as for the GL equation, see comments at the end of Section~\ref{sec:ISR}; the only qualitative difference is the possibility of $g\to gg$ splitting at leading order, due to non-Abelian nature of QCD. The key new feature is the boundary conditions at small $Q$. In the case of QED, the conditions~\leqn{BCs} are motivated by the observation that, if all ISR photons were detected, one would simply end up with a monochromatic electron beam. In a hadron collision, though, even if every collinear gluon is detected, one still needs to specify the probabilities of finding various partons in the proton. These cannot be computed within perturbation theory, and at present the only viable approach is to measure them from data. Thus, in applying the parton model, one must first perform a fit to a number of experimental observables which would provide the pdf's, and then use these pdf's to predict other observables. The reason this works
is that the pdf's are universal, that is, independent of the reaction that the partons enter. The proof of this universality follows along the same lines as the collinear factorization proof in Section~\ref{sec:ISR}, although details are considerably more involved. 

Which observables should be used to determine pdf's? They should obey three important criteria. Firstly, the set should be broad enough to achieve sensitivity to as many pdf's and in as broad a range of $x$ values as possible ($Q$ values are less of a concern, since evolution in $Q$ is calculable via the AP equations.) Secondly, an accurate theoretical prediction {\it at the parton level} should be available. Observables used in practice are typically known to NLO or NNLO in the QCD coupling constant. Finally, only processes where contamination from new physics beyond the SM is unlikely should be used; ideally, all data  included in the fit should be at energies well below the scale where the SM has already been extensively tested and verified.

\begin{table}
\tbl{The main processes included in the MSTW-2008 fit, ordered in three groups: fixed-target experiments, HERA, and the Tevatron. For each process, the table lists dominant partonic subprocesses, the partons whose pdf's are primarily probed, and the $x$ range constrained by the data. From Ref.~\cite{MSTW}.}
{\begin{tabular}{llll}
      \hline
      \hline
      Process & Subprocess & Partons & $x$ range \\ \hline
      $\ell^\pm\,\{p,n\}\to\ell^\pm\,X$ & $\gamma^*q\to q$ & $q,\bar{q},g$ & $x\gtrsim 0.01$ \\
      $\ell^\pm\,n/p\to\ell^\pm\,X$ & $\gamma^*\,d/u\to d/u$ & $d/u$ & $x\gtrsim 0.01$ \\
      $pp\to \mu^+\mu^-\,X$ & $u\bar{u},d\bar{d}\to\gamma^*$ & $\bar{q}$ & $0.015\lesssim x\lesssim 0.35$ \\
      $pn/pp\to \mu^+\mu^-\,X$ & $(u\bar{d})/(u\bar{u})\to \gamma^*$ & $\bar{d}/\bar{u}$ & $0.015\lesssim x\lesssim 0.35$ \\
      $\nu (\bar{\nu})\,N \to \mu^-(\mu^+)\,X$ & $W^*q\to q^\prime$ & $q,\bar{q}$ & $0.01 \lesssim x \lesssim 0.5$ \\
      $\nu\,N \to \mu^-\mu^+\,X$ & $W^*s\to c$ & $s$ & $0.01\lesssim x\lesssim 0.2$ \\
      $\bar{\nu}\,N \to \mu^+\mu^-\,X$ & $W^*\bar{s}\to\bar{c}$ & $\bar{s}$ & $0.01\lesssim x\lesssim 0.2$ \\\hline
      $e^\pm\,p \to e^\pm\,X$ & $\gamma^*q\to q$ & $g,q,\bar{q}$ & $0.0001\lesssim x\lesssim 0.1$ \\
      $e^+\,p \to \bar{\nu}\,X$ & $W^+\,\{d,s\}\to \{u,c\}$ & $d,s$ & $x\gtrsim 0.01$ \\
      $e^\pm p\to e^\pm\,c\bar{c}\,X$ & $\gamma^*c\to c$, $\gamma^* g\to c\bar{c}$ & $c$, $g$ & $0.0001\lesssim x\lesssim 0.01$ \\
      $e^\pm p\to\text{jet}+X$ & $\gamma^*g\to q\bar{q}$ & $g$ & $0.01\lesssim x\lesssim 0.1$ \\ \hline
      $p\bar{p}\to \text{jet}+X$ & $gg,qg,qq\to 2j$ & $g,q$ & $0.01\lesssim x\lesssim 0.5$ \\
      $p\bar{p}\to (W^\pm\to\ell^{\pm}\nu)\,X$ & $ud\to W,\bar{u}\bar{d}\to W$ & $u,d,\bar{u},\bar{d}$ & $x\gtrsim 0.05$ \\
      $p\bar{p}\to (Z\to\ell^+\ell^-)\,X$ & $uu,dd\to Z$ & $d$ & $x\gtrsim 0.05$
      \\
      \hline
      \hline
    \end{tabular} }
\label{tab:PDF_processes}
\end{table}

As an example, consider the recent MSTW-2008 pdf fit~\cite{MSTW}. The reactions used in the fit, along with the pdf's primarily constrained by each reaction, are listed in Table~\ref{tab:PDF_processes}. Most of the data
comes from {\it deep inelastic scattering} (DIS) experiments, including neutral-current scattering $e p \to e X$ and charged-current reactions $ep \to \nu X$, $\nu p\to \ell X$. The most accurate and complete data set on $ep$ scattering is provided by the experiments at the HERA $ep$ collider. Using DIS data has an important advantage that only one of the reacting particles is composite. A disadvantage is that only some linear combinations of pdf's are determined. For example, there is no direct sensitivity to the gluon pdf, since at tree level leptons do not interact with gluons. Measuring quark pdf's at different $Q$ scales provides an indirect measurement of the gluon pdf via AP evolution equations. This method works quite well at low $x$. The fit includes additional data, such as the Tevatron dijet rate (see Sec.~\ref{sec:ppjet}), to constrain the gluon pdf at high $x$. Large uncertainties remain at $x\gsim 0.1$, partly because $f_g$ is quite small in that range.
Additional constraints on quark pdf's are provided by the measurements of electroweak gauge boson rapidity distributions at the Tevatron, as discussed in Sec.~\ref{sec:Zrap}.

\begin{figure}
\begin{center}
\psfig{file=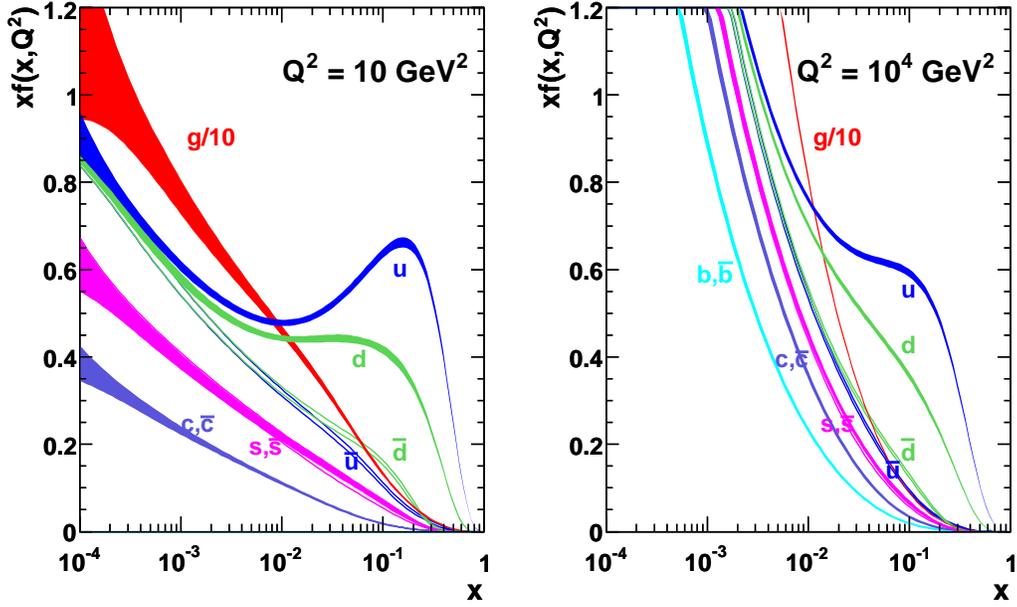,width=14cm}
\end{center}
\caption{MSTW-2008 pdf's at $Q^2=10$ GeV$^2$ and $10^4$ GeV$^2$. The width of the lines indicates the error bars. From Ref.~\cite{MSTW}.}
\label{fig:PDF_all}
\end{figure}

Figure~\ref{fig:PDF_all} shows the MSTW-2008 pdf's for two values of $Q$: $Q^2=10$ GeV$^2$ and $10^4$ GeV$^2$ (the latter corresponding roughly to $Q=M_Z$, a common choice for describing electroweak processes). Error bars are indicated by the width of the lines; roughly speaking, pdf's can be varied within these bands without spoiling consistency of the fit with the data. A few features of the pdf's are worth remembering for anyone wishing to build qualitative intuition about hadron collider physics. First of all, note that while in the naive quark model 
the proton is ``made of" $u$ and $d$ quarks, in reality they only dominate the pdf's at $x\gsim 0.1$. 
(The $u$ and the $d$ do carry all of the proton's electric charge, but the pdf's describe the distribution of {\it momentum}, not charge, among partons.) 
For lower $x$, the dominant component of the proton is the gluon, especially in the low-$x$ and high-$Q^2$ regions. (Note that in the figure, the gluon pdf is divided by 10!) This is especially relevant at the LHC, where most collisions with parton center-of-mass energies up to about 1 TeV will be between gluons; for this reason, the LHC is sometimes called a ``gluon collider".
Moreover, below $x\lsim 0.01$, the pdf's of all light quarks and antiquarks are roughly the same; this is because most of them come from gluon splittings, which are flavor-blind and produce equal number of $q$ and $\bar{q}$. (The strange and charm pdf's are suppressed due to their mass, but the effect is not strong. At $Q\gg m_b$, bottom pdf should be taken into account as well.)
The low-$x$ quarks and antiquarks are often referred to as the ``sea" partons, whereas $u$ and $d$ quarks carrying $x\gsim 0.01$ are called ``valence quarks". Of course, the distinction between  
sea and valence $u$ and $d$ quarks is not precise, so these terms can only be applied in a qualitative sense. As far as $Q$ evolution is concerned, for both quarks and gluons, pdf's migrate towards lower $x$ as $Q$ is increased. The reason is that, as $Q$ is increased, new partons are added that come from collinear splittings of the original partons. Since splitting always lowers $x$, this increases density at low $x$ and suppresses it at high $x$. This effect can be clearly seen in Fig.~\ref{fig:PDF_all}. Since splitting amplitudes are proportional to the QCD coupling constant evaluated at the scale $Q$, the speed of the evolution decreases with $Q$ due to asymptotic freedom of QCD.

\subsection{Electroweak Gauge Boson Production}

As an example of a hadron collider process, let us consider production of a single $Z$ boson. We will first compute the total production cross section of the $Z$ at the Tevatron and the LHC, and then proceed to discuss its kinematic distributions.

\subsubsection{Cross Section}

At leading (tree) level in perturbation theory, the $Z$ can only be produced in $q\bar{q}$ collisions, with cross section 
\beq
\sigma(q\bar{q}\to Z)\,=\,\frac{4\pi^2}{3}\,\frac{\Gamma(Z\to q\bar{q})}{M_Z}\,\delta(\hat{s}-M_Z^2)\,,
\eeq{Zparton_xsec}
where $\Gamma(Z\to q\bar{q})=\Gamma_Z \cdot\,{\rm Br}~(Z\to q\bar{q})$ is the partial decay width of the $Z$ in the $q\bar{q}$ channel. At the hadron level, this yields
\beq
\sigma(pp\to Z+X) \,=\, \frac{4\pi^2}{3}\,\frac{\Gamma_Z}{M_Z}\,\int_0^1 dx_1\,\int_0^1 dx_2 \, \sum_q 2 f_q(x_1, Q) f_{\bar{q}}(x_2,Q)\, {\rm Br}\,(Z\to q\bar{q}) \,\delta(x_1x_2s-M_Z^2)\,. 
\eeq{Zproton_xsec1}
The same formula applies to $p\bar{p}$ collisions, with the substitution
\beq
2 f_q(x_1, Q) f_{\bar{q}}(x_2,Q) \,\longrightarrow\,  f_q(x_1, Q) f_q (x_2,Q) +f_{\bar{q}}(x_1, Q) f_{\bar{q}}(x_2,Q) \,.
\eeq{ppbar_subst}

An alarming feature of Eq.~\leqn{Zproton_xsec1} is that the cross section seems to depend on the scale $Q$. Recall that $Q$ is defined as the minimal transverse momentum at which an ISR parton is registered as an extra jet. However, we are considering the process $Z+X$, where $X$ includes any hadronic activity, so events with any number of extra jets should be included and the cross section should be independent of $Q$. The reason for this apparent contradiction is that we used the {\it leading-order} parton-level cross section in our calculation. When radiative corrections are included, the parton-level cross section itself becomes $Q$-dependent. For example, at NLO in $\alpha_s$, the parton-level cross section contains a contribution from the process $q\bar{q}\to gZ$, but only gluons with $p_T>Q$ should be included. (The gluons with $p_T<Q$ have already been accounted for by the pdf's, so including them again in the parton-level cross section would be double-counting.) Including the NLO parton-level cross section cancels the $Q$ dependence of the pdf's up to terms of order $\alpha_s^2$; cancelling those terms would require an NNLO parton-level calculation, {\it etc.} The upshot is that while the true hadron-level cross section is $Q$-independent, in practice theoretical predictions always have a residual $Q$-dependence due to uncalculated higher order terms in perturbation theory. This raises two questions: How does one choose $Q$? And, since the $Q$ dependence should presumably be considered as a systematic uncertainty on theoretical predictions, how does one assign a conservative but reasonable value for this uncertainty? Not surprisingly, there are no precise answers. Several prescriptions have been proposed based on physical arguments. As more and more higher-order calculations become available, these prescriptions can be tested against ``data": For example, if both NLO and NNLO answers are known, as is the case for $pp\to Z+X$, one can ask whether the central value and the error bar based on an NLO calculation would be compatible with the NNLO result. Without going into details, let me simply state the simplest prescription: For the central value, take $Q$ to be the invariant mass of the final state, in our case, $M_Z$. (This is motivated since higher-order terms involving collinear emission are enhanced by $\log (s/Q^2)$, as we saw in Section~2; the choice $Q\sim M_Z$ avoids large logs.) For the error bar, take the variation of the cross section as $Q$ is varied between $M_Z/2$ and $2M_Z$. This recipe is often used in practice, and tends to roughly agree with more sophisticated prescriptions. 

\begin{figure}
\begin{center}
\psfig{file=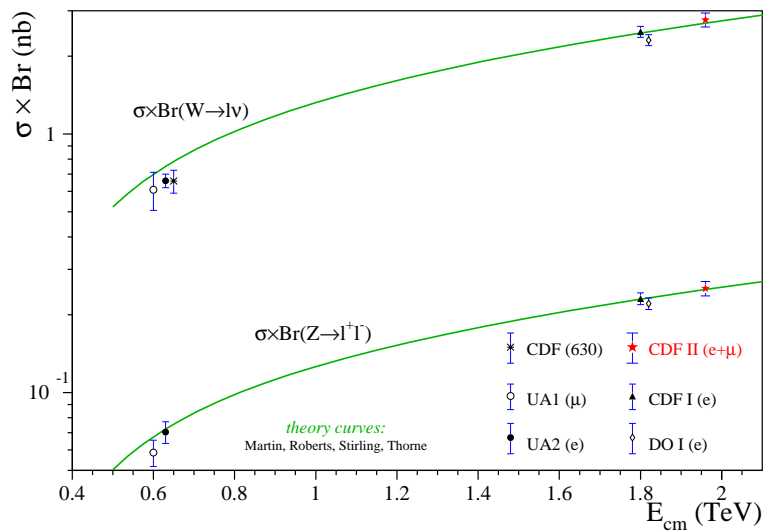,width=10cm}
\end{center}
\caption{$p\bar{p}\to Z+X$ and $p\bar{p}\to W+X$ cross sections, multiplied by leptonic branching ratios, as a function of center-of-mass energy. Data collected by CDF and D{\O} collaborations at the Tevatron Run-I and Run-II, as well as earlier data from the CERN SPS collider, are shown. Solid lines: SM prediction. From the CDF collaboration web site~\cite{Zxsec_CDF}.}
\label{fig:Zxsec}
\end{figure}

Returning to our calculation, we can easily perform one of the integrals:
\beq
\int_0^1 dx_2 \, \delta(x_1x_2 s-M_Z^2) \,=\, \frac{1}{x_1s} \,\theta(x_1s-M_Z^2)\,,
\eeq{int1}
yielding
\beq
\sigma(pp\to Z+X) \,=\, \frac{4\pi^2}{3}\,\frac{\Gamma_Z}{M_Z}\,\frac{1}{s}\int_{M_Z^2/s}^1 \frac{dx_1}{x_1}\,\sum_q 2 f_q(x_1, M_Z) f_{\bar{q}}(\frac{M_Z^2}{x_1s},M_Z)\, {\rm Br}\,(Z\to q\bar{q}) \,. 
\eeq{Zproton_xsec2}
Note that only partons carrying momentum above a certain threshold can participate in $Z$ production: 
\beqa
x_{\rm min}\,=\,\frac{M_Z^2}{s} &\approx& 2.5\times 10^{-3}~{\rm at~the~Tevatron}\,,\CR
&\approx& 4.0\times 10^{-5}~{\rm at~the~LHC}\,.
\eeqa{xmin}
Comparing with Table~\ref{tab:PDF_processes}, we confirm that pdf measurements at HERA probe the entire range of $x$ values necessary for the Tevatron calculation, and almost the entire range (except for the extreme low-$x$ corner) needed for the LHC. (In practice, the cross section only receives a small contribution from $x\sim x_{\rm min}$, since this requires the other parton to be close to $x=1$, where pdf's are suppressed. Most of the cross section is typically contributed by the region with $x_1\sim x_2 \sim \sqrt{x_{\rm min}}$, although this is not a sharp peak.) The next step is to perform the $x_1$ integral. This can be done numerically using the pdf's downloaded from the web site~\cite{DurhamPDF} in {\tt C++}, {\tt Fortran}, or {\tt Mathematica}-compatible versions. To get an estimate, we can use rough log-linear extrapolations of the pdf's: for example ({\it cf.} Fig.~\ref{fig:PDF_all})
\beq
xu(x, M_Z) \,=\,\begin{cases} -1.2 -0.9 \log_{10} x\,,& \text{$10^{-3}<x\leq10^{-2}$;}\\ 
			  0.7\,,& \text{~~~$10^{-2}<x\leq0.1$;}\\
                             -0.2 -0.9 \log_{10} x\,,& \text{~~~$0.1<x\leq0.6$;}\\
0\,,& \text{~~~$x>0.6$.}\end{cases}
\eeq{updf_extrap} 
For the Tevatron parameters, this gives
\beqa
\int_{x_{\rm min}}^1\,\frac{dx_1}{x_1}\,u(x_1)\,u\left(\frac{x_{\rm min}}{x_1}\right) &\approx & 600\,,\CR
\int_{x_{\rm min}}^1\,\frac{dx_1}{x_1}\,\bar{u}(x_1)\,\bar{u}\left(\frac{x_{\rm min}}{x_1}\right) &\approx & 5\,.
\eeqa{Tev_pdfs}
So, $Z$ production at the Tevatron is mostly due to valence quark collisions, with sea partons contributing only about 1\% of the cross section. Plugging in the numbers gives the cross section (from $u\bar{u}$ collisions alone) of about 3 nb. The actual SM prediction, including all quark flavors as well as NLO-QCD corrections, is 8.2 nb. Note that the NLO (${\cal O}(\alpha_s)$) corrections are not small, enhancing the cross section by about 30\%. (This is not unique to $Z$ production: many hadron-initiated processes receive an NLO-QCD correction of a similar size or even larger.) Next-to-next-to-leading order (NNLO) corrections have also been computed~\cite{Ztot_NNLO,Zrap_NNLO} and are numerically much smaller, indicating convergence of the perturbation series. The SM predictions for inclusive $Z$ production, and for a very similar process $p\bar{p}\to W+X$, are in excellent agreement with data, as shown in Fig.~\ref{fig:Zxsec}.

At the LHC, we obtain
\beq
2 \int_{x_{\rm min}}^1\,\frac{dx_1}{x_1}\,u(x_1)\,\bar{u}\left(\frac{x_{\rm min}}{x_1}\right) \,\approx\, 6\times 10^5\,,
\eeq{LHCpdf}
and the total cross section is about 60 nb. $Z$ production at the LHC is dominated by sea parton collisions, and the cross section is {\it larger} than that at the Tevatron, thanks to the growth of pdf's at low $x$. The growth of cross section with $\sqrt{s}$ is due to the composite nature of hadrons, and is in sharp contrast with the decrease of cross sections with $\sqrt{s}$ in {\it elementary} particle collisions, see Eq.~\leqn{xsec_scaling}.

\subsubsection{$Z$ Rapidity Distribution}
\label{sec:Zrap}

\begin{figure}
\begin{center}
\begin{picture}(150,70)(0,0)
\SetOffset(20,-50)
\LongArrow(41,90)(1,90)
\LongArrow(41,90)(81,90)
\DashLine(-10,90)(121,90){3}
\LongArrow(115,90)(121,90)
\LongArrow(1,90)(1,60)
\LongArrow(81,90)(81,120)
\LongArrow(41,90)(81,120)
\LongArrow(41,90)(1,60)
\Vertex(41,90){3}
\Text(43,87)[tl]{$Z$}
\Text(117,87)[tc]{$z$}
\Text(68,87)[tc]{$\hat{{\bf p}}_{\parallel 1}$}
\Text(21,94)[bc]{$\hat{{\bf p}}_{\parallel 2}$}
\Text(83,105)[cl]{$\hat{{\bf p}}_{\perp 1}$}
\Text(-1,75)[cr]{$\hat{{\bf p}}_{\perp 2}$}
\Text(59,107)[br]{$\hat{{\bf p}}_1$}
\Text(23,73)[tl]{$\hat{{\bf p}}_2$}
\end{picture}

\end{center}
\caption{Kinematics of $Z\to \ell^+\ell^-$ decay in the parton frame.}
\label{fig:Lkin}
\end{figure}
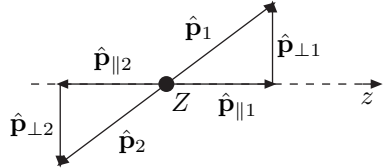

The produced $Z$ boson is at rest in the parton frame; in the lab frame, it has no transverse momentum, and is simply moving with velocity $\beta$, see Eq.~\leqn{beta}, along the beamline.
The $Z$ decays promptly; let us consider the leptonic decay $Z\to \ell^+\ell^-$, where $\ell=e$ or $\mu$. (The {\it ``Drell-Yan"} process $pp\to Z\to \ell^+\ell^-$ provides the cleanest  signature of the $Z$ at hadron colliders due to low backgrounds. Hadronic $Z$ decays must be distinguished from a large background of 2-jet events from pure QCD processes, which is challenging.) The kinematics of the leptons in the parton frame is shown in Fig.~\ref{fig:Lkin}. For each lepton, its {\it pseudorapidity} $\heta_i$ is defined by
\beq
\tanh \hat{\eta}_i \,=\, \frac{\hat{p}_{\parallel i}}{ \hat{E}_i}\,,
\eeq{eta_def}
where hats on all symbols indicate that they refer to parton-frame values of the variables. (The term ``pseudorapidity" is often abbreviated to simply ``rapidity" in collider physics applications, since true rapidity is never used; we will follow this practice below.) With this definition,
\beqa
\hat{p}_{\parallel i} &=& \hat{p}_{\perp i} \sinh \heta_i\,,\CR
\hat{E}_i = \sqrt{(\hat{p}_{\parallel i})^2+(\hat{p}_{\perp i})^2} &=& \hat{p}_{\perp i} \cosh \heta_i\,,
\eeqa{heta}
where $\hat{p}_{\perp i}=|{\hat{\bf p}}_{\perp i}|$.  Note that $\hat{p}_{\perp 1}= \hat{p}_{\perp 2}$, and $\heta_1=-\heta_2$. To find lepton momenta in the lab frame, we must boost by $-\beta_z=\frac{x_2-x_1}{x_2+x_1}$. Defining 
\beq
\eta_z \,=\, \tanh^{-1} \beta_z \,\equiv\, \frac{1}{2}\,\log \frac{1+\beta_z}{1-\beta_z}\,,
\eeq{boost_rap}
it is easy to show that 
\beq
p_{\perp i} = \hat{p}_{\perp i},~~~\eta_i = \heta_i + \eta_z\,,
\eeq{boost}
where the symbols with no hats refer to the lab-frame values. The variables $p_\perp$ and $\eta$ (in addition to the asymuthal angle $\phi$) fully define the lepton momentum, and have an important advantage of very simple transformations under boosts along the beamline: $p_\perp$ and $\phi$ are invariant, and $\eta_i$ transforms additively. This feature makes this set of kinematic variables extremely convenient, and they are in fact widely used not just for the process at hand but throughout hadron collider physics. Their relation to the directly measured lab-frame energy $E$ and scattering angle $\theta$ is 
\beq
p_\perp = E \sin \theta\,,~~\sinh\eta_i = \frac{1}{\tan\theta}\,.
\eeq{rela} 

\begin{figure}
\begin{center}
\psfig{file=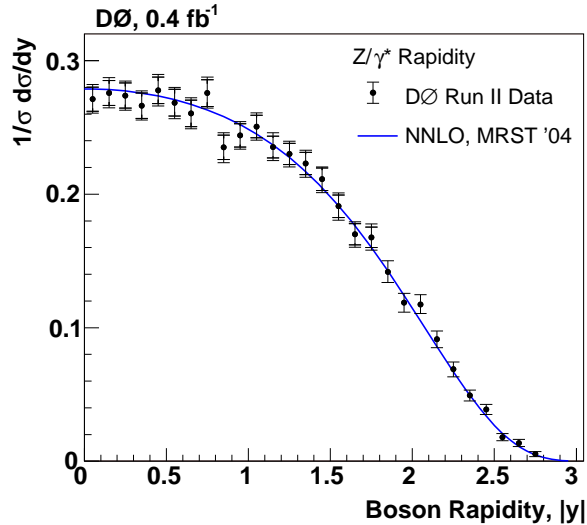,width=8cm}
\end{center}
\caption{Data points: $Z$ boson rapidity distribution measured in $p\bar{p}\to Z\to e^+e^-$ at the Tevatron by D{\O}. Solid line: NNLO theoretical prediction by Anastasiou {\it et.al.}~\cite{Zrap_NNLO}. From Ref.~\cite{D0_Zeta}.}
\label{fig:Zeta}
\end{figure}

Note that the quantity $\eta_z$ is simply the $Z$-boson rapidity in the lab frame. In each event, $\eta_z$ can be determined experimentally: since $\heta_1=-\heta_2$, we have $\eta_z=\eta_1+\eta_2$. On the other hand, 
\beq
\eta_z \,=\,\frac{1}{2}\,\log \frac{1+\beta_z}{1-\beta_z}\,=\,\frac{1}{2}\log\frac{x_1}{x_2}\,=\,
\frac{1}{2}\log\left(\frac{s}{M_Z^2}x_1^2\right)\,.
\eeq{Zrap} 
Thus, measuring the $Z$ rapidity distribution in effect provides the distribution of $x$ values of partons contributing to $Z$ production, giving an indirect but powerful constraint on pdf's. The distribution measured by the D{\O} collaboration at the Tevatron, along with the NNLO theoretical prediction of Ref.~\cite{Zrap_NNLO}, is shown in Fig.~\ref{fig:Zeta}. The agreement between theory and experiment is spectacular. Note that, while the shape of the curve depends on the pdf's, the location of the endpoints is easy to understand: $x_1\in [x_{\rm min}, 1]$ implies $|\eta_z|\leq \frac{1}{2}\log s/M_Z^2=3.0$ at the Tevatron. (At the LHC, $\eta_{\rm max}=5.0$.)

The inclusive processes whose cross sections we computed here are sometimes called ``$W/Z+\geq 0$ jets". Cross sections with 1 or more jets in the final state are also of interest. In general, cross sections for $W/Z+\geq 1$ jet, $W/Z+\geq 2$ jets, etc. are easier to compute and compare to experiment than those with {\it exactly} specified number of jets, due to the ambiguities in counting jets arising from collinear gluon emission (see Section~\ref{sec:jets}). These cross sections depend sensitively on the minimum $p_\perp$ required for the jet(s), since low-$p_\perp$ jet rate is large due to the collinear singularity in ISR gluon emission. Very roughly speaking, each extra jet in the final state with $p_\perp\gsim M_Z$ reduces the cross section by $\alpha_s\sim 0.1$, but the suppression is smaller if $p_\perp$ is lowered.

\subsubsection{$W$ Mass Measurements}

It is easy to measure the $Z$ mass at a hadron collider: One simply needs to select the Drell-Yan events (those with with two opposite-charge leptons in the final state), and compute the {\it dilepton invariant mass}, $s_{12} = (p_1+p_2)^2$, in each event. Most Drell-Yan events with large invariant mass come from $Z$ decays, with a small contribution from off-shell photon exchanges. 
While the $Z$ velocity along the beam line varies between events, it does not affect $s_{12}$, which is Lorentz invariant. So, plotting the number of events vs. $s_{12}$ will produce a Breit-Wigner curve, with a peak at $M_Z$ and width $\Gamma_Z$. This method provides a simple and unambiguous determination of the mass of any unstable particle, provided that it has at least one decay channel where {\it all} decay products can be detected, and backgrounds in that channel are manageable.

Unfortunately, this is not the case for the $W$. Hadronic $W$ decays produce dijet final states, which are buried under a large background from pure-QCD events (see Section~\ref{sec:ppjet}). Leptonic decays are clean, but involve a weakly-interacting particle, the neutrino, whose momentum cannot be measured. Thus, the simple invariant-mass measurement is impossible, and other techniques have to be used. I will outline two useful approaches below. 

\begin{figure}
\begin{center}
\psfig{file=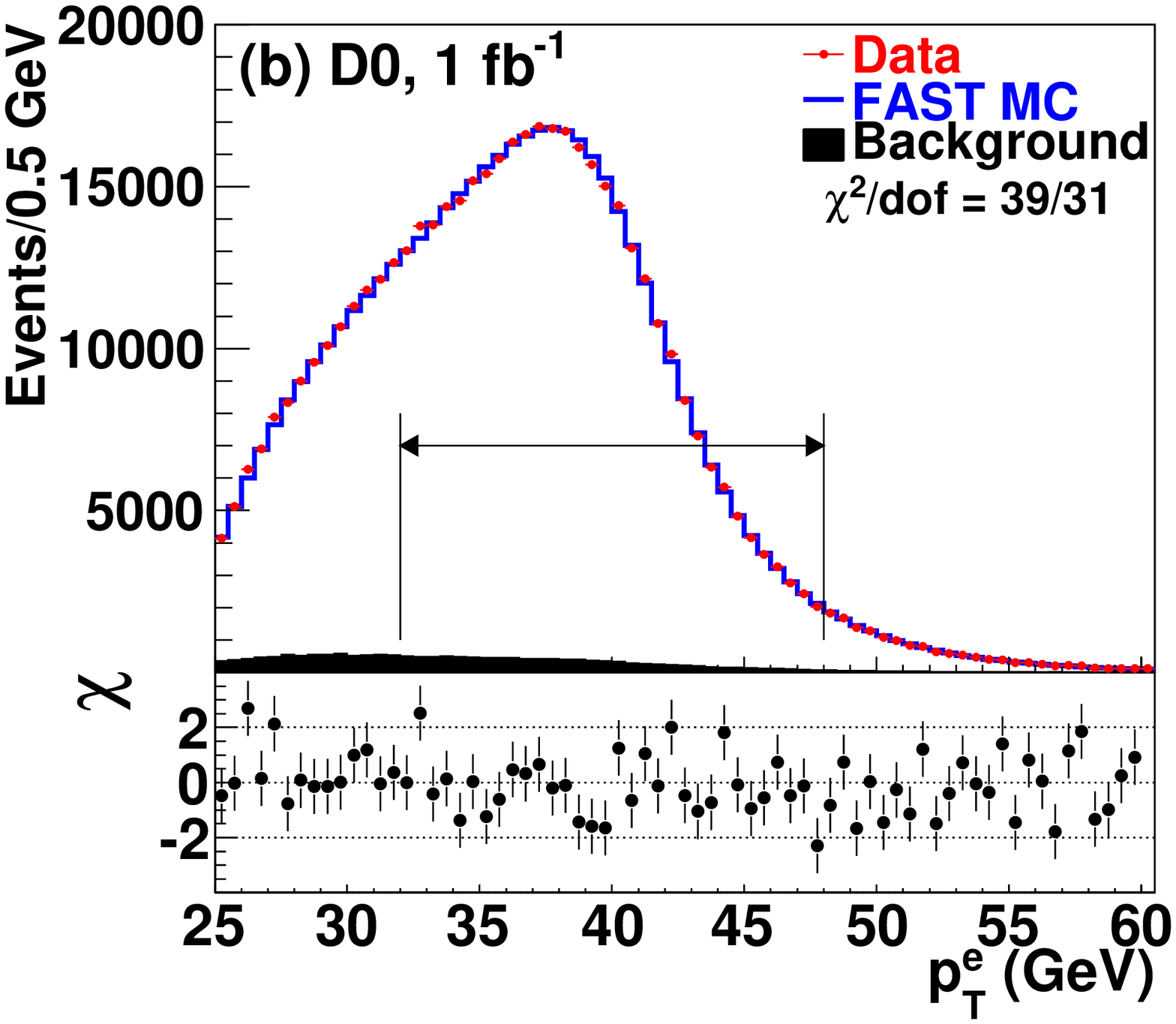,width=7.5cm}
\hskip.5cm
\psfig{file=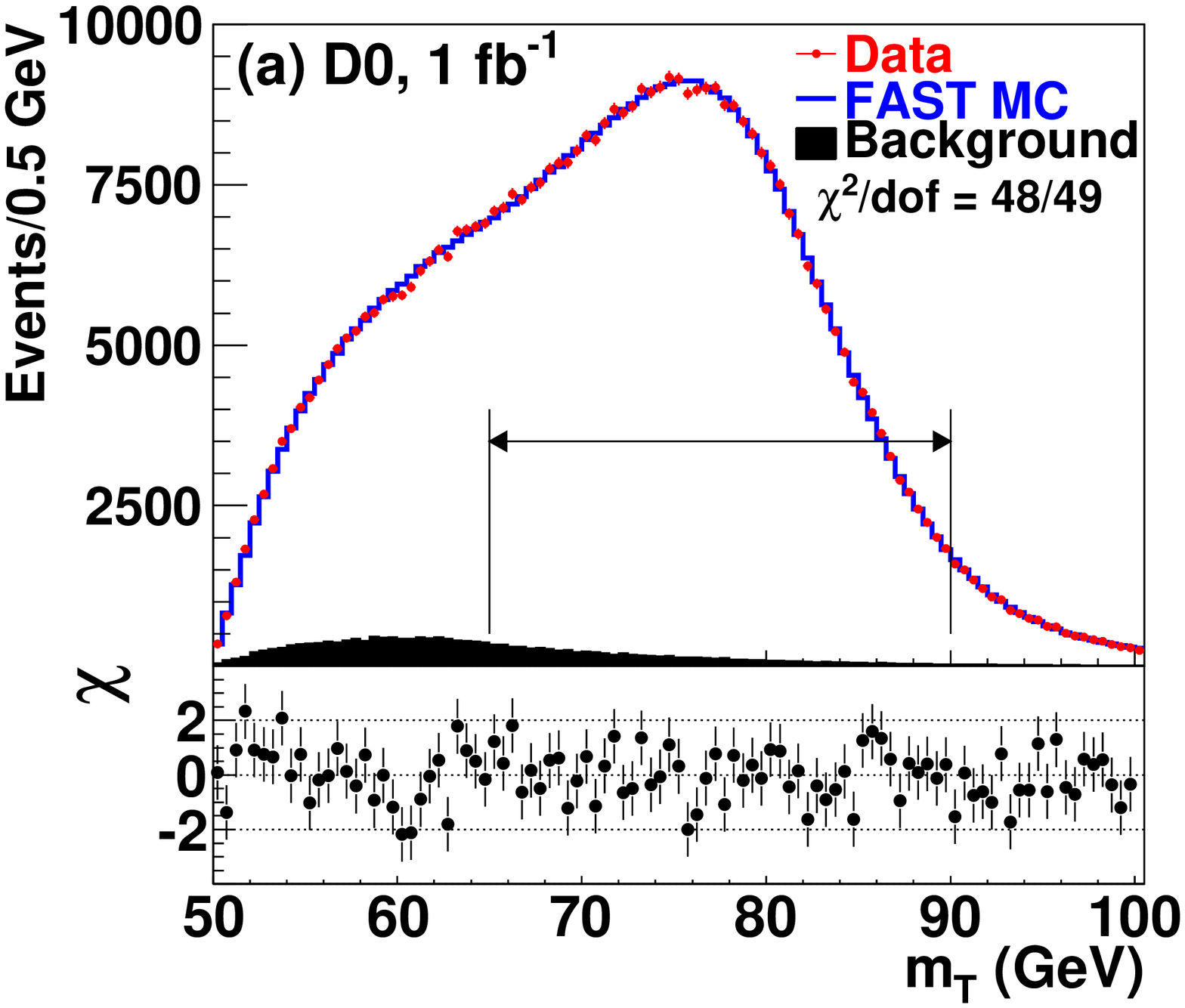,width=7.5cm}
\end{center}
\caption{Distributions of the electron $p_T$ (left panel) and transverse mass (right panel) in the reaction $p\bar{p}\to W\to e\nu$, measured by the D{\O} collaboration at the Tevatron. From Ref.~\cite{D0_mw}.}
\label{fig:mw}
\end{figure}

The first approach is to look at the $p_\perp$ distribution of the charged lepton. Since (at least to leading order in $\alpha_s$) the produced $W$ is at rest in the parton frame, we have
\beq
p_\perp \,=\, \hat{p}_\perp \,=\, \hat{E} \sin \hat{\theta} \,=\, \frac{M_W}{2}\,\sin\hat{\theta}\,,
\eeq{pperp}
where hats indicate quantities evaluated in the parton frame, and the lepton mass has been set to zero. It follows that 
\beq
0 \leq p_\perp \leq \frac{M_W}{2}\,.
\eeq{pperp1}
Moreover,
\beq
\frac{d\sigma}{dp_\perp} \,=\, \frac{d\cos\hat{\theta}}{dp_\perp}\,\frac{d\sigma}{d\cos\hat{\theta}}\,=\, \frac{p_\perp}{\sqrt{\left( \frac{M_W}{2} \right)^2 - p_\perp^2}}\, \frac{d\sigma}{d\cos\hat{\theta}}\,.
\eeq{pperp2}
It is easy to show that $\frac{d\sigma}{d\cos\hat{\theta}}$ does not vanish at $\hat{\theta}=\pi/2$, corresponding to $p_\perp=M_W/2$. Eqs.~\leqn{pperp1},~\leqn{pperp2} then imply that the lepton-$p_\perp$ distribution blows up at its upper boundary, $M_W/2$, and then abruptly drops to zero. 
This discontinuity is tempered by the $W$ width effects, but a sharp peak at $M_W/2$, called {\it Jacobean peak} to reflect its origin in the variable change $\hat{\theta}\to p_\perp$, remains. 
The peak is clearly seen in the distribution of electron $p_\perp$ from $W$ decays measured by the D{\O} collaboration at the Tevatron, shown on the right panel of Fig.~\ref{fig:mw}. 
Measuring the position of this peak determines the $W$ mass.

A difficulty in using the lepton $p_\perp$ is that the transverse motion of the $W$, for example arising from recoils against an extra jet which can be emitted at NLO in $\alpha_s$, must be carefully taken into account before an accurate mass determination can be made. An alternative variable, which does not suffer from this problem, is the {\it transverse mass}. To define it, note that while neutrino cannot be detected, its momentum in the direction {\it transverse} to the beam can be reconstructed, since the sum of all transverse momenta in the event must be zero by momentum conservation:
\beq
{\bf p}_\perp^\nu \,=\, - {\bf p}_\perp^\ell - \sum_j {\bf p}_\perp^j\,,
\eeq{nu_pT}
where the sum is over all jets in the event. (The neutrino momentum along the beamline cannot be reconstructed, since the momentum carried by remnants of the colliding protons along this direction cannot be measured.) The transverse mass is then defined as
\beq
m_T^2 \,=\, (|{\bf p}_\perp^\nu|+|{\bf p}_\perp^\ell|)^2\,-\,({\bf p}_\perp^\nu+{\bf p}_\perp^\ell)^2\,.
\eeq{mT_def} 

~\\
{\bf Homework Problem 3:} Show that $0\leq m_T^2\leq M_W^2$. Hint: compare $m_T^2$ with the (unobservable) $e\nu$ invariant mass, $s_{e\nu}=(|{\bf p}^\nu|+|{\bf p}^\ell|)^2\,-\,({\bf p}^\nu+{\bf p}^\ell)^2 = M_W^2$.
~\\

Just like for the lepton $p_\perp$, the change of variables $\hat{\theta}\to m_T$ introduces a Jacobean factor that blows up at the upper boundary $m_T=M_W$, leading to a Jacobean peak around this value. This peak is clearly visible in the experimental data, shown in the left panel of Fig.~\ref{fig:mw}.

A combination of these techniques yields an amazingly accurate measurement of $m_W$ at the Tevatron: at present, the error is only about 30 MeV, or $\Delta m_W/m_W\approx  4\times 10^{-4}$! This is comparable to the precision achieved at LEP. The lesson is clear: with clever variable choices and thorough understanding of systematic issues, hadron colliders may be capable of matching the precision measurements possible in $e^+e^-$, even for observables which at first glance suffer from the lack of kinematic information. 

\subsection{Dijet Production}
\label{sec:ppjet}

The most common processes at hadron colliders are QCD-mediated reactions with only strongly interacting particles in the final state. Conservation of energy-momentum and fermion number imply that there are no $2\to1$ processes of this kind; the dominant reactions are $2\to2$ processes, which upon hadronization result in two-jet, or {\it dijet}, final states. We will briefly consider dijet production in this subsection.  

In general, the only variable needed to describe the kinematics of a $2\to2$ scattering process in the parton frame is the scattering angle $\theta$, or, equivalently, the Mandelstam variable $t$ (for massless particles, $t=-s(1-\cos\theta)/2$). In the case of hadron collisions, there are two additional variables, $x_1$ and $x_2$, which determine the parton c.o.m. energy as well as the motion of the parton frame with respect to the lab frame. To compare with data, it is convenient to change variables from $(\hat{t}, x_1, x_2)$ to the directly observable $(p_\perp, \eta_1, \eta_2)$. In terms of these variables, the triple-differential cross section has the form
\beq
\frac{d^3\sigma}{dp_\perp d\eta_1 d\eta_2} \,=\, 2p_\perp x_1 x_2 \, \sum_{i_1,i_2} f_{i_1}(x_1) f_{i_2}(x_2)\,\frac{d\sigma}{d\hat{t}} (i_1+i_2\to 1+2)\,,
\eeq{triple_dif}
where
\beqa
x_1 &=& \frac{2p_\perp}{\sqrt{s}}\,\cosh \eta_- \,e^{\eta_+}\,,\CR
x_2 &=& \frac{2p_\perp}{\sqrt{s}}\,\cosh \eta_- \,e^{-\eta_+}\,,\CR
\hat{t} &=& -2p_\perp^2\,\cosh \eta_- e^{-\eta_-}\,,
\eeqa{var_change}
with $\eta_\pm = (\eta_1\pm\eta_2)/2$.

~\\
{\bf Homework Problem 4:} Derive the relations~\leqn{var_change} and the triple-differential distribution~\leqn{triple_dif}.
~\\

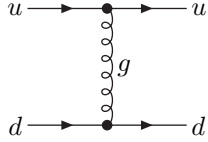
\begin{figure}
\begin{center}
\begin{picture}(90,65)(0,0)
\SetOffset(20,-55)
\ArrowLine(1,109)(31,109)
\ArrowLine(1,65)(31,65)
\Gluon(31,109)(31,65){2}{6}
\ArrowLine(31,109)(61,109)
\ArrowLine(31,65)(61,65)
\Vertex(31,109){2}
\Vertex(31,65){2}
\Text(-1,65)[cr]{$d$}
\Text(-1,109)[cr]{$u$}
\Text(63,65)[cl]{$d$}
\Text(63,109)[cl]{$u$}
\Text(35,87)[cl]{$g$}
\end{picture}
\end{center}
\caption{The Feynman diagram for the process $ud\to ud$ at the tree level.}
\label{fig:udud}
\end{figure}

\begin{figure}
\begin{center}
\psfig{file=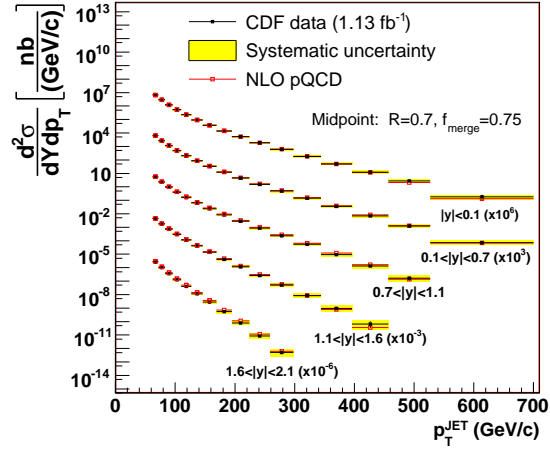,width=8cm}
\end{center}
\caption{Jet cross sections in five rapidity regions, as a function of jet $p_\perp$, measured by the CDF collaboration at the Tevatron and compared with the NLO-QCD calculation. The cross sections for the five regions are scaled by a factor of $10^3$ from each other for presentation purposes. From Ref.~\cite{CDF_dijet}.}
\label{fig:CDF_dijet}
\end{figure}

To compute the dijet rates, we need to know parton-level differential cross sections $d\sigma/d\hat{t}$ for each possible reaction with two strongly interacting particles in the final state. There is a large number of such reactions. Here is a partial list: $q\bar{q}\to q\bar{q}$, where $q$ can be any of the quark flavors; $q\bar{q}\to gg$; $q\bar{q}\to q^\prime\bar{q}^\prime$ with $q^\prime\not=q$; $qq\to qq$; $qq^\prime\to qq^\prime$; $gq\to gq$; $qq\to gg$; etc. As an illustrative example, consider the subprocess $ud\to ud$. The single Feynman diagram that contributes to this process at leading order in $\alpha_s$ is shown in Fig.~\ref{fig:udud}. The cross section is
\beq
\frac{d\sigma}{d\hat{t}} \,=\, \frac{4\pi \alpha_s^2}{9\hat{s}^2} \,\Bigl[ \frac{\hat{s}^2+\hat{u}^2}{\hat{t}^2} \Bigr]\,.
\eeq{udud_xsec}
Using $\hat{s}=x_1x_2s$ and $\hat{s}+\hat{t}+\hat{u}=0$, together with Eqs.~\leqn{var_change}, it is easy to show that 
\beq
\frac{d\sigma}{d\hat{t}}\,\propto\, \frac{1}{p_\perp^4},~~~{\rm as}~p_\perp\to 0\,.
\eeq{udud_limit} 
At the hadron level, the triple-differential cross section (for fixed jet rapidities) behaves as
\beq
\frac{d^3\sigma}{dp_\perp d\eta_1 d\eta_2} \propto p_\perp^3 \cdot f_u (c_up_\perp) f_d(c_d p_\perp) \cdot \frac{1}{p_\perp^4}+ \ldots\,,
\eeq{triple_limit}
where $c_u$ and $c_d$ are constants independent of $p_\perp$, and dots indicate contributions from other subprocesses. Since the pdf's grow faster than $1/x$ as $x\to 0$, the right-hand side of Eq.~\leqn{triple_limit} grows faster than $1/p_\perp^3$, and the hadron-level cross section diverges strongly as $p_\perp\to 0$. This singularity is of a different nature than the soft and collinear singularities encountered so far: It is due to an exchange of a massless particle in the $t$-channel, and is exactly analogous to the Rutherford singularity in the elastic scattering by Coulomb interaction in QED. The singularity (and similar divergences that occur in other subprocesses involving $t$ or $u$-channel exchanges) is effectively regulated by the IR divergence of the strong coupling constant: When the momentum exchanged in the $t$ channel is of order $\Lambda_{\rm QCD}$, perturbation theory breaks down and the leading-order result is no longer valid. As a result, the total rate for pure-QCD events is simply given by the geometric cross section 
\beq
\sigma_{\rm QCD} \sim \frac{4\pi}{\Lambda_{\rm QCD}^2}\,\approx 0.1~{\rm bn}.
\eeq{allQCD}
This cross section is huge: at current luminosity, there are over $10^7$ pure-QCD events per second at the Tevatron. This rate is far too high for the experiments to be able to record all events on tape, let alone analyze them. However, the vast majority of pure-QCD events involve only hadrons with $p_\perp\sim \Lambda_{\rm QCD}$, whereas events of interest for probing the SM at high energy scales and searching for new physics  involve much higher transverse momenta. This leads to a strategy called {\it triggering}: only events with at least one jet above some minimal $p_\perp$ value are recorded and analyzed. (Events with non-strongly-interacting objects, such as leptons or photons, are far less numerous than pure-QCD, and most of them are also recorded.) Thus, the experimentally relevant cross section is that for producing jets with $p_\perp$ above the threshold required by the trigger (or a larger value set by analysis cuts). For example, for $p_\perp\geq 100$ GeV, a typical value for the LHC analyses, the total jet cross section at the LHC is about 1 mb, about a factor of 20 larger than the inclusive $Z$ production cross section.

Comparing dijet differential cross sections with data provides a valuable test of QCD at high energy scales, as well as the most accurate determination of gluon pdf at high $x$. Within limitations due to theoretical and experimental uncertainties in jet definitions, uncalculated higher-order perturbative corrections, pdf uncertainties, and other systematic errors, all available data is in agreement with the SM. An example is provided by a recent CDF measurement of dijet rates at the Tevatron, see Fig.~\ref{fig:CDF_dijet}.

\subsection{Hadron Collider Cross Section Summary}

\begin{figure}
\begin{center}
\psfig{file=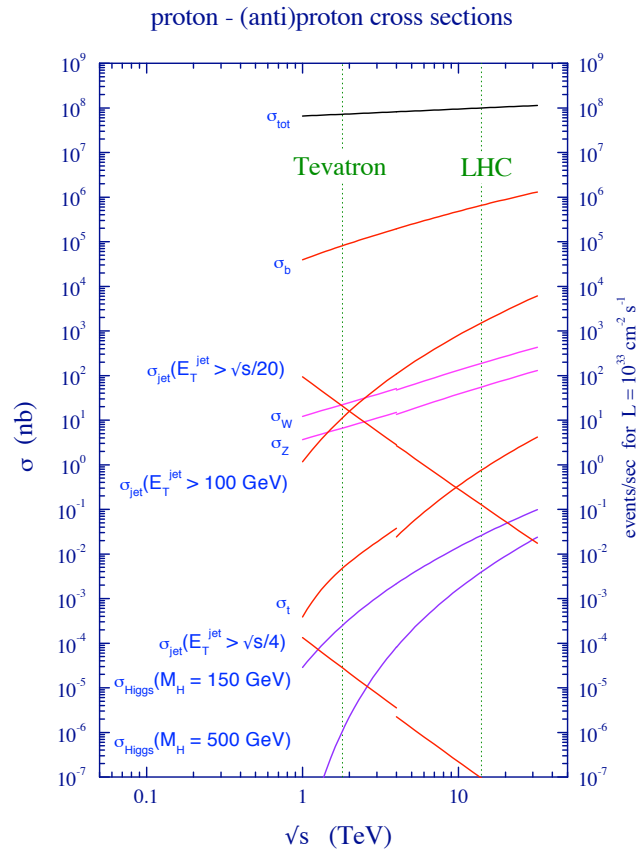,width=10cm}
\end{center}
\caption{Cross sections for a variety of SM reactions, as a function of center-of-mass energies, in $p\bar{p}$ collisions (for $\sqrt{s}<4$ TeV) and $pp$ collisions (for $\sqrt{s}>4$ TeV).}
\label{fig:xsec_all}
\end{figure}

A useful summary of cross sections for the most important SM processes in hadron collisions is shown in Fig.~\ref{fig:xsec_all}. The total scattering cross section $\sigma_{\rm tot}\sim 0.1$ bn is completely dominated by low-$p_\perp$ pure-QCD scattering, and depends only weakly (logarithmically) on $\sqrt{s}$. The inclusive cross section for $b$ quark production is 2-3 orders of magnitude lower, reflecting the effective $p_\perp$ cutoff provided by the $b$ mass. The strongest electroweak processes, inclusive $W$ and $Z$ production, have cross sections of order 10-100 nb: in other words, there is 1 electroweak boson produced per 1-10 million pure-QCD events. Of course, as discussed above, the vast majority of pure-QCD events are not recorded on tape, and experimentally relevant QCD events are those with at least one jet with $p_\perp$ above a trigger threshold. While specific thresholds depend on the experiment and the analysis, the figure shows the total jet cross sections for a few representative values: $\sqrt{s}/20$, $\sqrt{s}/4$, and 100 GeV. As a rough guidance, it is useful to remember that at the Tevatron, the jet rate with $p_\perp\geq 100$ GeV is roughly comparable to the electroweak boson production rates, while at the LHC it is larger by 1-2 orders of magnitude. Going further down the cross section scale, the top pair-production cross section is in the few-pb range at the Tevatron, and close to a nb at the LHC. The sharp rise with $\sqrt{s}$ is due to the fact that top pairs are typically produced by partons with $x\sim 0.2$ at the Tevatron and $x\sim 0.02$ at the LHC; the higher parton density at low $x$, especially for gluons, increases the cross section. At design luminosity, the LHC will produce about 10$^7$ top pairs a year, allowing for detailed studies of top properties (but also providing an important background for many new physics searches). New physics cross sections are generally further suppressed: as an example, cross sections for the SM Higgs production, for two representative values of the Higgs mass, are shown in the figure. The main challenge in new physics searches at hadron colliders is to distinguish the rare new physics events from the much more common SM processes. To do this, one must identify specific observable features, or {\it signatures}, of the new physics events that distinguish them from the SM.

\section{Searches for New Physics}
\label{sec:newphys}

In this Lecture, I will first outline the general strategy for new physics searches in collider experiments. I will then consider two specific examples: searches for the Higgs boson, and supersymmetry.

\subsection{General Strategy}

Suppose that you are given a model that extends the SM and predicts new particles at energy scales potentially within reach of current or near-future colliders. Your task is to plan a search for this model: You need to identify its experimental signatures, devise a data analysis strategy in order to maximize sensitivity to new physics, and evaluate how much data would be required to convincingly establish, or rule out, the theory. (Parenthetically, note that these tasks fall right on the boundary of what theorists and experimentalists do. Typically, a first-round study would be done by theorists. If the results look promising, a more detailed analysis, taking account of experimental conditions, would be conducted by experimentalists, who will then implement the designed search with real data.) Of course, the detailed strategy would depend on the precise nature of the predicted new particles and their interactions, as well as on the experiment. However, most studies of this kind roughly follow the ``twelve-step" blueprint outlined below.

\begin{enumerate}

\item Suppose that our model predicts a particle $X$, and we would like to design a search for this particle. Start with the Lagrangian of the model. Identify all couplings of $X$ to the SM. There is no need to derive precise Feynman rules yet; simply list the vertices involving $X$ and SM states. 

\item Given the vertices, identify $X$ production mechanisms at the collider of interest. Draw Feynman diagrams with $X$ in the final state. Remember that cross sections generally decrease with the number of interaction vertices and with the number of particles in the final state, so focus on drawing diagrams as simple as possible. If considering a hadron collider, remember that pdf's drop fast with $x$, and therefore cross sections decrease with the invariant mass of the final state. This implies that the final states with highest cross sections will involve as few heavy particles as possible. If the new particle can be singly produced, this will be the dominant process; if single production is forbidden by symmetries (e.g. R-parity in supersymmetric models), pair-production will dominate. 

\item Based on the above considerations, identify the most promising production processes for $X$. You can also order diagrams according to the strength of couplings that enter: for example, $\alpha_s>\alpha_w$, so, everything else being equal, QCD-mediated processes will have higher cross sections than electroweak reactions.

\item Carefully derive the Feynman rules for all vertices that enter the promising production processes. 

\item Compute the total cross section for each of the promising processes. Do not impose cuts, unless necessary to avoid singularities; if necessary, impose very loose cuts roughly consistent with experimental conditions (if unsure, consult your experimental colleagues). Given the integrated luminosity (actual or expected) of your collider, compute the number of events $N_{\rm ev}$ expected in your data set for each of the processes. (See Table~\ref{tab:colliders} for luminosity values.)

\item If $N_{\rm ev}<10$ for a given process, the process is probably uninteresting -- do not consider it further. Note that $N_{\rm ev}$ typically depends on {\it a priori} unknown model parameters, such as the $X$ mass. As long as there is a range of parameter values, consistent with theoretical considerations and with constraints from prior experiments, where the $N_{\rm ev}>10$ condition is satisfied, it is worthwhile to continue with the analysis. 

\item Identify all vertices through which $X$ can decay (both to SM particles and to other exotics). For each possible decay channel, check whether it is kinematically allowed. If there are no viable decays, $X$ is a stable particle. If $X$ is stable, is it charged under strong and/or electromagnetic interactions?

\begin{enumerate}

\item If yes, you need to think carefully about the signature of $X$ in the detector. If only EM charge is present, $X$ would behave like a heavy muon, and will leave a trace in the muon detector. If $X$ is strongly coupled, it will quickly hadronize, and the resulting hadrons with interact in the hadron calorimeter, as well as in the tracker if electrically charged. (As an aside, note that there are significant cosmological constraints on charged and/or strongly-interacting stable or long-lived massive particles.) See K.~Zurek's lectures at this school for more details. 

\item If no, $X$ can at best interact with matter in the detector via weak interactions, or possibly even weaker. In this case, $X$ escapes the detector without interacting, like the SM neutrino, and is ``invisible". Invisible particles can be detected as an apparent missing momentum (transverse momentum in the case of hadron collisions) recoiling against visible particles. So, in this case, make sure that there is at least one observable ``tag" particle in the final state of your process. A jet or a photon, which can always be emitted as ISR, can serve as tags.

\end{enumerate}

\item If $X$ is unstable (as is usually the case for heavy new particles), list all $X$ decay channels, and compute their branching ratios. (Again, focus on the simplest possible decays since those with more interaction vertices are suppressed.) If some of the $X$ decay products are themselves unstable, repeat this step for those particles, and iterate until you only have stable particles left. List all final states in terms of ``detector objects" - jets, SM leptons, photons, invisible particles (neutrinos and possibly exotic stable invisibles), and possibly exotic charged/colored states. For each final state, compute the probability to obtain it from $X$ decay, multiplying the relevant branching ratios. Multiply by $N_{\rm ev}$ to obtain the expected signal rate in each channel, $N_{\rm sig}$.

\item For each detector-level final state with a sizable expected signal rate, list all SM processes that can give rise to this final state - the {\it Standard Model background}. (Again, focus on the simplest possible processes. You can use Fig.~\ref{fig:xsec_all} as a guide to the most common processes and their cross sections, but remember that it is incomplete.)

\item Compute the cross section for each SM background process, and multiply by the luminosity to get $N_{\rm SM}$.

\item A meaningful search ({\it i.e.} one that can lead to a discovery or a publishable constraint on new physics) must satisfy three conditions:

\begin{enumerate}

\item Events: $N_{\rm sig}>$~a few. Detector efficiency should be taken into account when evaluating $N_{\rm sig}$; for a first-round, ``theory-level" analysis, a rough estimate of the efficiency would suffice. Note that $N_{\rm sig}\propto L_{\rm int}$, so this condition can at least in principle always be satisfied by running the experiment long enough.

\item Statistics: $N_{\rm sig}/\sqrt{N_{\rm SM}}>$~a few. ``A few" is typically 3 or 5, depending on how confident one wants to be. In case of an actual discovery, 5 would probably be required before most people are convinced, given the long history of ``3-sigma" effects disappearing with increased statistics. In this case, $N_{\rm sig}/\sqrt{N_{\rm SM}} \propto L_{\rm int}^{1/2}$, so this condition in principle can also be always fulfilled by increasing the data sample size, although the improvement is slower than for (a).

\item Systematics: $N_{\rm sig}/N_{\rm SM}>$~(a few)$\,\times\,\sigma_{\rm sys}$, where $\sigma_{\rm sys}$ is the fractional systematic uncertainty of the SM prediction. In the case of hadron collisions, this uncertainty includes residual scale dependence from uncalculated higher-order corrections, pdf uncertainties, as well as experimental errors, e.g. uncertainty in the detector efficiency. (Talk to your experimental colleagues to get a reasonable estimate for the latter.) Note that $N_{\rm sig}/N_{\rm SM}\propto L_{\rm int}^0$, so no amount of luminosity will help if this condition is not satisfied. Thus, it is crucially important to understand all sources of systematic uncertainty, and to eliminate them as far as possible. 

\end{enumerate}

\item If the three conditions are {\it not} satisfied, one can usually improve the situation by imposing {\it selection cuts}: Look for differences in signal and background kinematic distributions, and devise criteria to select only events in the signal-rich and/or background-poor part of the phase space. This improves signal/background ratio, making conditions (b) and (c) easier to satisfy.(However, one must be careful since (a) can fail if the selection cuts are too tight!)

\end{enumerate}

Once the three conditions are satisfied, the proposed search strategy can be studied in more detail by experimentalists, and, if the conclusions hold up, implemented with real data. You have succeeded!

\subsection{Higgs Searches}

The remainder of this Lecture is a brief overview of some of the most important searches for new physics which are the focus of the current experimental program at the Tevatron, as well as the upcoming one at the LHC.

\begin{figure}
\begin{center}
\begin{picture}(310,90)(0,0)
\SetOffset(20,-40)
\DashLine(1,87)(21,87){3}
\Vertex(21,87){2}
\ArrowLine(21,87)(51,109)
\ArrowLine(51,109)(51,65)
\Gluon(51,109)(76,109){2}{4}
\Gluon(51,65)(76,65){2}{4}
\ArrowLine(51,65)(21,87)
\Vertex(51,109){2}
\Vertex(51,65){2}
\Text(78,109)[cl]{$g$}
\Text(78,65)[cl]{$g$}
\Text(-1,87)[cr]{$h$}
\Text(34,100)[br]{$f$}
\Text(38,54)[cc]{(a)}

\DashLine(101,87)(121,87){3}
\Vertex(121,87){2}
\ArrowLine(121,87)(151,109)
\ArrowLine(151,109)(151,65)
\Photon(151,109)(176,109){2}{4}
\Photon(151,65)(176,65){2}{4}
\ArrowLine(151,65)(121,87)
\Vertex(151,109){2}
\Vertex(151,65){2}
\Text(178,109)[cl]{$\gamma$}
\Text(178,65)[cl]{$\gamma$}
\Text(99,87)[cr]{$h$}
\Text(134,100)[br]{$f$}
\Text(138,54)[cc]{(b)}

\DashLine(201,87)(221,87){3}
\Vertex(221,87){2}
\Photon(221,87)(251,109){2}{6}
\Photon(251,109)(251,65){2}{6}
\Photon(251,109)(276,109){2}{4}
\Photon(251,65)(276,65){2}{4}
\Photon(251,65)(221,87){2}{6}
\Vertex(251,109){2}
\Vertex(251,65){2}
\Text(278,109)[cl]{$\gamma$}
\Text(278,65)[cl]{$\gamma$}
\Text(199,87)[cr]{$h$}
\Text(234,100)[br]{$W$}
\Text(238,54)[cc]{(c)}

\end{picture}
\end{center}
\caption{One-loop Feynman diagrams which induce the Higgs coupling to gluons (a) and photons (b, c). In diagrams (a) and (b) contributions of all quarks and leptons in the loop must be summed over, but in practice the top quark contribution dominates due to its large Yukawa coupling.}
\label{fig:hgg}
\end{figure}
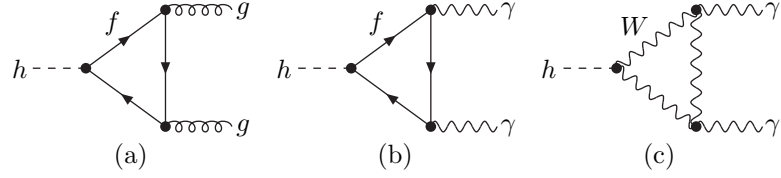

We begin with the search for the Higgs. I will focus on the simplest scenario incorporated in the SM, a single Higgs doublet field yielding one physical Higgs boson $h$. The Higgs potential has two free parameters, the mass parameter $\mu^2$ and the quartic coupling $\lambda$; these can be traded for the Higgs vacuum expectation value (vev) $v=246$ GeV and the physical Higgs mass $m_h$. The Higgs is coupled to SM fermions via Yukawa couplings $y_f h \bar{\psi}\psi$, where $y_f=\sqrt{2}m_f/v$, and to electroweak gauge bosons via the usual gauge vertices. Note that tree-level triple-boson couplings of the form $hVV$ exist for massive vector bosons, $V=W/Z$, but not for the photon. The vertices $h\gamma\gamma$ and $hgg$ arise from the one-loop diagrams in Fig.~\ref{fig:hgg}; while loop-suppressed, these couplings play an important role in Higgs phenomenology. Note that all Higgs couplings can be expressed in terms of known SM parameters; the only unknown parameter in this model is $m_h$. 

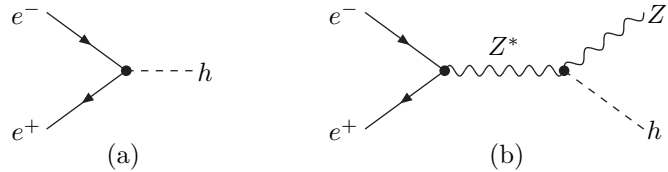
\begin{figure}
\begin{center}
\begin{picture}(270,85)(0,0)
\SetOffset(20,-40)

\ArrowLine(1,109)(31,87)
\ArrowLine(31,87)(1,65)
\DashLine(31,87)(56,87){3}
\Vertex(31,87){2}
\Text(-1,109)[cr]{$e^-$}
\Text(-1,65)[cr]{$e^+$}
\Text(58,87)[cl]{$h$}
\Text(30,55)[cc]{(a)}

\ArrowLine(121,109)(151,87)
\ArrowLine(151,87)(121,65)
\Photon(151,87)(196,87){2}{6}
\Photon(196,87)(226,109){2}{4}
\DashLine(226,65)(196,87){3}
\Vertex(151,87){2}
\Vertex(196,87){2}
\Text(119,109)[cr]{$e^-$}
\Text(119,65)[cr]{$e^+$}
\Text(228,109)[cl]{$Z$}
\Text(228,65)[cl]{$h$}
\Text(174,100)[tc]{$Z^*$}
\Text(175,55)[cc]{(b)}

\end{picture}
\end{center}
\caption{Leading-order diagrams for Higgs production in $e^+e^-$ collisions.}
\label{fig:eeHprod}
\end{figure}

\begin{figure}
\begin{center}
\psfig{file=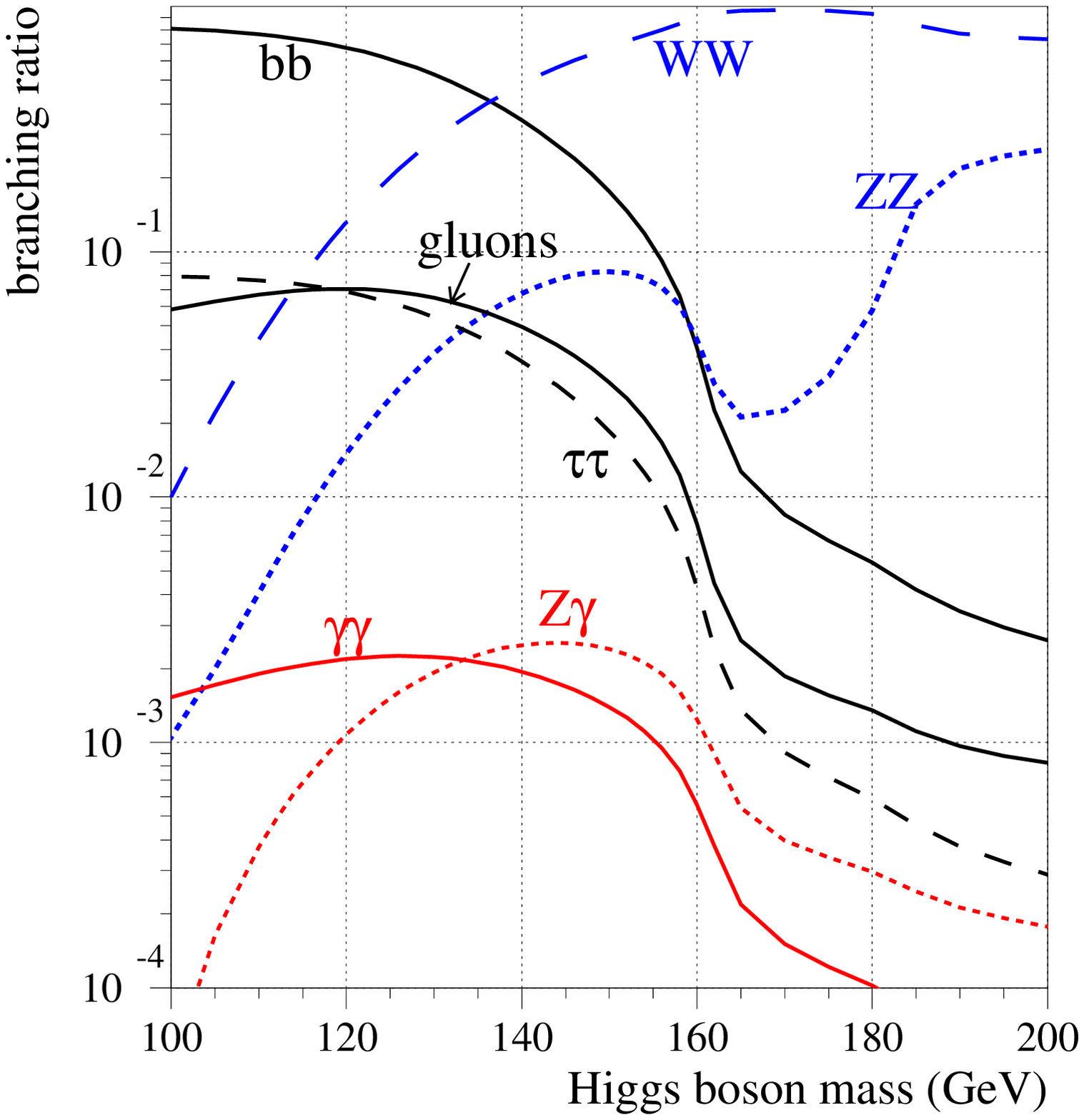,width=7.5cm}
\vskip-1.6cm
\psfig{file=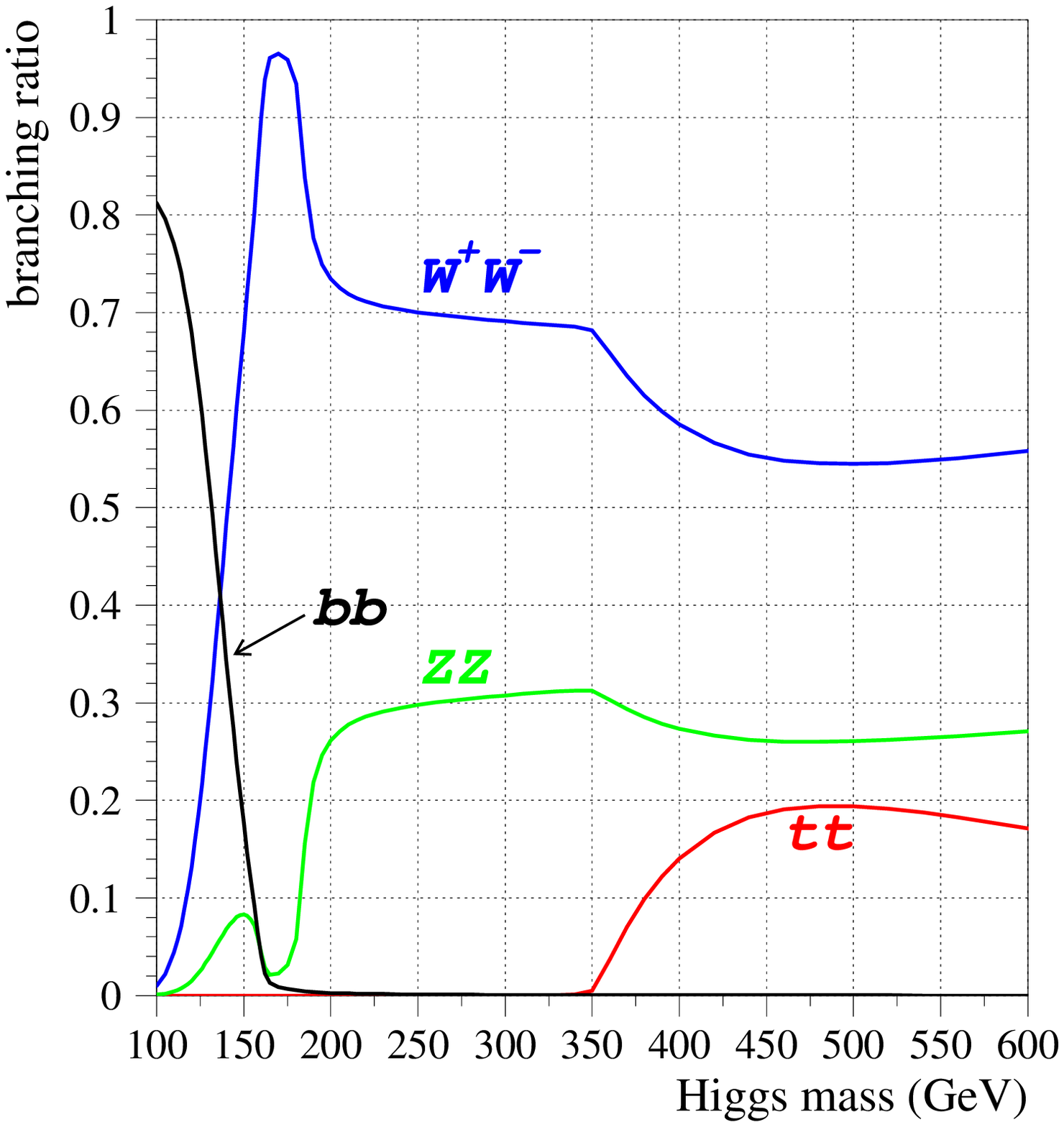,width=6.3cm}
\end{center}
\caption{Branching ratios of the SM Higgs. From the web site~\cite{Higgs_Web}.}
\label{fig:Higgs_BR}
\end{figure}

Having listed the vertices, we can now identify Higgs production processes at colliders. First, consider $e^+e^-$ collisions. The simplest reaction would be $s$-channel resonant production, $e^+e^-\to h$, shown in Fig.~\ref{fig:eeHprod} (a) ; however, the cross section of this process is tiny due to suppression by $y_e^2\sim 10^{-12}$. The strongest production channel is the so-called {\it Bjorken process} shown in Fig.~\ref{fig:eeHprod} (b), which utilizes the gauge coupling of the Higgs. This reaction is kinematically allowed if $\sqrt{s} > m_h + m_Z$. At LEP-2, this implies that the maximum Higgs mass that can in principle be probed is about 118 GeV. In practice, the Higgs mass must be a bit lower to get a sizable production cross section: LEP-2 expected to produce 10 or more Higgs events for $m_h\leq 115$ GeV. To identify the Higgs signature, we next need to 
consider Higgs decays. The SM Higgs branching ratios, as a function of its mass, are presented in Fig.~\ref{fig:Higgs_BR}. Throughout the parameter space accessible at LEP, the dominant Higgs decay mode is $b\bar{b}$, with a sizable (order few-\%) contribution from $\tau^+\tau^-$. Thus, the interesting final states include $b\bar{b}$ or $\tau^+\tau^-$ pairs in association with lepton or jet pairs from $Z$ decays\footnote{Such pairs can be identified in the data by the fact that their invariant mass is close to $m_Z$.}, or missing momentum from invisible $Z$ decays $Z\to\nu\bar{\nu}$. SM backgrounds for all these channels primarily come from $WW$ and $ZZ$ final states, and can be accurately computed. Combining the channels, LEP-2 experiments place the lower bound on the Higgs mass~\cite{PDG}:
\beq
m_h \geq 114.4~{\rm GeV,~~~95\%~c.l.}
\eeq{LEP2_hbound}    
Interestingly, the ALEPH collaboration reported 5 candidate events that may be viewed as a hint for the Higgs boson at about 115 GeV~\cite{ALEPH_Higgs}; however, low statistics, and lack of confirmation by other LEP experiments, make the situation inconclusive.

\begin{figure}
\begin{center}
\begin{picture}(450,90)(0,0)
\SetOffset(20,-35)
\ArrowLine(1,109)(31,87)
\ArrowLine(31,87)(1,65)
\DashLine(31,87)(56,87){3}
\Vertex(31,87){2}
\Text(-1,109)[cr]{$q$}
\Text(-1,65)[cr]{$\bar{q}$}
\Text(58,87)[cl]{$h$}
\Text(30,50)[cc]{(a)}

\Gluon(101,109)(131,109){2}{4}
\Gluon(131,65)(101,65){2}{4}
\ArrowLine(131,65)(131,109)
\ArrowLine(131,109)(151,89)
\ArrowLine(151,89)(131,65)
\DashLine(151,89)(176,89){3}
\Vertex(131,109){2}
\Vertex(131,65){2}
\Vertex(151,89){2}
\Text(99,109)[cr]{$g$}
\Text(99,65)[cr]{$g$}
\Text(178,89)[cl]{$h$}
\Text(138,50)[cc]{(b)}

\ArrowLine(221,109)(241,87)
\ArrowLine(241,87)(221,65)
\Photon(241,87)(276,87){2}{6}
\Photon(276,87)(296,109){2}{4}
\DashLine(296,65)(276,87){3}
\Vertex(241,87){2}
\Vertex(276,87){2}
\Text(219,109)[cr]{$q$}
\Text(219,65)[cr]{$\bar{q}/\bar{q}^\prime$}
\Text(298,109)[cl]{$W/Z$}
\Text(298,65)[cl]{$h$}
\Text(259,100)[tc]{$W/Z$}
\Text(260,50)[cc]{(c)}

\ArrowLine(351,109)(371,109)
\ArrowLine(351,65)(371,65)
\Vertex(371,109){2}
\Vertex(371,65){2}
\Photon(371,109)(391,87){2}{4}
\Photon(371,65)(391,87){2}{4}
\Vertex(391,87){2}
\DashLine(411,87)(391,87){3}
\ArrowLine(371,109)(411,114)
\ArrowLine(371,65)(411,60)
\Text(349,109)[cr]{$q$}
\Text(349,65)[cr]{$q$}
\Text(379,96)[tr]{$W$}
\Text(413,87)[cl]{$h$}
\Text(381,50)[cc]{(d)}

\end{picture}

\end{center}
\caption{Higgs production processes in hadron collisions: (a) $s$-channel resonant production (suppressed by small Yukawas); (b) gluon fusion; (c) associated production with a vector boson; and (d) vector boson fusion.}
\label{fig:ppHprod}
\end{figure}
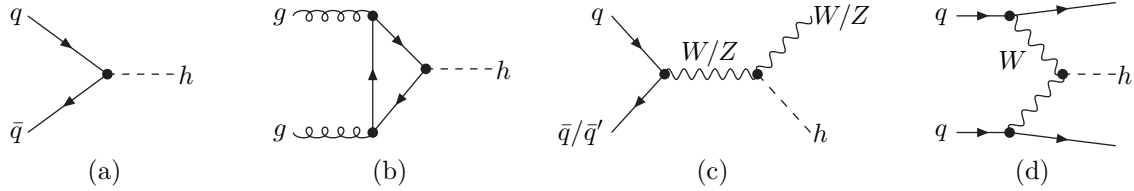

\begin{figure}
\begin{center}
\psfig{file=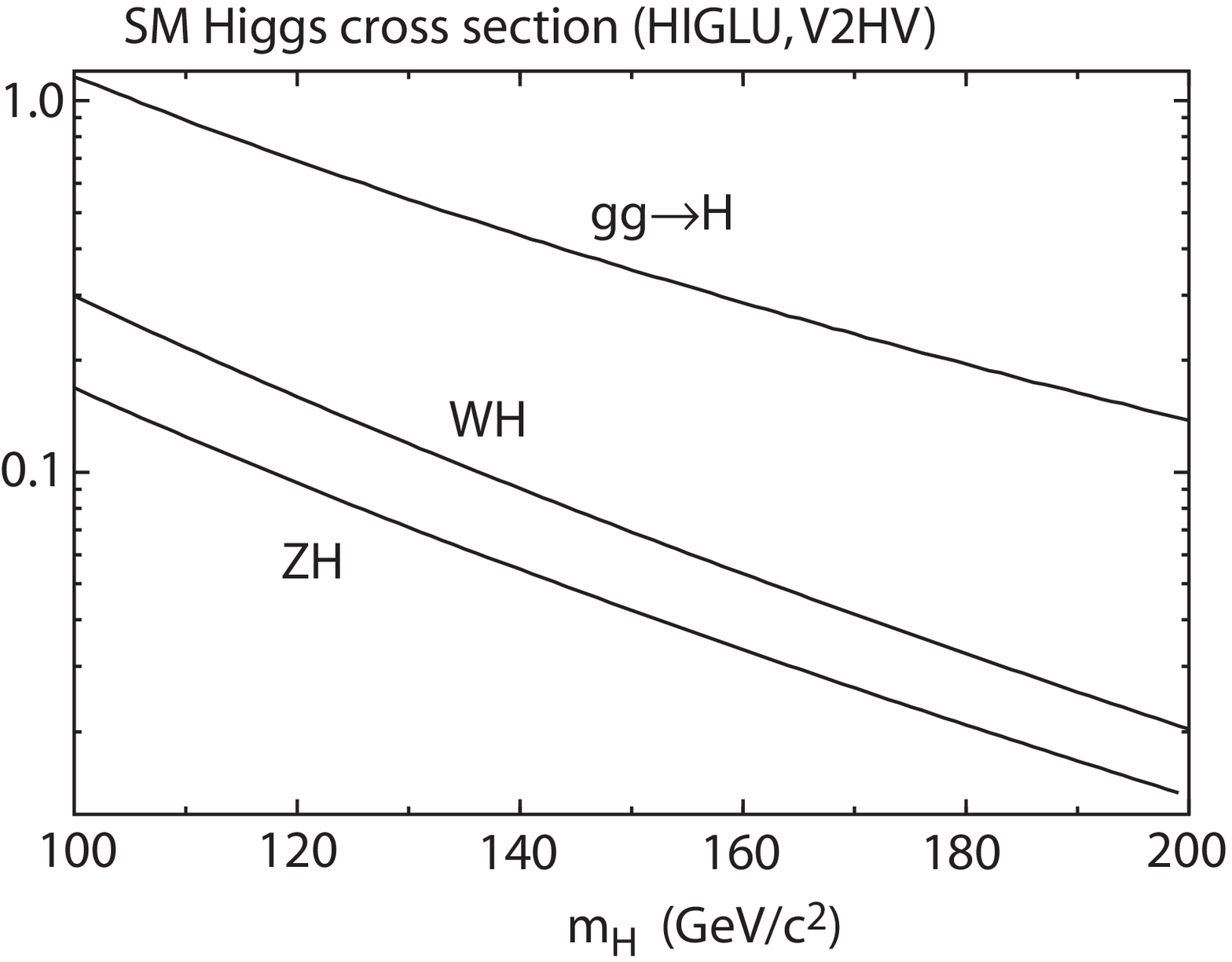,width=7.0cm}
\end{center}
\end{figure}
\vskip-0.5cm
\begin{figure}
\begin{center}
\psfig{file=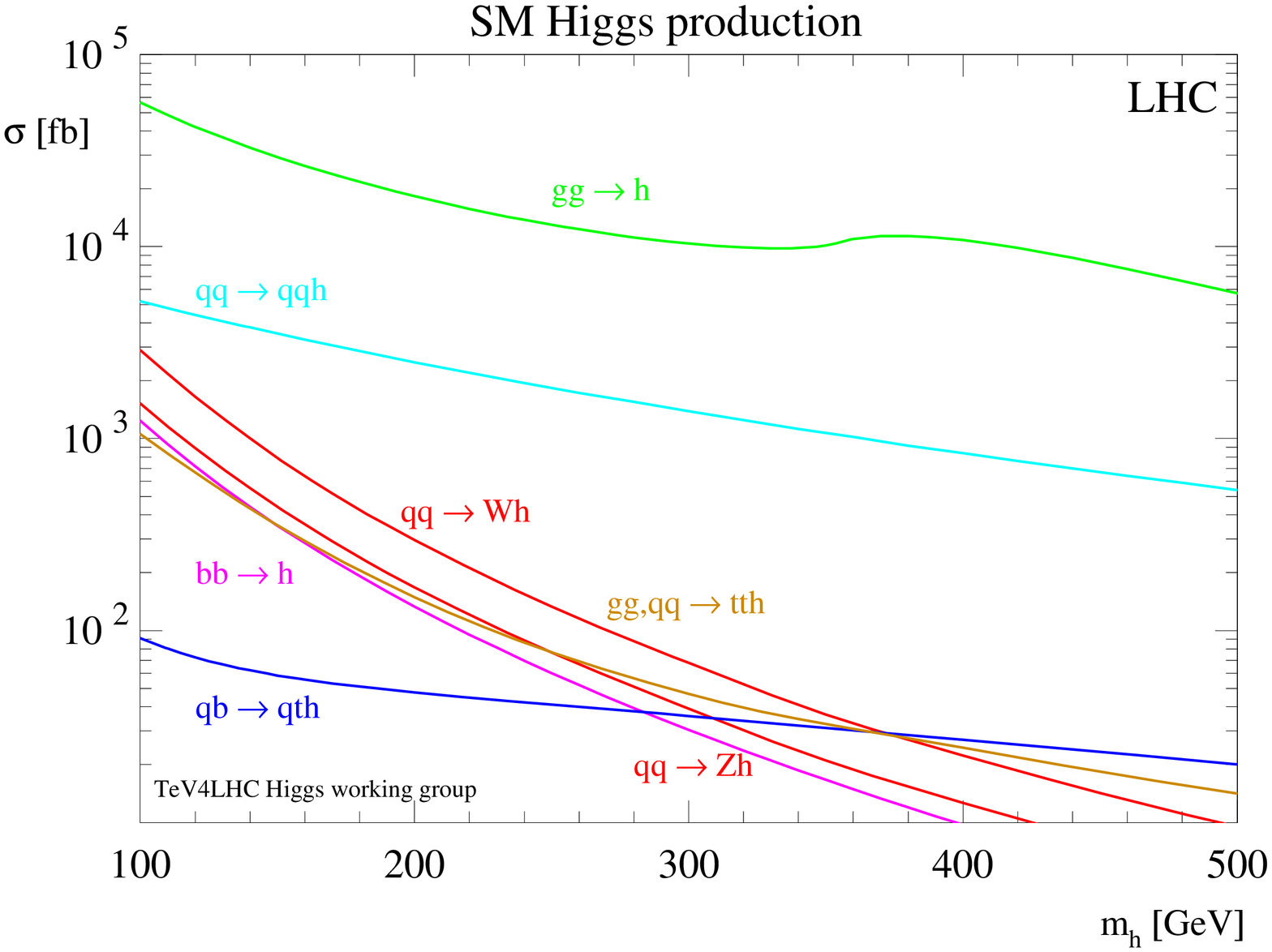,width=5.5cm,angle=-90}
\end{center}
\caption{SM Higgs production cross section at the Tevatron, in pb (top) and the LHC, in fb (bottom), as a function of the Higgs mass. The VBF cross section at the Tevatron is very small, and is not shown here. The Tevatron plot is from the web site~\cite{conway}, and the LHC plot is from Ref.~\cite{LHC_Higgs}.}
\label{fig:Hxsec}
\end{figure}

The Higgs searches at hadron colliders are somewhat more complicated. Resonant production by quark-antiquark pairs, Fig.~\ref{fig:ppHprod} (a), is again strongly suppressed by weak Yukawas. The dominant process, at both the Tevatron and the LHC, is the one-loop {\it gluon fusion} process shown in Fig.~\ref{fig:ppHprod} (b). However, several other processes also lead to useful signatures: the most important ones are the {\it associated production} with an electroweak gauge boson, Fig.~\ref{fig:ppHprod} (c), and the {\it vector boson fusion (VBF)} reaction, Fig.~\ref{fig:ppHprod} (d). The cross sections for these processes at the Tevatron and the LHC are  shown in Fig.~\ref{fig:Hxsec}. The Tevatron should expect to collect a sizable sample of Higgs bosons for $m_h\lsim 200$ GeV, whereas the LHC at design luminosity should produce plenty of Higgses throughout the theoretically sensible range, $m_h\lsim 1$ TeV. The signature of the Higgs depends crucially on its mass. Roughly speaking, one can divide the parameter space into two regions: the {\it low-mass} region, below about 140 GeV, and the {\it high-mass} region, from 140 GeV to 1 TeV. In the low-mass region, the dominant decay mode is $b\bar{b}$. The problem is that this mode, combined with gluon fusion production, gives a signature of 2 jets, at a rate tiny compared to the SM pure-QCD dijet rate. It is impossible to identify the Higgs events on top of this huge background, and one is forced to focus on either subleading production processes, or subdominant decay modes. At the Tevatron, the most promising strategy is to focus on $h\to b\bar{b}$ but look for associated production events, $Zh$ and $Wh$, with leptonic vector boson decays. The SM backgrounds in these channels, predominantly $Z/W+2$ jets, are much more manageable. At the LHC, where the expected Higgs production rate is much higher, one can concentrate instead on a rare but very clean decay $h\to\gamma\gamma$. In the high-mass region, on the other hand, the dominant decay mode is $h\to WW$ (with one of the $W$s off-shell for $m_h\lsim 160$ GeV), and $h\to ZZ$ also makes a significant contribution for $m_h\gsim 180$ GeV. If the vector boson decays leptonically, these modes do not suffer from large QCD backgrounds, and provide a clean signature. The ``golden" $h\to ZZ \to 4\ell$ mode is especially advantageous since it contains no invisible particles, so the Higgs boson shows up as a clear peak in the 4-lepton invariant mass distribution. Combined with the strong gluon fusion production channel, the vector boson decays make the search in the high-mass region relatively straightforward, giving the Tevatron and especially the LHC sensitivity to high Higgs masses in spite of the decreasing production cross section. 

\begin{figure}
\begin{center}
\psfig{file=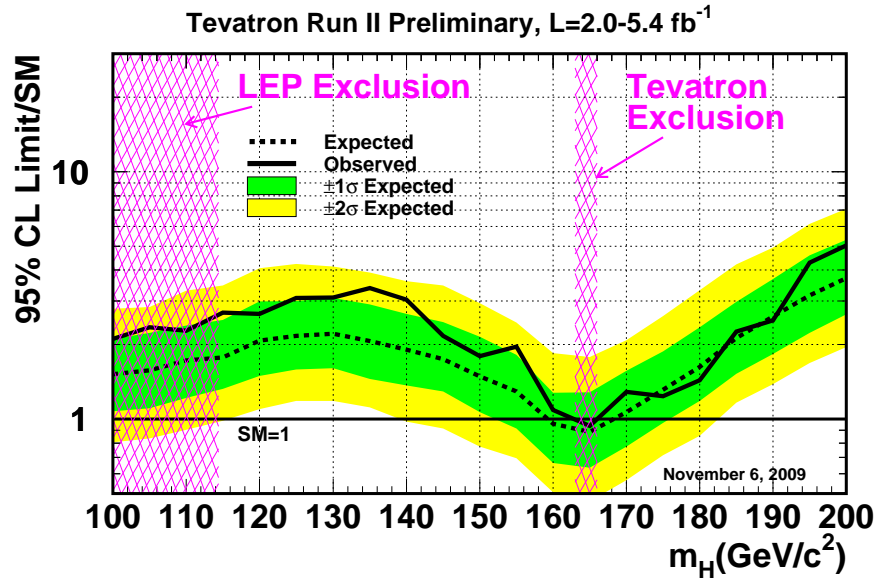,width=13cm}
\end{center}
\caption{Tevatron bounds on the SM Higgs production cross section, as a function of the Higgs mass. The bound combines searches for a variety of signatues (see text for details), at both CDF and D{\O}. From Ref.~\cite{TeV_Hbound}.}
\label{fig:TeV_Hbound}
\end{figure}

The latest published results of the SM Higgs searches at the Tevatron are summarized in Fig.~\ref{fig:TeV_Hbound}. To present the negative results of the searches, the experiments assume that the Higgs production cross section in each channel is a free parameter, and put an upper bound on this cross section. (If several production channels are used in a search, the {\it ratios} of their cross sections are assumed to be the same as in the SM.) This bound, expressed in units of the SM production cross section for each Higgs mass, is plotted on the figure. The SM Higgs can be said to be ruled out if the cross section bound in these units is below 1. Currently, the Tevatron experiments only exclude the SM Higgs at the 95\% c.l. in a narrow mass window just above the $h\to WW$ decay threshold, between 163 and 166 GeV. As the integrated luminosity increases, the Tevatron is expected to expand this window, unless a deviation from the SM is seen. 

\begin{figure}
\begin{center}
\psfig{file=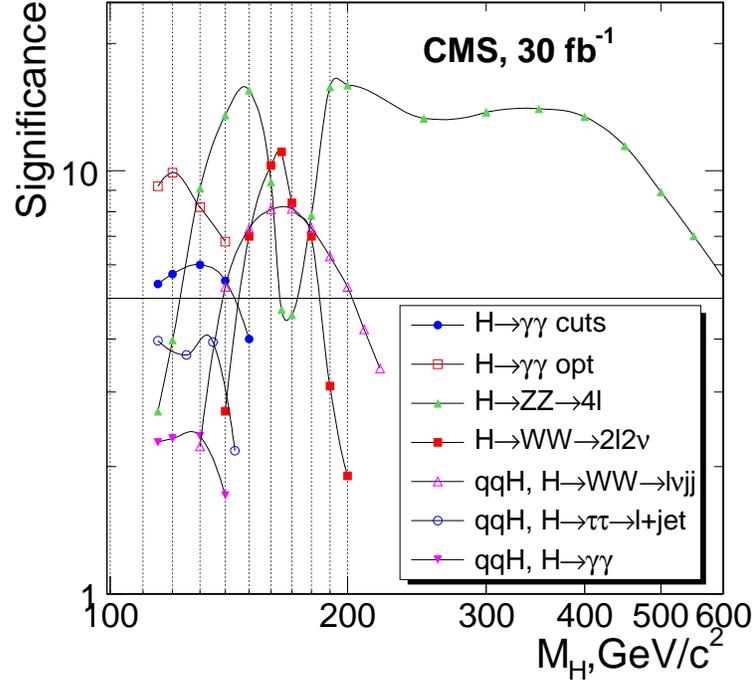,width=10cm}
\end{center}
\caption{Expected sensitivity of the CMS detector at the LHC to the SM Higgs, as a function of the Higgs mass. From Ref.~\cite{CMS_TDR}.}
\label{fig:LHC_Hsens}
\end{figure}

LHC experiments are expected to be sensitive to the entire theoretically consistent range of the Higgs masses at the 5-sigma level with about 30 fb$^{-1}$ of data, see Fig.~\ref{fig:LHC_Hsens}. This corresponds to 3 years of running at design luminosity, although of course depending on the Higgs mass such sensitivity may be achieved much sooner. Note that, just as at the Tevatron, the LHC sensitivity is not a monotonic function of the mass: in fact, a 300-GeV Higgs is substantially easier to discover, via the $h\to ZZ \to 4\ell$ mode, than a 115-GeV Higgs where one must rely on the rare decay $h\to\gamma\gamma$. 

\subsection{Searches for Supersymmetry}

As every TASI student knows, the discovery of the Higgs would mark the beginning, not the end, of the hunt for new physics at the TeV scale. Strong theoretical arguments indicate that the SM Higgs cannot live in isolation, requiring other new particles to stabilize its mass and complete the electroweak symmetry breaking sector. The nature of these new particles can at present only be guessed at. One popular guess is that physics at the TeV scale is {\it supersymmetric}: for each particle in the SM, a new ``superpartner" particle appears at this scale, with the same quantum numbers but spin different by $\hbar/2$. Appearing in loops, these particles would cancel the notorious quadratic divergence of the Higgs mass parameter present in the SM, restoring naturalness to the Higgs sector. I will not have the opportunity for a thorough review of the model here; a reader in need of such a review is referred to the set of lectures by P.~Fox at this school, as well to the excellent review article by Martin~\cite{Martin} and the comprehensive book by Baer and Tata~\cite{BT}. Instead, I will simply very briefly mention the features of the model crucial for collider searches.

\renewcommand{\arraystretch}{1.4}
\begin{table}
\tbl{The undiscovered particles in the Minimal Supersymmetric Standard Model. From Ref.~\cite{Martin}.}
{\begin{tabular}{|c|c|c|c|c|}
\hline
Names & Spin & $P_R$ & Gauge Eigenstates & Mass Eigenstates \\
\hline\hline
Higgs bosons & 0 & $+1$ & 
$H_u^0\>\> H_d^0\>\> H_u^+ \>\> H_d^-$ 
& 
$h^0\>\> H^0\>\> A^0 \>\> H^\pm$
\\ \hline
& & &${\stilde u}_L\>\> {\stilde u}_R\>\> \stilde d_L\>\> \stilde d_R$&(same)
\\
squarks& 0&$-1$& ${\stilde s}_L\>\> {\stilde s}_R\>\> \stilde c_L\>\>
\stilde c_R$& (same) \\
& & &
$\stilde t_L \>\>\stilde t_R \>\>\stilde b_L\>\> \stilde b_R$ 
&
${\stilde t}_1\>\> {\stilde t}_2\>\> \stilde b_1\>\> \stilde b_2$
\\ \hline
& & &${\stilde e}_L\>\> {\stilde e}_R \>\>\stilde \nu_e$&(same) 
\\
sleptons& 0&$-1$&${\stilde \mu}_L\>\>{\stilde \mu}_R\>\>\stilde\nu_\mu$&(same)
\\
& & &
$\stilde \tau_L\>\> \stilde \tau_R \>\>\stilde \nu_\tau$ 
&
${\stilde \tau}_1 \>\>{\stilde \tau}_2 \>\>\stilde \nu_\tau$
\\
\hline
neutralinos & $1/2$&$-1$ & 
$\stilde B^0 \>\>\>\stilde W^0\>\>\> \stilde H_u^0\>\>\> \stilde H_d^0$   
&
$\stilde N_1\>\> \stilde N_2 \>\>\stilde N_3\>\> \stilde N_4$ 
\\
\hline
charginos & $1/2$&$-1$ & 
$\stilde W^\pm\>\>\> \stilde H_u^+ \>\>\>\stilde H_d^-$ 
&
$\stilde C_1^\pm\>\>\>\stilde C_2^\pm $ 
\\
\hline
gluino & $1/2$&$-1$ &$\stilde g$  &(same) \\
\hline
$\frac{{\rm goldstino}}{{\rm (gravitino)}}$ & $\frac{1/2}{(3/2)}$&$-1$&$\stilde 
G$  &(same) \\
\hline
\end{tabular} }
\label{tab:MSSM_particles}
\end{table}
\renewcommand{\arraystretch}{1.0}

The particle content at the TeV scale predicted by the Minimal Supersymmetric Standard Model (MSSM) is summarized in Table~\ref{tab:MSSM_particles}. There are 34 new particles waiting to be discovered! Notice the appearance of 3 additional Higgs boson states not present in the SM; this is due to the presence of a second Higgs doublet field, required in the MSSM for theoretical consistency. All other listed particles are simply superpartners of known SM states. If supersymmetry were an exact symmetry of nature, all superpartners would have exactly the same mass as their SM counterparts; of course, this possibility is already completely ruled out by data. Thus, SUSY must be broken, lifting the superpartner masses into the TeV domain. (Larger masses would render SUSY irrelevant for solving the hierarchy problem.) Many models have been proposed to describe SUSY breaking (some promising directions were described by P.~Meade and D.~Shih at this school). However, despite over two decades of intense theoretical work, no completely compelling model has yet emerged. In the absence of clear theoretical predictions, a conservative approach is to simply consider a completely general SUSY breaking Lagrangian, imposing only the restriction that SUSY breaking must be ``soft", to not interfere with the SUSY solution to the gauge hierarchy problem. Such general soft Lagrangian contains about 100 free parameters, including all superpartner masses, as well as certain trilinear couplings among scalar superpartners (so-called ``A-terms"). The superabundance of free parameters is a curse on the study of SUSY signatures at colliders: Search strategies should ideally be designed that cover as much of the huge parameter space as possible, and, equally daunting, the results of each search should be presented as exclusion regions in this space. In practice, this challenge has not been met: most searches are planned and executed in terms of simplified models that impose relations among parameters, with the 5-parameter {\it ``minimal supergravity",} or mSUGRA, model being the most common choice.  

In spite of the many free parameters, it is possible to identify generic qualitative features of collider events with superparticles. To avoid rapid proton decay, the MSSM postulates an additional discrete symmetry, the R-parity. This is a $Z_2$ symmetry, {\it i.e.} any physical particle is either odd or even under it. Every term in the Lagrangian must be even, so that an even number of R-odd particles must appear in each interaction vertex predicted by the theory. All known SM particles, as well as all Higgs scalars, are postulated to be R-even. The R-parity is embedded in the SUSY algebra in such a way that superpartners must have opposite R-charges; in other words, all superpartners of the SM particles are odd. This has two consequences, both crucial for collider searches:

\begin{itemize}

\item Superpartners must be {\it pair-produced} in collisions of SM particles.

\item The lightest superpartner (LSP) must be {\it stable}, and any other superpartner must decay to the LSP plus (any number of) R-even particles, either SM or extra Higgses. 

\end{itemize}

Strong cosmological bounds on electrically charged and colored stable particles indicate that, if the MSSM is indeed realized in nature, the LSP should be one of the uncharged, uncolored particles of the model. There are three possibilities: the lightest neutralino $\tilde{N}_1$, the lightest sneutrino $\tilde{\nu}$, or the gravitino $\tilde{G}$. Of the three, the neutralino is an especially interesting possibility, since it can easily have the correct relic abundance to explain the observed dark matter, without conflict with direct dark matter search experiments. (The neutralino also naturally emerges as the LSP in several popular models of SUSY breaking, including mSUGRA.)
All these particles interact with ordinary matter too weakly to leave a trace in collider detectors, and so they would appear invisible, like the SM neutrino. Thus, the generic signature of SUSY at colliders is the excess of events with ``missing momentum", accompanied by the SM particles from decays of the originally produced superpartners to the LSP\footnote{One should keep in mind that if the LSP is the gravitino, the next-to-lightest superpartner (nLSP) may be sufficiently long-lived to decay outside the detector. If the nLSP is charged or colored, in this scenario one would look for exotic heavy particles leaving traces in the detector, instead of the usual missing energy signature.}. This qualitative prediction depends only on two basic assumptions --exactly conserved R-parity and neutral LSP -- and is independent of the precise values of the model parameters. Thus, quite generally, an inclusive search for events with missing energy in association with jets and/or charged leptons is the best model-independent search strategy.

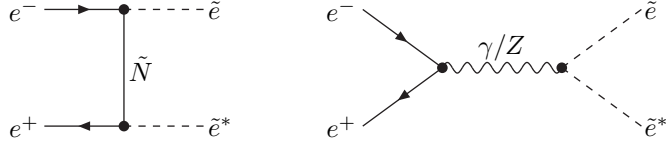
\begin{figure}
\begin{center}
\begin{picture}(265,70)(0,0)
\SetOffset(20,-55)
\ArrowLine(1,109)(31,109)
\ArrowLine(31,65)(1,65)
\Line(31,65)(31,109)
\DashLine(31,109)(61,109){3}
\Vertex(31,109){2}
\DashLine(31,65)(61,65){3}
\Vertex(31,65){2}
\Text(-1,109)[cr]{$e^-$}
\Text(-1,65)[cr]{$e^+$}
\Text(63,65)[cl]{$\tilde{e}^*$}
\Text(63,109)[cl]{$\tilde{e}$}
\Text(33,87)[cl]{$\tilde{N}$}

\ArrowLine(121,109)(151,87)
\ArrowLine(151,87)(121,65)
\Photon(151,87)(196,87){2}{6}
\DashLine(196,87)(226,109){3}
\DashLine(226,65)(196,87){3}
\Vertex(151,87){2}
\Vertex(196,87){2}
\Text(119,109)[cr]{$e^-$}
\Text(119,65)[cr]{$e^+$}
\Text(228,109)[cl]{$\tilde{e}$}
\Text(228,65)[cl]{$\tilde{e}^*$}
\Text(174,100)[tc]{$\gamma/Z$}

\end{picture}

\end{center}
\caption{Leading-order diagrams for the MSSM scalar electron production in $e^+e^-$ collisions.}
\label{fig:FD_sel}
\end{figure}

\begin{figure}
\begin{center}
\psfig{file=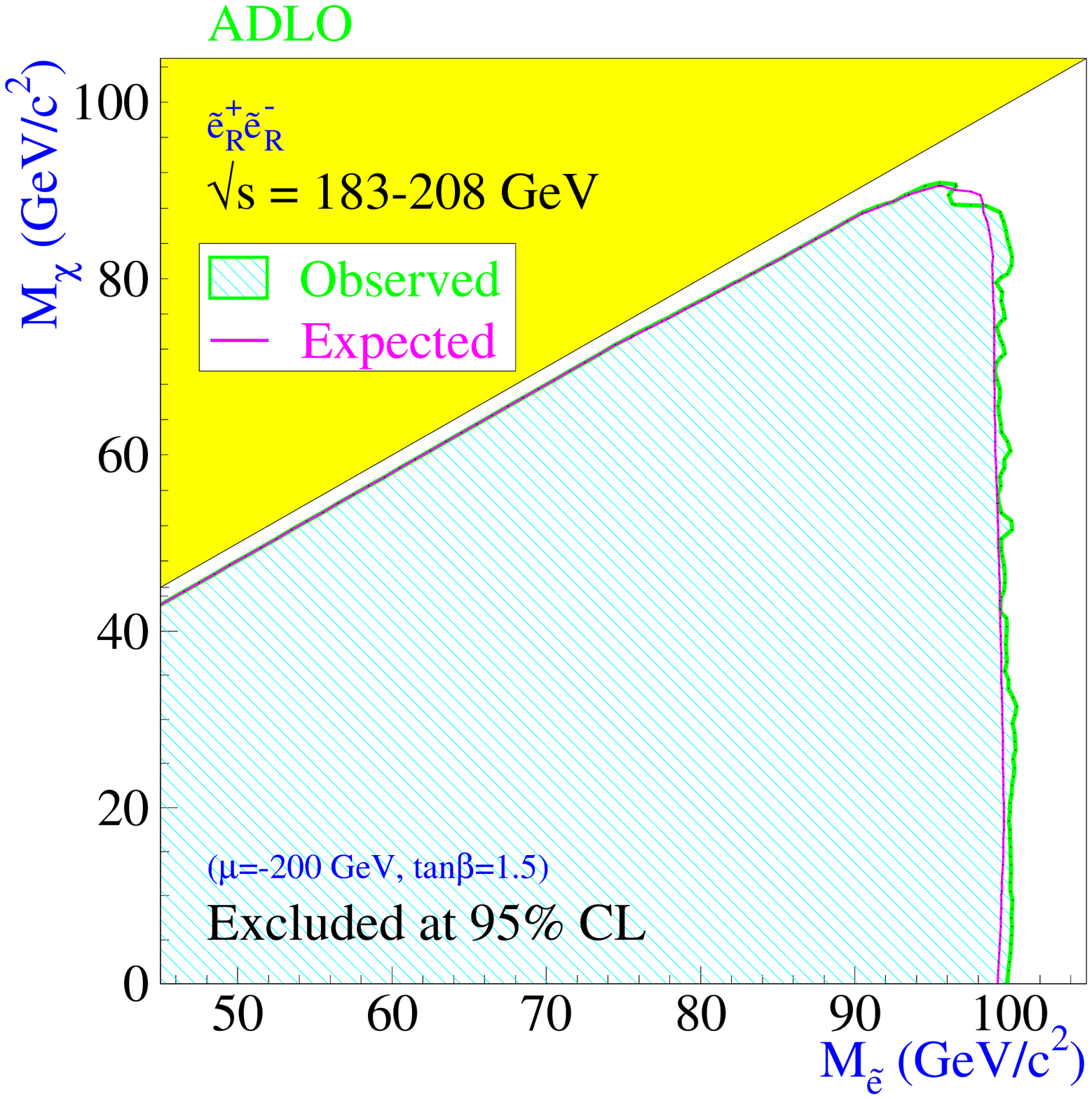,width=7cm}
\psfig{file=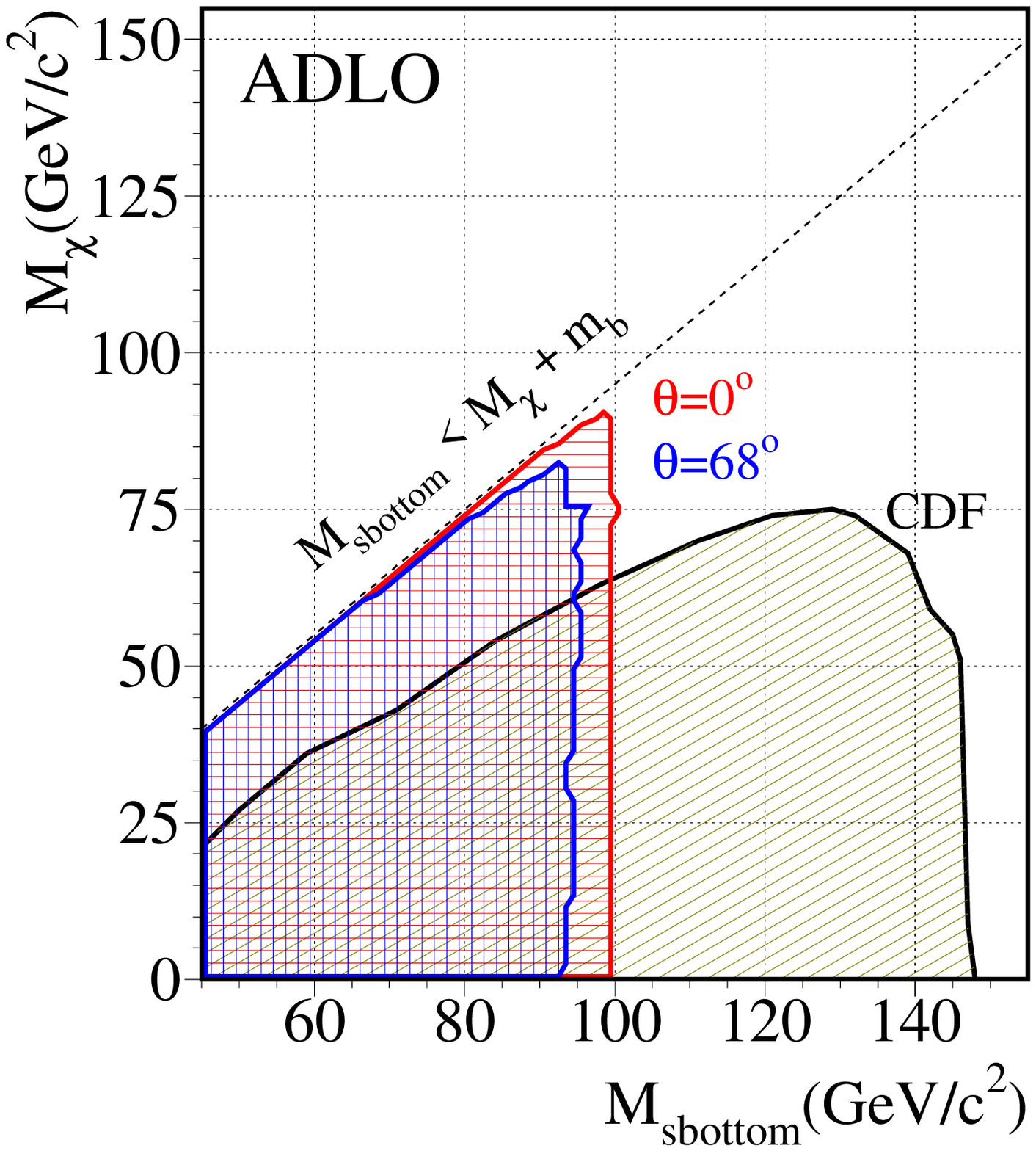,width=7cm}
\end{center}
\caption{LEP-2 bounds on the selectron mass (left) and the sbottom mass (right), plotted vs. the neutralino LSP mass. From the LEP2 Joint SUSY Working Group web site~\cite{LEP2_SUSY}.}
\label{fig:LEP2_susy_mind}
\end{figure}

In $e^+e^-$ collisions, sleptons, squarks and charginos can be produced via $s$-channel $\gamma^*/Z$ exchanges. In addition, selectrons, charginos and netralinos can be produced via $t$-channel exchange of neutralinos, sneutrinos and selectrons, respectively. As an example, consider the selectron search. The Feynman diagrams for selectron production are shown in Fig.~\ref{fig:FD_sel}. Produced selectrons can decay directly to electrons and the lightest neutralino, assumed to be the LSP. This leads to the $e^+e^-+$missing energy final state, which can be easily identified at LEP. The SM backgrounds in this channel are not large, and the search is sensitive as long as the selectron production cross section is significant. Kinematically, selectron pair production is possible if $2m(\tilde{e})< \sqrt{s}$; this means that LEP-2 could in principle search for selectrons up to about 104 GeV. In practice, as usual, the bound is somewhat lower:
\beq
m(\tilde{e})\geq 100~{\rm GeV}\,.
\eeq{LEP_sel_bound}
However, there is a small caveat.
The search requires that the electrons and positrons from selectron decays be identified in the detector. If the selectron mass is close to the LSP mass, these particles are very soft and cannot be registered, so the search sensitivity is lost for $m(\tilde{e})-m(\tilde{N}_1)\lsim 1$ GeV. The region of the $m(\tilde{e})/m(\tilde{N}_1)$ plane excluded by the LEP-2 experiments is shown in Fig.~\ref{fig:LEP2_susy_mind}. Note that this plot is quite robust with respect to variations of the other MSSM parameters, since the bound is primarily determined by the kinematics of selectron production and decay. The searches for all other electrically charged superpartners (sleptons, squarks, and charginos) proceed in a similar way, and all yield bounds close to 100 GeV. For example, the LEP-2 bound on the sbottom mass (along with, for comparison, the CDF bound on the same parameter) is shown in Fig.~\ref{fig:LEP2_susy_mind}. 

\begin{figure}
\begin{center}
\psfig{file=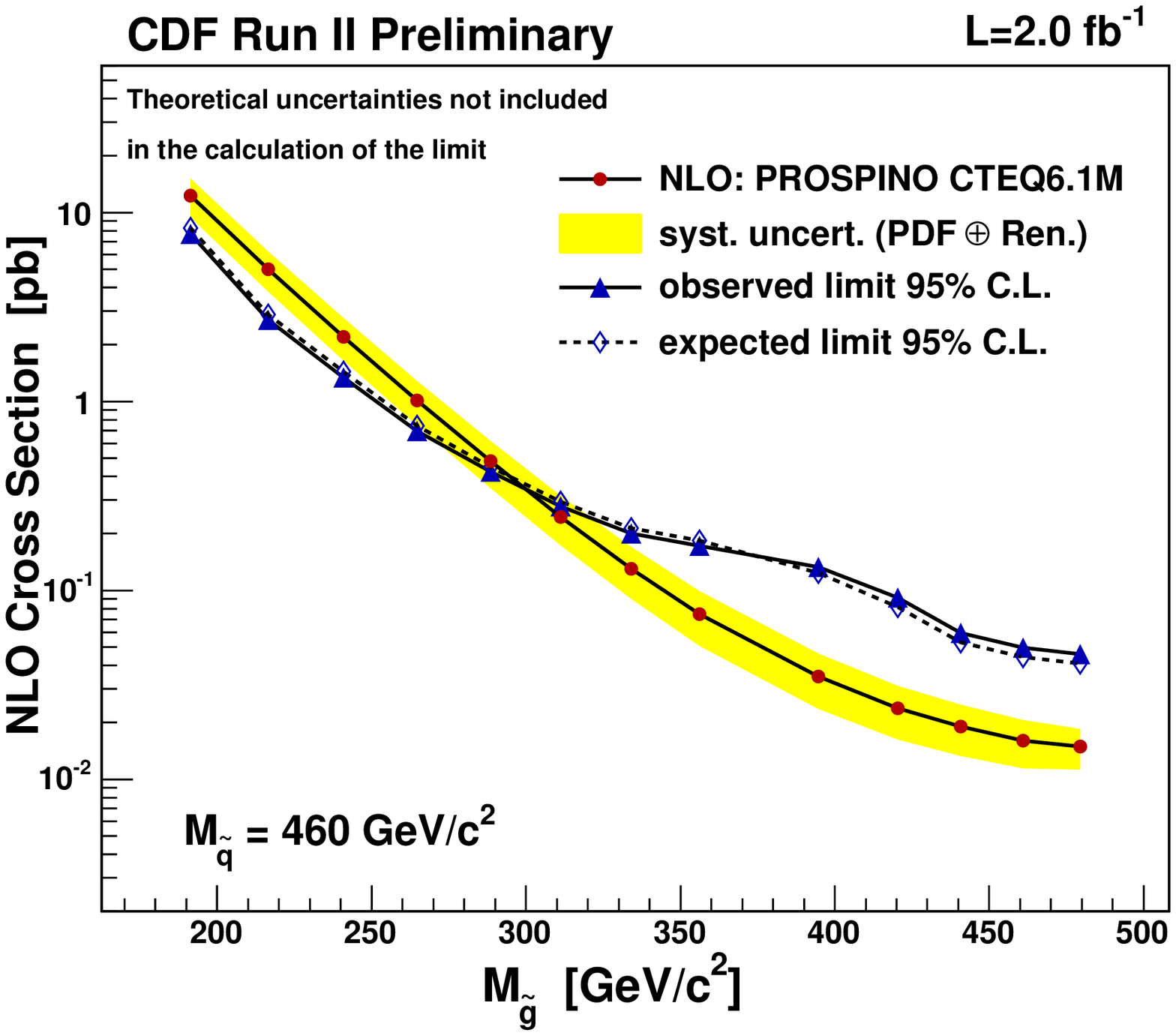,width=8cm}
\psfig{file=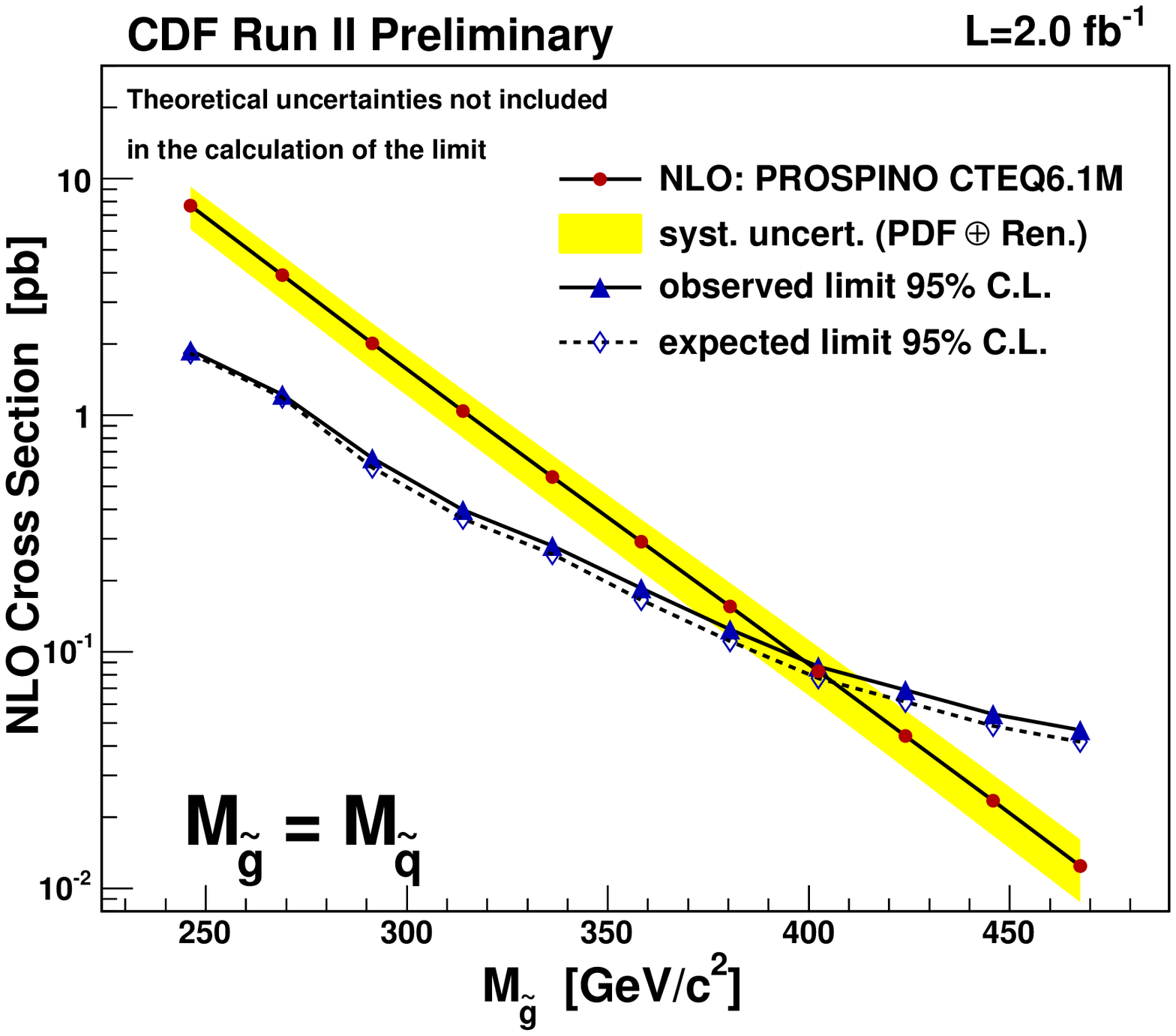,width=8cm}
\end{center}
\caption{Yellow (shaded) bands: Theoretical cross section for total squark/gluino production at the Tevatron, as a function of the gluino mass, assuming squark mass of 460 GeV (left panel), or equal squark and gluino masses (right panel). Triangles: the upper bound on the cross section observed by the CDF collaboration at the Tevatron. The bound assumes that squark/gluino decays follow the pattern predicted by the mSUGRA model. From the web page~\cite{TeV_susy_web} (see also Ref.~\cite{TeV_susy}).}
\label{fig:TeV_susy_xsec}
\end{figure}

\begin{figure}
\begin{center}
\psfig{file=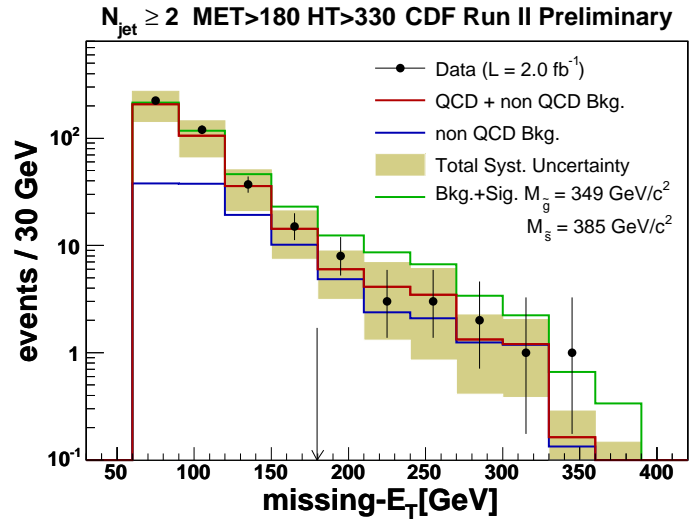,width=10cm}
\end{center}
\caption{The MET distribution of events with $\geq 2$ jets and MET. Solid line: SM prediction. Dashed line: SM+SUSY (mSUGRA) prediction for 349 GeV gluino and 385 GeV squark. Data points: measurement by the CDF collaboration at the Tevatron. From the web page~\cite{TeV_susy_web} (see also Ref.~\cite{TeV_susy}).}
\label{fig:TeV_susy_MET}
\end{figure}

\begin{figure}
\begin{center}
\psfig{file=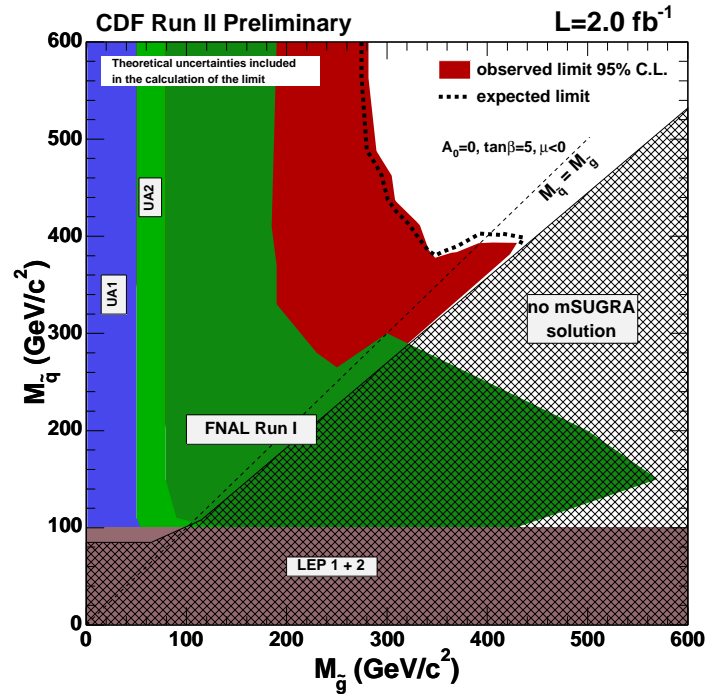,width=10cm}
\end{center}
\caption{Exclusion plane at 95\% c.l. as a function of squark and gluino masses in an mSUGRA scenario, reported by the CDF collaboration at the Tevatron. From the web page~\cite{TeV_susy_web} (see also Ref.~\cite{TeV_susy}).}
\label{fig:TeV_susy_massbound}
\end{figure}

At a hadron collider, the dominant superpartner production process is the production of squarks and gluinos via strong interactions: $\tilde{q}\tilde{q}^*, \tilde{q}\tilde{q}, \tilde{q}^*\tilde{q}^*, \tilde{q}\tilde{g}, \tilde{q}^*\tilde{g}$, and $\tilde{g}\tilde{g}$ final states are all possible. The cross sections for these reactions depend only on squark and gluino masses, and thus can be robustly predicted. The combined cross section for all squark/gluino production channels at the Tevatron is shown by the yellow bands in Fig.~\ref{fig:TeV_susy_xsec}. A large sample of squark and gluinos could in principle be produced at the Tevatron: for example, for $m(\tilde{g})=200$ GeV, the current data sample would contain of order $10^5$ such particles. As usual, the produced squarks and gluinos would decay promptly; the simplest decays are $\tilde{q}\to q\tilde{N}$ and $\tilde{g}\to q\tilde{q}\tilde{N}$, resulting in 2-jet, 3-jet, and 4-jet final states associated with missing transverse momentum. (In this context, missing transverse momentum is often referred to, somewhat misleadingly, as ``missing transverse energy", or MET. I will reluctantly follow this practice.) The SM background to this signature primarily comes from the reaction $Z+n$ jets, with invisible $Z$ decays, $Z\to\nu\bar{\nu}$. In addition, instrumental backgrounds, from pure-QCD multi-jet events where apparent MET results from mis-measurement of one or more of the jets, is also important due to very large pure-QCD event rates. All backgrounds, however, fall much faster than the expected SUSY signal with increased MET, as shown in Fig.~\ref{fig:TeV_susy_MET}. (The amount of MET in SUSY events is controlled by the squark-neutralino or gluino-neutralino mass difference, of order few hundred GeV; the backgrounds peak at 0.) Imposing a lower cut on MET in the analysis improves sensitivity to the SUSY signal. Optimizing this cut, the CDF collaboration~\cite{TeV_susy} obtained the bounds on the squark/gluino production cross section shown in Fig.~\ref{fig:TeV_susy_xsec}. This was translated into a bound on squark and gluino masses, Fig.~\ref{fig:TeV_susy_massbound}. 
Roughly speaking, gluinos below 300 GeV, and squarks below 350 GeV, are excluded in this scenario. However, note that while the squark and gluino production cross sections are rather robust, their decay chains, and the kinematics of the decay products, depend on numerous MSSM parameters beyond their masses: most notably, the LSP mass is important, since it controls both the amount of MET and the typical jet momenta in SUSY events. The CDF analysis~\cite{TeV_susy} assumed mSUGRA relations between the MSSM parameters; this implies, among other things, the ratio of the LSP and gluino masses of about 1:7. The cuts were optimized with this assumption. It should be noted that, if the MSSM is correct but the mSUGRA is not, such optimization could result in a loss of sensitivity: in other words, it is possible that the data contains a hint for SUSY that would be missed by this analysis~\cite{Jay}. Such considerations should be taken into account when designing future SUSY searches.

\begin{figure}
\begin{center}
\psfig{file=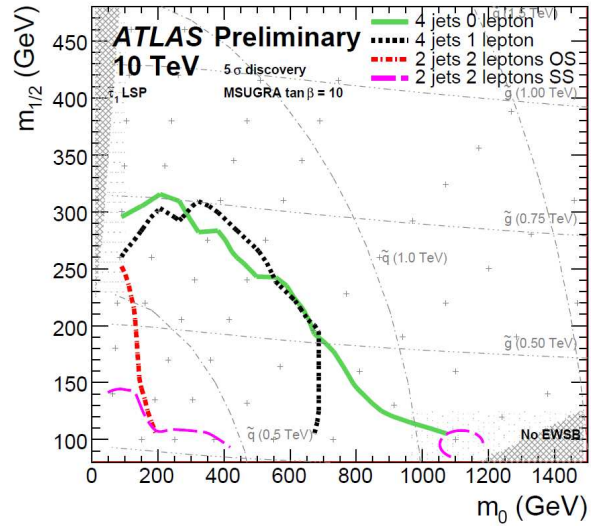,width=8cm}
\end{center}
\caption{Discovery reach for SUSY projected by the ATLAS collaboration at the LHC. The mSUGRA model is assumed, and the reach is shown in terms of the $m_0$ and $m_{1/2}$ parameters, with the other parameters fixed: $\tan\beta=10$, $A_0=0$, $\mu>0$. The reach is shown for $n$-jet+MET signatures. Integrated luminosity of 200 pb$^{-1}$ (1/50 of a year at design luminosity) is assumed. From Ref.~\cite{LHC_susy}.}
\label{fig:LHC_susy}
\end{figure}

The same search can be repeated at the LHC. It will benefit from a much larger production cross section: inclusive SUSY cross sections in the tens of pb range (or 100,000's events/year at design luminosity) are possible in parameter regions not already ruled out by the Tevatron. The sensitivity of the search projected by the ATLAS collaboration is shown in Fig.~\ref{fig:LHC_susy}. Even with only 200 pb$^{-1}$ of data, the LHC will dramatically increase the reach of the Tevatron, so very early discoveries are possible. (It should be kept in mind, however, that unambiguous measurement of MET requires that all potential instrumental sources of MET, {\it e.g.} cracks in the detector, mis-calibrated calorimeter cells, etc., must be identified first. This task may require substantial additional amount of data.) 
Eventually, the LHC is expected to cover the entire range where SUSY provides an attractive solution to the hierarchy problem: If it is not discovered, then either electroweak symmetry breaking is severely fine-tuned, or some other mechanism for stabilizing the Higgs mass must be found.

\section{Further Reading}
\label{readings}

Collider physics is a vast subject, and in four lectures it is only possible to touch the tip of the iceberg. Luckily, a variety of excellent resources is available to students wishing to delve deeper into the subject. The popular Quantum Field Theory textbook by Peskin and Shroeder~\cite{PS}, which I referred to throughout these lectures, contains an insightful discussion of the key theoretical concepts crucial to the field. Sadly, some of these concepts are often not given sufficient emphasis in a typical one-year QFT course: Infrared divergences (chapter 6) and evolution equations (section 17.5) are two examples that come to mind. I tried to give a flavor of these topics in these lectures, but interested students are strongly encouraged to study them further. At a slightly more advanced level, the text by Ellis, Stirling and Webber~\cite{ESW} contains a wealth of material on applications of QCD to $e^+e^-$ and hadron collisions, as well as on electroweak processes and Higgs searches at hadron colliders. The text by Barger and Philips~\cite{BP} covers similar subjects, in addition to collider phenomenology of a few extensions of the SM. For readers primarily interested in physics beyond the SM, the book by Baer and Tata~\cite{BT} provides a comprehensive discussion of SUSY searches. Another useful source is the recent review article by Feng {\it et. al.}~\cite{Feng}. Pedagogical introduction to collider phenomenology of non-SUSY extensions of the SM is harder to find, although there are useful review articles on at least some of the popular models (e.g.~models of dynamical electroweak symmetry breaking~\cite{DEWSB_review}, Little Higgs~\cite{LH_review}, and Randall-Sundrum and Higgsless models~\cite{RS_review}). 
In addition to theoretical work, anyone interested in the subject ought to pay attention to results coming out of the currently running Tevatron experiments. Both CDF and D{\O} collaborations maintain web pages~\cite{CDF_web,D0_web} that provide easy and convenient access to both published and preliminary public results, grouped by subject. For the LHC, ATLAS and CMS collaborations have published Technical Design Reports (TDRs)~\cite{ATLAS_TDR,CMS_TDR}, which describe detailed physics studies performed by the collaborations to assess their potential for a large number of measurements and searches. For a theory student, studying these Reports is an excellent preparation for understanding the LHC data.  

{\it Note Added:}~~In the Fall semester of 2009, I taught a class on Collider Physics at Cornell, which covered many of the topics discussed here in much more detail. All (handwritten) lecture notes can be downloaded from the course web page~\cite{P7661_web}.

\end{document}